\documentclass{aa}  

\usepackage{graphicx}
\usepackage{txfonts}
\usepackage{graphicx}	
\usepackage{amsmath,thmtools}	
\usepackage{amssymb}	
\usepackage{xspace} 
\usepackage{epstopdf}
\usepackage{hyperref}
\usepackage{enumitem}
\usepackage{xcolor}
\usepackage{soul}
\sethlcolor{red}
\usepackage{comment}
\usepackage{layouts}
\usepackage{placeins}
\usepackage{float}
\usepackage{afterpage}
\usepackage{booktabs}
\usepackage{multirow}
\usepackage{orcidlink}

\newcommand{\msun}{\hbox{$\hbox{\rm M}_{\odot}$}\xspace}

\newcommand{\lya}{Ly$\alpha$\xspace}
\newcommand\T{\rule{0pt}{2.6ex}}       
\newcommand\B{\rule[-1.2ex]{0pt}{0pt}} 

\begin{document} 

   \title{QSO MUSEUM III: the circumgalactic medium in Ly$\alpha$ emission around 120 $z\sim3$ quasars covering the SDSS parameter space}
   \subtitle{Witnessing the instantaneous AGN feedback on halo scales}
   \author{Jay Gonz\'alez Lobos\inst{1,2}\orcidlink{0000-0002-6056-3425}
          \and
          Fabrizio Arrigoni Battaia\inst{1}
          \orcidlink{0000-0002-4770-6137}
          \and
          Aura Obreja\inst{1,3,4}
          \orcidlink{0000-0003-4196-8555}
          \and
          Guinevere Kauffmann\inst{1}
          \and
          Emanuele Paolo Farina\inst{5}\orcidlink{0000-0002-6822-2254}
          \and
          Tiago Costa\inst{6}\orcidlink{0000-0002-6748-2900}
          }
          \titlerunning{QSO MUSEUM III}

   \institute{
              Max-Planck-Institut f\"ur Astrophysik, Karl-Schwarzschild-Str 1, 85748 Garching bei M\"unchen, Germany\\
              \email{jagonzalez@mpia.de}
              \and
              Max-Planck-Institut f\"ur Astronomie, K\"onigstuhl 17,
              69117 Heidelberg, Germany
              \and
              Interdisziplin\"ares Zentrum f\"ur Wissenschaftliches Rechnen, Universit\"at Heidelberg, Im Neuenheimer Feld 205, D-69120 Heidelberg, Germany
             \and
             Zentrum für Astronomie, Institut f\"ur Theoretische Astrophysik, Universit\"at Heidelberg, Albert-Ueberle-Straße 2, D-69120 Heidelberg, Germany
            \and
            International Gemini Observatory/NSF NOIRLab, 670 N A’ohoku Place, Hilo, Hawai'i 96720, USA
            \and 
            School of Mathematics, Statistics and Physics, Newcastle University, Newcastle upon Tyne, NE1 7RU, UK
             }


  \abstract
   {Recent surveys have shown that $z\gtrsim2$ quasars are surrounded by Hydrogen Lyman-$\alpha$ (Ly$\alpha$) glows with diverse emission levels and extents. These characteristics seem to depend on the activity of embedded quasars, on the number of active galactic nucleus (AGN) photons able to reach the halo gas or circumgalactic medium (CGM) and the physical properties of the CGM. In this framework, we present VLT/MUSE snapshot observations ($45$~minutes/source) of 
   59 $z\sim3$ quasars extending the long-term QSO MUSEUM campaign to the fainter SDSS sources. 
   The whole survey -- homogeneously reduced and analyzed here -- now targets 120 quasars with a median redshift $z=3.13$, and bolometric luminosities, black hole masses and Eddington ratios in the ranges $45.1 < {\rm log}(L_{\rm bol}/[{\rm erg\ s^{-1}]}) < 48.7$, $7.9 < {\rm log}(M_{\rm BH}/[{\rm M_{\odot}]}) < 10.3 $ and $0.01< \lambda_{\rm Edd} <1.8$, respectively. We detected extended Ly$\alpha$ emission in 110/120 systems, with all the non-detections in the newly added fainter sample. Indeed, the surface brightness of the CGM Ly$\alpha$ emission (SB$_{\rm Ly\alpha}$) increases with quasar luminosity. Stacking of the non-detections unveils emission just below our individual field detection limit. Moreover, the Ly$\alpha$ linewidth increases in the central regions (projected radius $R<40$~kpc or $\sim40$~\% $R_{\rm vir}$) of the CGM around quasars with stronger radiation. Both the variation in surface brightness and velocity dispersion as a function of quasar luminosity indicate that we are witnessing the instantaneous AGN feedback in action on CGM scales. 
   Assuming that all targeted quasars sit in halos of $M_{\rm DM}\sim10^{12.5}\,M_\odot$ independent of luminosity, as suggested by clustering studies, 
   the trend in SB$_{\rm Ly\alpha}$ can be naturally explained by a larger fraction of cool gas mass being illuminated, implying that brighter quasars have larger ionization cone opening angles. Similarly, brighter AGN seem to perturb the cool ($T\sim10^4$~K) gas more strongly. 
   We show that QSO MUSEUM starts to have enough statistics to study this instantaneous AGN feedback while controlling for black hole properties (e.g., mass), which will be key to constraining AGN models. 
   } 
    
   \keywords{quasars:emission lines -- quasars:general -- galaxies:halos -- galaxies:high redshift -- (galaxies:)intergalactic medium
               }

   \maketitle

\section{Introduction}\label{sec:intro}

A large fraction of the baryons in the universe are thought to reside in regions between the interstellar medium of galaxies and their host dark-matter halo virial radius. This material, currently referred to as the CGM 
(\citealt{Tumlinson2017}), has been determined to be multiphase, including cold (10-100~K; e.g., \citealt{Emonts2016,Vidal-Garcia2021,Emonts2023}), cool ($\sim10^4$~K; e.g., \citealt{Werk2014,Wisotzki2016,Nateghi2024}) and warm/hot ($>10^5$~K; e.g., \citealt{Predehl2020,DiMascolo2023,Zhang2024}) gas. The importance of the CGM in galaxy evolution has become clearer in the past two decades and its phases are currently under intense scrutiny (\citealt{FG-Oh2023}). Indeed, the CGM stores information on the complex interplay of several processes that happen during and successively shape galaxy formation and evolution, including gas inflows from the larger scales of the intergalactic medium (IGM; e.g., \citealt{Keres2005,Decataldo2024}), galactic or active galactic nucleus (AGN) winds/outflows (e.g., \citealt{Springel2005,Stinson2006,Wright2024}), radiation (e.g., \citealt{Ciotti1997,Obreja2019}) and interactions with and gas stripping of satellite galaxies 
(e.g., \citealt{Hopkins2006,AnglesAlcazar2017}).

The CGM of AGN is the optimal case study to encompass all of the aforementioned processes. AGN, more specifically those identified by the observation of their broad emission lines, i.e. quasars, are known to reside in relatively massive halos up to $z\sim6$ ($M_{\rm DM}\sim10^{12.5}$~M$_{\odot}$; \citealt{White2012,Timlin2018,Farina2019,Fossati2021,deBeer2023,Costa2024}). This mass range should be characterized by (i) the presence of overdense environments around these systems which could contribute to their growth trough infall and mergers (e.g., \citealt{KauffmannHaehnelt2000}), and (ii) the presence of several satellite galaxies, some of which 
have large star formation rates (SFR$\gtrsim$100$\,{\rm M_\odot\,yr^{-1}}$) as demonstrated by recent observations (e.g., \citealt{Decarli2017,Fossati2021,Chen2021,Bischetti2021,FAB2022,Nowotka2022,FAB2023b}). Also, quasar host galaxies are more massive and more star forming than typical galaxies at the same cosmic time (e.g., \citealt{Walter2009,Pitchford2016,Molina2023}), implying that their CGM and environment need to provide enough material to sustain such activity.

Importantly, quasars, accreting super-massive black holes (SMBHs), have extreme luminosities which are expected to expel, enrich and highly ionize the gas reservoir within the host and its surroundings, a process known as AGN feedback (e.g., \citealt{Fabian2012,King2015}). To reach their exceptional masses ($M_{\rm BH}\sim10^8-10^{10}$~M$_{\odot}$), SMBHs are expected to be fueled in different regimes, including 
major mergers of gas-rich galaxies giving rise to the most extreme black hole growth and assembly (e.g., \citealt{Ni2022}), and 
self-regulated growth mostly at the center of their host galaxies in a cycle of accretion and feedback 
(e.g., \citealt{DiMatteo2005}). Indeed, quasars have been observed to be highly variable (e.g., \citealt{MacLeod2012}), with typical variability timescales of the order of 10$^{5-6}$~years (\citealt{Schawinski2015,Eilers2017}) and ages of $10^6-10^8$~years (\citealt{Martini2004,Khrykin2021}). The quasar's ``shut-down'' phase is loosely constrained to $10^4-10^5$~years based on the geometry and extent of few quasar light echoes and the recombination time scale of narrow line emission (e.g., \citealt{Lintott2009,Schirmer2013,Davies2015}). The impact of quasars on their surrounding reservoir and environment is therefore not only almost instantaneous, but also cumulative (e.g., \citealt{HRA2024}). 

The quasars number density, and therefore their activity is at its apex at $z\sim2$ (e.g., \citealt{Shen2020}). This is fortunate as the quasar's CGM at their peak activity can therefore be probed both in absorption against bright background sources and directly in emission.
Absorption line studies of the CGM of hundreds of $z\sim2-3$ quasars 
revealed the cool ($T\sim10^4$\,K) massive ($M\sim10^{11}\,\msun$) and metal rich ($Z\sim0.5$~Z$_\odot$) gas reservoirs as traced by optically thick absorbers, whose spatial distribution is highly anisotropic likely due to quasar illumination mainly along our line-of-sight \citep[e.g.,][]{Hennawi2007,Prochaska2014,Lau2016}. Such studies have been able to provide information on the average properties of the CGM gas despite 
the sparseness of bright background quasars. On the other hand, detailed information on the morphology, physical properties and kinematics of the CGM around individual quasars requires direct observations. 

The direct detection of the CGM of $z\sim2-3$ quasars has been a long-sought aim since the early predictions of the possible presence of extended glows of Hydrogen \lya emission around AGN (\citealt{Rees1988,HR2001}), which should result due to the quasar illuminating its surrounding distribution of infalling gas. Several spectroscopic and narrow-band studies were successful in unveiling their  
CGM gas out to distances of $<50$\,kpc (e.g., \citealt{HuCowie1987,Weidinger2005,Christensen2006,Hennawi2013,FAB2016}), but were hampered by the shallow sensitivity of past instrumentation as well as the lack of accurate quasar systemic redshifts to gather statistical samples of detections. 
Notwithstanding these difficulties, these pioneering works were able to (i) unveil the Ly$\alpha$ signal out to even $\sim500$~kpc for the most exceptional systems (\citealt{Cantalupo2014,Hennawi2015}), (ii) constrain, in a photoionization scenario, the cool gas emitting Ly$\alpha$ to be dense, and metal enriched in some cases (\citealt{Heckman1991,Heckman1991b}), (iii) show that the kinematics of the CGM are relatively quiescent, possibly dominated by infall (\citealt{Weidinger2004}), and (iv) start to investigate
correlations between nebulae and QSO properties (\citealt{Christensen2006}).

The development of integral field spectrographs, such as the Multi Unit Spectroscopic Explorer \citep[MUSE,][]{Bacon2010} at the Very Large Telescope (VLT) and the Keck Cosmic Web Imager \citep[KCWI,][]{Morrisey2018} at the Keck Observatory revolutionized this field of research by allowing to map with unprecedented surface brightness limits (${\rm SB_{Ly\alpha}\sim10^{-18}\,erg\,s^{-1}\,cm^{-2}\,arcsec^{-2}}$) the CGM of $2<z<6$ bright quasars.  It was therefore possible to uncover a large diversity of \lya nebulae (sample of $\sim100$ at $z\sim3$) with sizes up to $\sim100$\,kpc in only one hour of telescope time per object \citep{Borisova2016,Farina2017,FAB2018,Ginolfi2018,Cai2018,FAB2019,FAB2019b,Farina2019,Cai2019,Travascio2020,Drake2020,Lau2022,GonzalezLobos2023}. 

The origin of the extended \lya emission is subject of debate, even if quasars provide enough ionizing photons to keep in principle the surrounding gas ionized. Indeed, important parameters are frequently not known, including fraction of volume illuminated by each quasar or quasar ionization cones opening angle, geometry of the host galaxy and its position with respect to the quasar ionization cones, presence of winds/ouftlows. Several mechanisms can act together and, through their combination, produce the observed \lya glow: recombination radiation following gas ionization by the quasar radiation (e.g., \citealt{Cantalupo2005,Kollmeier2010,Costa2022}), shocks (e.g., \citealt{Mori2004}), resonant scattering of
\lya photons from the quasar broad line regions (BLRs, e.g., \citealt{Cantalupo2014,Costa2022}), gravitational cooling radiation (e.g., \citealt{Haiman2000,Dijkstra2006}). Furthermore, \lya photons produced on CGM scales could resonantly scatter shaping the nebulae morphology and the spectral shape of the \lya emission (\citealt{Costa2022}). Overall, the aforementioned observational studies more frequently highlighted the importance of photoionization due to the quasar radiation followed by recombination in optically thin gas. In this framework, the \lya SB scales as the product of the gas mass and density \citep[SB$_{\rm Ly\alpha}^{\rm thin}\propto n_{\rm H}M_{\rm gas}$;][]{Hennawi2013}. In this scenario, the observed \lya SB implies densities of the cool gas of $n_{\rm H}\gtrsim1\,{\rm cm}^{-1}$ \citep{Hennawi2015,FAB2015a,FAB2019b}, which are comparable to star-forming regions in the interstellar medium of galaxies. Moreover, extended \ion{He}{ii}$\lambda1640$ and \ion{C}{iv}$\lambda1549$ emission has been detected in some individual cases or using stacking analysis of tens of systems  \citep{FAB2018,Cantalupo2019,Guo2020,Travascio2020,Fossati2021,Lau2022,Sabhlok2024}, indicating that the cool CGM of $2<z<6$ quasars is metal enriched and ionized, but not all illuminated by the quasar radiation (\citealt{Obreja2024}). 

Most of these studies focused on the CGM of the brightest quasars, however recent work by \citet{Mackenzie2021} targeted with MUSE 12 $z\sim3$ quasars selected to be fainter ($-23.8<M_{\rm i}(z=2)<-27.1$) than previous studies and detected \lya nebulae in all of them. 
The \lya nebulae of fainter quasars reported in that work  were on average fainter and less extended than those around brighter systems, indicating a correlation between the \lya SB of the nebula and the quasar UV and \lya luminosities.  
The authors discussed that there were several possible explanations for the observed relation, 
but no clear observational evidence to distinguish among these. 
The possible explanations that were discussed were: 
(i) a dependency with halo mass with brighter quasars sitting in more massive halos; 
(ii) varying opening angles with fainter quasars having smaller opening angles;
(iii) a different dominant powering mechanism for extended emission around the brighter and fainter quasars with resonant scattering dominating the emission around bright quasars while recombination dominating at the faint end; 
(iv) unresolved inner regions of the nebulae or an unresolved component from the interstellar medium of the host galaxy seen in faint quasars. 

In this framework, we report on the continuation of the Quasar Snapshot Observations with MUse: Search for Extended Ultraviolet eMission (QSO MUSEUM, \citealt[][]{FAB2019,Herwig2024}) survey.  
\citet{FAB2019}, hereafter QSO MUSEUM I, detected with MUSE a diversity of \lya nebulae around 61 bright quasars at $z=3.17$. 
QSO MUSEUM I showed that (i) quasars with very similar bolometric luminosities can be surrounded by very different extended \lya emission (in extent and SB level; see also \citealt{FAB2023}), (ii) the motions traced by the \lya emission have amplitudes consistent with gravitational motions expected in dark matter haloes hosting $z\sim3$ quasars, (iii) there is an  evolution of the level of \lya emission around $z\sim2$ and $z\sim3$ quasars, and (iv) the nebulae are likely powered by a combination of photoionization and resonant scattering of \lya photons. 
\citet{Herwig2024}, hereafter QSO MUSEUM II, focused on the CGM of 
physically associated quasar pairs at different projected distances ($50-500$\,kpc) and found that among the detected \lya nebulae, the most extended stretch along the line of direction of each pair, suggesting that the extended \lya emission traces the direction of interactions (close pairs) or  cosmic web filaments (distant pairs).

In this work (QSO MUSEUM III), 
we extend the sample presented in the QSO MUSEUM I survey to a total of 120 single quasar fields by targeting with MUSE 59 fainter systems ($-27<M_{\rm i}(z = 2)<-24$) at $z=3.1$, making it the largest effort to date to map the CGM of quasars at $z\sim3$. 
With this large sample we can now test the findings presented in \citet{Mackenzie2021} in addition to the link between SMBH properties, AGN feedback and extended \lya nebulae. This paper is organized as follows: in Section~\ref{sec:data} we provide an overview of the sample selection and data reduction, Section~\ref{sec:results} highlights our main observational findings (SB levels, luminosities, morphology, kinematics), while in Section~\ref{sec:discussion} we discuss the relationship between nebular emission and quasar properties, the implications of the observations in light of quasar variability, powering mechanisms, CGM physical properties and propose a simple model to address the observed trends. 
We adopt a flat $\Lambda$CDM cosmology with $H_0=70\,{\rm km\,s^{-1}\,Mpc^{-1}}$, $\Omega_{\rm m}=0.3$ and $\Omega_\Lambda = 0.7$. In this cosmology, 1~arcsec corresponds to $\sim7.5$\,kpc at the median redshift of the sample ($z=3.1$).

\section{Observations and Data Reduction} \label{sec:data}

\subsection{Sample selection}

\begin{figure}[t]
    \centering
    \includegraphics[width=\linewidth]{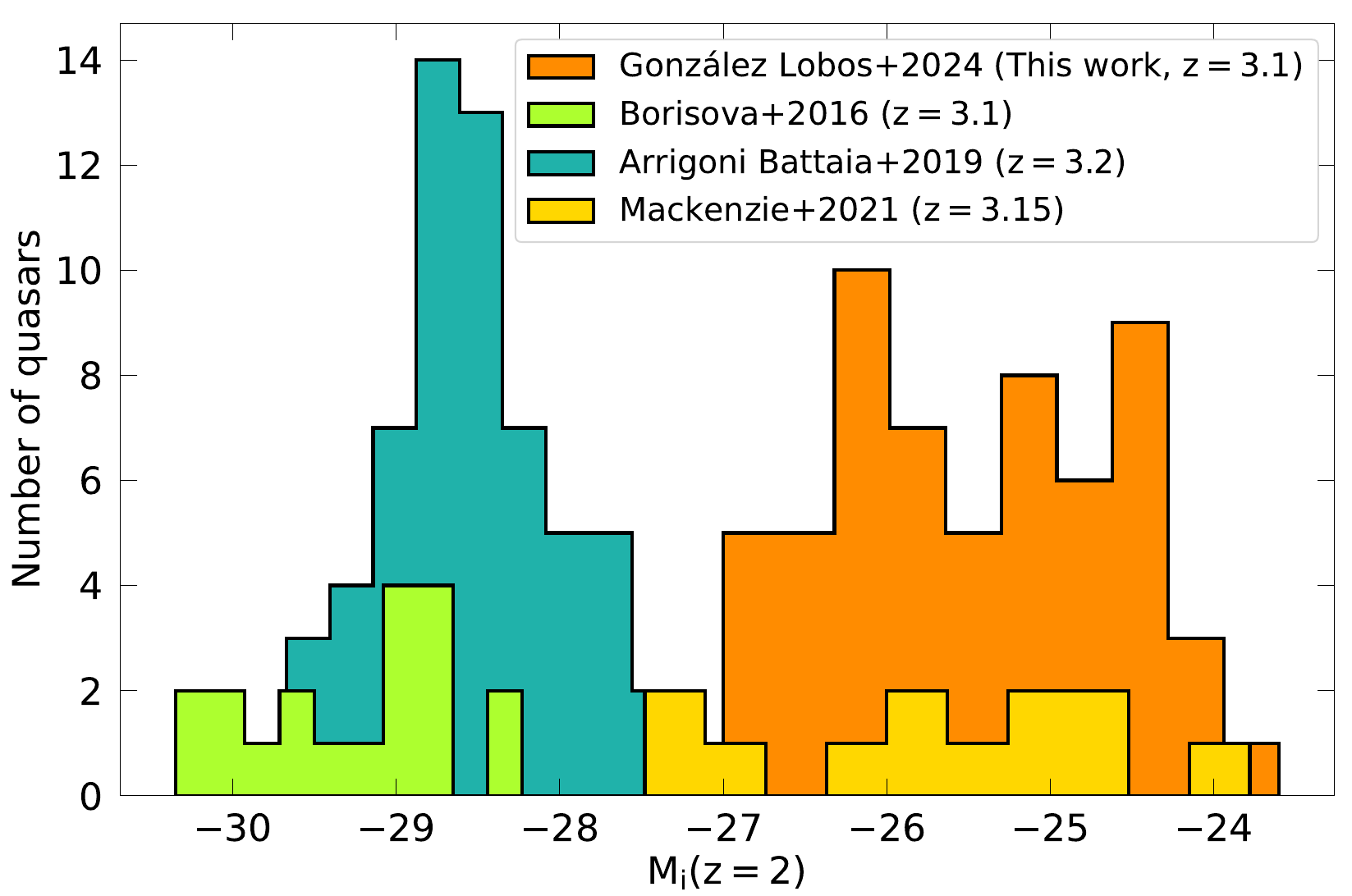}
    \caption{Overview of $z\sim3$ quasar samples with MUSE observations. Histograms of the absolute $i$-band magnitude normalized at $z=2$ \citep[following][]{Ross2013} 
     of the QSO MUSEUM III survey: 59 faint quasars from this study (orange), and 61 bright from  
     QSO MUSEUM I  
     (dark green). For comparison, we show the 19 bright quasars from \citet{Borisova2016} (light green) and the 12 faint from \citet{Mackenzie2021} (yellow). The median redshift of each sample is indicated in the legend.
     }
    \label{fig:i-band}
\end{figure}

The QSO MUSEUM I survey 
targeted with MUSE 61 quasars covering a range of absolute $i$-band magnitudes normalized at $z=2$ of $-29.67<M_i(z=2)<-27.03$ and redshift $3.03<z<3.46$. These quasars represent the brightest objects not targeted by the MUSE Guaranteed Time Observation
(GTO) team (\citealt{Borisova2016}). In this work, we extend the survey by targeting 58 additional quasars covering a fainter range of magnitudes $-27<M_i(z=2)<-24$. Additionally, we include the $z=3.4$ quasar discovered around ID 31 in QSO MUSEUM I 
(see their Appendix~C).
This sample has been constructed from the public spectroscopic quasar catalog of the Sloan Digital Sky Survey / Baryon Oscillation Spectroscopic Survey (SDSS/BOSS) data release 14 (DR14) \citep{Paris2018}, selecting objects as in QSO MUSEUM I, but now in a more limited redshift range $3<z<3.2$ and with the condition of having 20 targets per unit magnitude $M_i(z=2)$ bin in the aforementioned range.
The redshift range is chosen to target the lowest redshift accessible by MUSE in \lya emission (close to the quasars' activity peak), and to avoid contamination from stronger and more frequent sky lines at 
longer wavelengths.  
Figure~\ref{fig:i-band} shows the $M_i(z=2)$ distribution of the QSO MUSEUM I sample  
(dark green) and the 59 fainter quasars presented in this work (orange). This extended sample makes up the largest search to date for nebular emission around $z\sim3$ quasars (120 systems). Additionally, we show in the same figure, the similarly faint quasar sample from \citet{Mackenzie2021} (yellow) and the bright quasars targeted by \citet{Borisova2016} (light green).

\subsection{Physical properties of the targeted quasars}
\label{sec:qso_phys_prop}

\begin{figure}
    \centering
    \includegraphics[width=\linewidth]{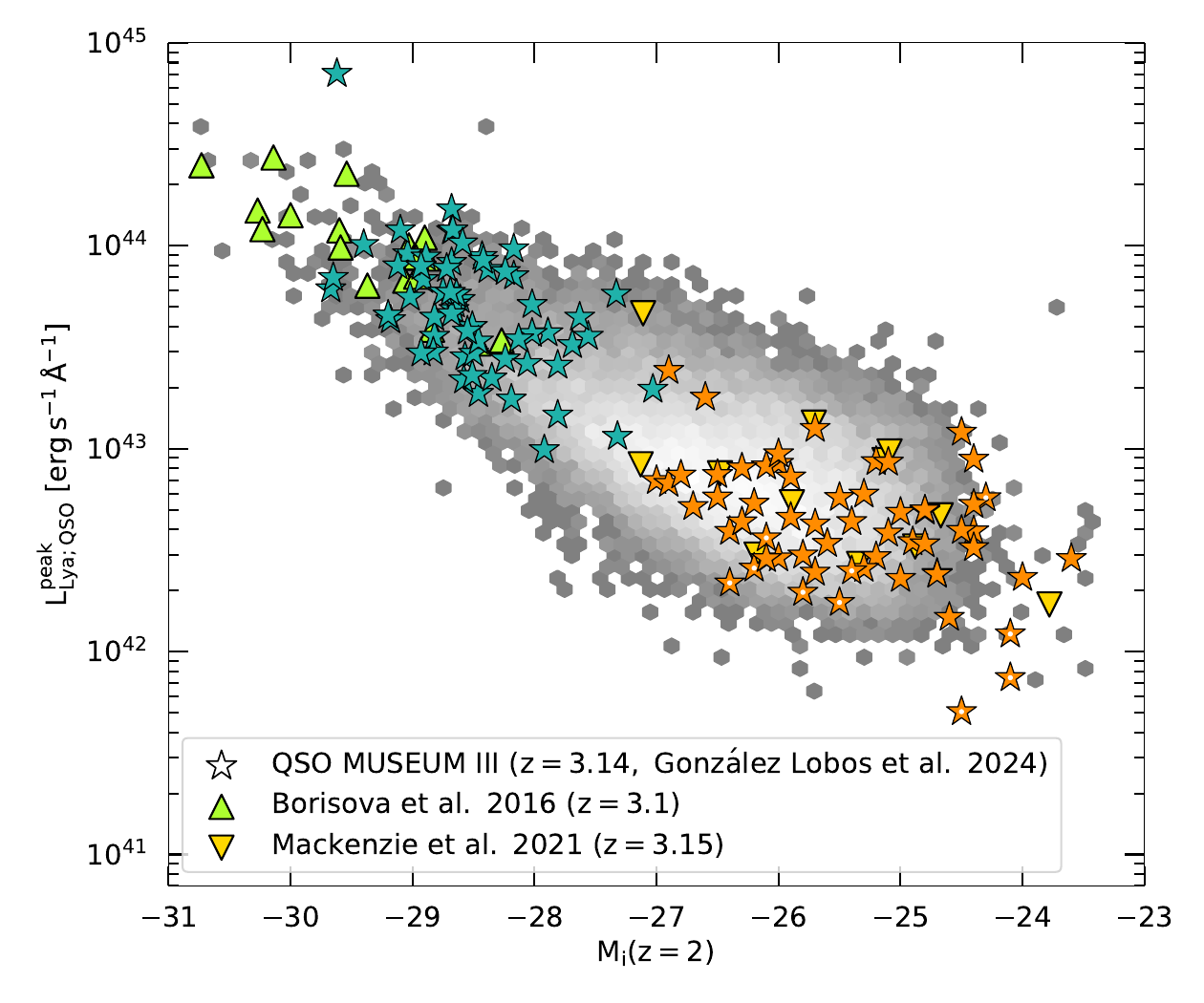}
    \caption{Overview of the luminosity distribution of observed $z\sim3$ quasars. The figure shows the quasar peak \lya luminosity density as a function of 
    absolute $i$-band magnitude normalized at $z=2$ \citep[$M_{\rm i}(z=2)$, following][]{Ross2013}. 
    The QSO MUSEUM III faint and bright quasars are shown with orange and dark green stars, respectively. The systems with no \lya nebula detected are marked with a white dot (see Section~\ref{sec:detection}). In addition, we 
    show the location of the 17 brighter quasars targeted in \citep{Borisova2016} (light green triangles) and the 12 
    fainter objects from \citet{Mackenzie2021} (yellow triangles). 
    The 2D number density of $3.0 < z < 3.46$ quasars from SDSS DR17 (25806 quasars, \citealt{Abdurro2022}) is shown in logarithmic grey scale, white marking the highest densities. 
    Section~\ref{sec:qso_phys_prop} explains how the luminosities are derived.}
    \label{fig:Lpeak}
\end{figure}
\begin{figure}
    \centering
    \includegraphics[width=\linewidth]{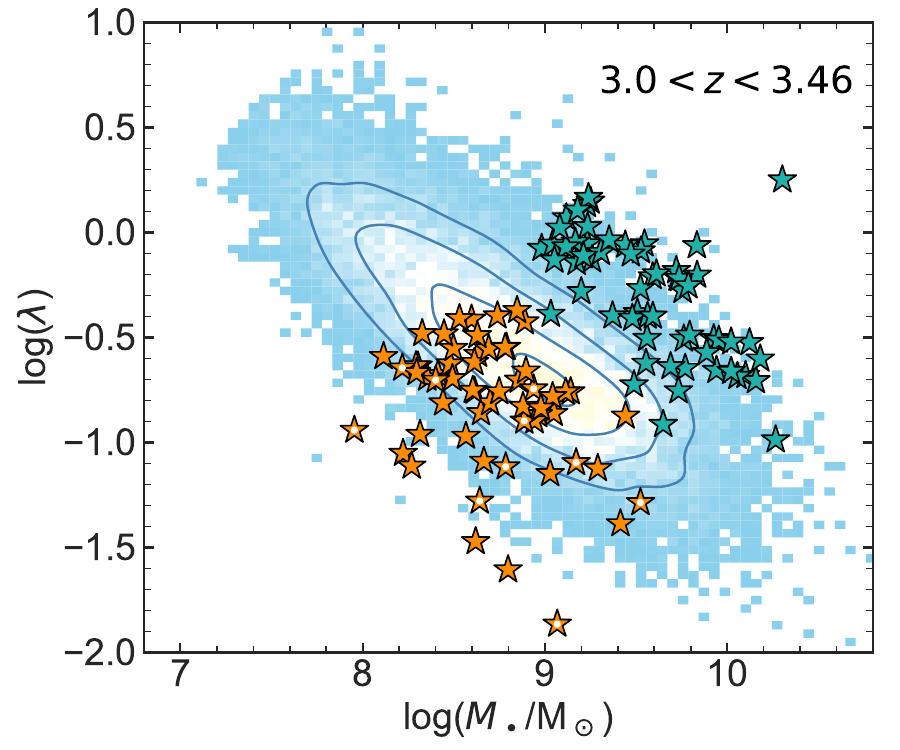}
    \caption{Eddington ratio versus black hole mass for the targeted sample. The datapoints for the QSO MUSEUM III sample (same symbols as in Figure~\ref{fig:Lpeak}) are compared with the values for SDSS quasars in the same redshift range (blue, 2-D number density histogram; \citealt{Rakshit2020}). The contours indicate the iso-proportions of the density at 0.2, 0.4, 0.68 and 0.95 levels, indicating that 20\%, 40\%,68\%, and 95\% of the quasars are outside that contour, respectively.}
    \label{fig:Mbh_lambda}
\end{figure}
\begin{figure}
    \centering
    \includegraphics[width=0.8\linewidth]{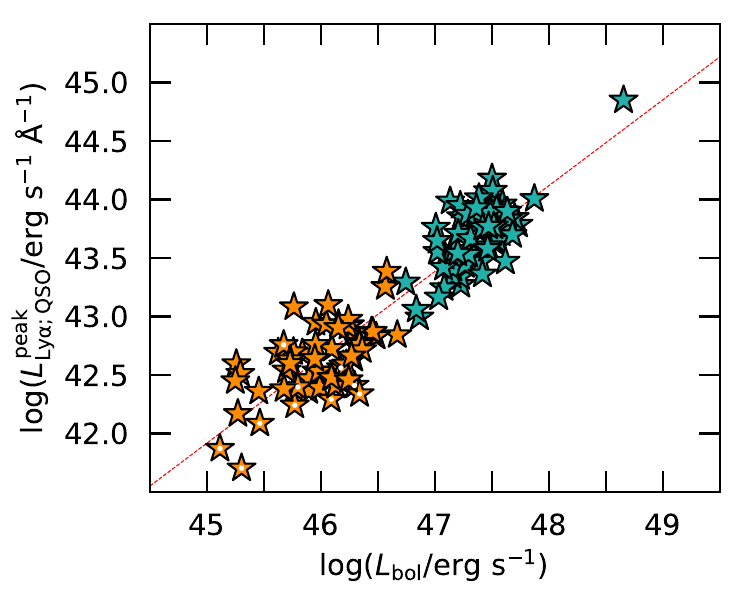}
    \caption{Peak Ly$\alpha$ luminosity density versus bolometric luminosity of the targeted quasars. The same symbols as Figure~\ref{fig:Lpeak} indicate the bolometric luminosity of the quasar, which is is computed using the monochromatic luminosity $L_{\lambda}(1350\,\AA)$ (see~Appendix~\ref{app:QSOfit}). Systems where no \lya nebulae were detected are marked with a white dot (see Section~\ref{sec:SBmaps}). The red dashed line represents a power law fit to the data of the form $\log(L_{\rm Ly\alpha;\,QSO}^{\rm peak}/{\rm erg\,s^{-1}\,\AA^{-1}})~=~8.96+0.73\log(L_{\rm bol}/{\rm erg\,s^{-1}})$.}
    \label{fig:LbolLpeak}
\end{figure}

The physical properties of the targeted quasars are summarized in Figures~\ref{fig:Lpeak},~\ref{fig:Mbh_lambda}, and \ref{fig:LbolLpeak}.
The distribution of the peak \lya luminosity density as a function of $i-$band absolute magnitude normalized at $z=2$ \citep[following][]{Ross2013} of the bright and faint quasars is shown in Figure~\ref{fig:Lpeak} with dark green and orange stars, respectively. The peak \lya luminosity density is computed from the MUSE datacube by integrating a spectrum inside a 1.5'' radius aperture centered at the quasar location. In addition, we compute using the MUSE datacubes the values from the sample of 17  quasars targeted in \citet{Borisova2016} and 12 quasars from \citet{Mackenzie2021} and show them with light green and yellow triangles, respectively.
For comparison, we compute the values from the SDSS DR17 \citep{Abdurro2022} $3.0<z<3.46$ quasars. Figure~\ref{fig:Lpeak} illustrates how the QSO MUSEUM survey now covers the parameter space of the SDSS quasar population. Specifically, the 120 quasars in QSO MUSEUM III encompass a wide range of quasar UV luminosities (six orders in $M_{\rm i}(z=2)$) and in peak \lya quasar luminosity $L_{\rm Ly\alpha; QSO}^{\rm peak}$ (three orders). 

Further, for all the QSO MUSEUM III targeted quasars we estimated their black-hole mass, bolometric luminosity and accretion rates. 
The first is calculated using the only available broad emission line observed for all the sample, \ion{C}{iv}, and that allow us to derive single-epoch black hole masses using the \citet{VP2006} estimator: 
\begin{multline}\label{eq:logMBH}
{\rm log}M_{\rm BH}(\ion{C}{iv})={\rm log}\left\{\left[\frac{{\rm FWHM(\ion{C}{iv})}}{1000\, {\rm km\, s^{-1}}}\right]^{2} \left[ \frac{\lambda L_{\lambda}(1350\,\AA)}{10^{44}\, {\rm erg\, s^{-1}}} \right]^{0.53} \right\}\\
+6.66,
\end{multline}
where FWHM(\ion{C}{iv}) is the width of the \ion{C}{iv} line and $L_\lambda(1350\,\AA)$ is the monochromatic luminosity of the quasar at rest-frame 1350\,\AA. 
The values of FWHM(\ion{C}{iv}) and $L_\lambda(1350\,\AA)$ are obtained using the
output from the fit 
to 
the quasar spectra using 
the public Python code \textsc{PyQSOFit}\footnote{\url{https://github.com/legolason/PyQSOFit}} \citep{Guo2018} as described in Appendix~\ref{app:QSOfit}. 
The typical uncertainties on the estimate of the black hole masses (Equation~\ref{eq:logMBH}) are of the order of 0.5\,dex. 

With the $M_{\rm BH}$, we can compute the theoretical maximum luminosity for the case when radiation pressure and gravity are in equilibrium in a spherical geometry, i.e., the Eddington luminosity \citep{Eddington1926} as:
\begin{equation}\label{eq:LEdd}
L_{\rm Edd} = \frac{4\pi G M_{\rm BH} m_{\rm p}c}{\sigma_{\rm T}}
= 1.26\times10^{38} \left(\frac{M_{\rm BH}}{M_{\odot}}\right) {\rm erg\, s^{-1}},
\end{equation}
where $\sigma_{\rm T}$ is the Thomson scattering cross section, $G$ is the gravitational constant, $m_{\rm p}$ is the proton mass, and $c$ is the speed of light. 
The Eddington luminosity is usually compared with the quasar bolometric luminosity ($L_{\rm bol}$) to assess the activity and hence accretion rate of the SMBH. A so-called Eddington ratio is usually defined as:
\begin{equation}\label{eq:ldaEdd}
\lambda_{\rm Edd}=L_{\rm bol}/L_{\rm Edd}
\end{equation}
We computed $L_{\rm bol}$ using the monochromatic luminosity $L_\lambda(1350\,\AA)$ as done in \citet{Rakshit2020} using the bolometric correction factor given in \citet{Shen2011} and adapted from \citet{Richards2006}:
\begin{equation}\label{eq:Lbol}
L_{\rm bol}=3.81 \times \lambda L_{\lambda}(1350\,\AA).
\end{equation}
This estimate can have up to 0.3 dex of instrinsic uncertainty. 
The values computed of $L_{\rm bol}$, together with $M_{\rm BH}$ and $\lambda_{\rm Edd}$ are listed in Tables~\ref{tab:QSOfit-bright}~and~\ref{tab:QSOfit-faint} together with their statistical uncertainties, which are much smaller than the aforementioned intrinsic uncertainties.

Figure~\ref{fig:Mbh_lambda} shows the Eddington ratio as a function of black hole mass for the QSO MUSEUM III sample with the same symbols as in Figure~\ref{fig:Lpeak}.  
Additionally, we show the number density for SDSS quasars in the same redshift range from \citet{Rakshit2020}, who used almost the same fitting method to estimate such quantities (see details in Appendix~\ref{app:QSOfit}).
Finally, we show the relation between the quasar peak \lya luminosity density and their bolometric luminosity 
for the 120 quasars in Figure~\ref{fig:LbolLpeak}.

\subsection{Observations}

\begin{table*}[!htbp]
\footnotesize \centering
\caption{Summary of the MUSE observations for the faint sample, which is first presented in this work.}
\label{tab:Data}
\begin{tabular}{|c|c|c|c|c|c|c|c|c|}
 \hline
 \T
 ID & Name & RA & Dec & ${M_i}\,(z=2)$\tablefootmark{(a)} & $\rm{Seeing_{Ly\alpha}}$ & ${\rm SB\,  limit}$\tablefootmark{(b)} & Sky \\
 & & [J2000] & [J2000] & [mag] & ${\rm [arcsec]}$ & $[{\rm 10^{-18}\,cgs}]$ & Conditions\tablefootmark{(c)}
 \B \\
\hline
\T
 62 & J0203-0443 & 02:03:57.00 & -04:43:10.2 & -24.6 & 0.69 & 2.28 & PH\\
 63 & J0229-0029 & 02:29:56.55 &-00:29:53.9 & -26.1 & 2.02 & 2.26 & CL\\
 64 & J0159-0032 & 01:59:45.45 & -00:32:28.4 & -24.7 & 0.77 & 2.28 & CL\\
 65 & J0829+1426 & 08:29:39.91 & +14:26:25.7 & -25.0 & 1.08 & 2.33 & PH\\
 66 & J0200-0606 & 02:00:12.60 & -06:06:15 & -27.0 & 0.58 & 1.81 & CL\\
 67 & J0247-0023 & 02:47:46.86 & -00:23:53.4 & -26.9 & 0.71 & 1.94 & PH\\
 68 & J0809+0643 & 08:09:49.16 & +06:43:10.8 & -24.5 & 0.93 & 1.85 & CL\\
 69 & J0834+1012 & 08:34:15.33 & +10:12:31.2 & -25.4 & 0.83 & 2.06 & PH\\
 70 & J0923+0011 & 09:23:00.29 & +00:11:56.7 & -24.8 & 0.86 & 1.89 & PH\\
 71 & J0823+0531 & 08:23:07.22 & +05:31:35.4 & -26.2 & 0.97 & 1.97 & CL\\
 72 & J0227-0113 & 02:27:21.91 & -01:13:24.6 & -24.1 & 0.93 & 2.04 & PH\\
 73 & J1038+0919 & 10:38:40.81 & +09:19:14.3 & -25.0 & 1.40 & 2.17 & CL\\
 74 & J0203-0153 & 02:03:33.76 & -01:53:08.4 & -25.6 & 1.52 & 1.92 & PH\\
 75 & J0210-0945 & 02:10:03.52 & -09:45:20.5 & -26.0 & 0.79 & 2.37 & TN\\
 76 & J0155-0732 & 01:55:25.44 & -07:32:16.3 & -26.6 & 1.08 & 1.97 & PH\\
 77 & J0151+0023 & 01:51:25.83 & +00:23:32.7 & -24.4 & 0.77 & 2.00 & PH\\
 78 & J0030+0047 & 00:30:42.91 & +00:47:43.3 & -24.0 & 0.91 & 2.82 & TN\\
 79 & J0148-0055 & 01:48:09.00 & -00:55:08.8 & -24.4 & 0.78 & 1.87 & PH\\
 80 & J0017+0316 & 00:17:44.80 & +03:16:06.1 & -26.9 & 0.87 & 2.31 & CL\\
 81 & J0925+0344 & 09:25:41.80 & +03:44:37.0 & -25.2 & 0.96 & 2.26 & CL\\
 82 & J0125-0005 & 01:25:29.50 & -00:05:13.5 & -24.1 & 1.23 & 4.75 & TN\\
 83 & J0843+1916 & 08:43:19.90 & +19:16:28.8 & -24.8 & 1.15 & 2.29 & CL\\
 84 & J0752+1244 & 07:52:54.22 & +12:44:51.5 & -25.1 & 0.94 & 2.26 & PH\\
 85 & J0018-0026 & 00:18:17.73 & -00:26:59.0 & -26.2 & 0.88 & 2.00 & TN\\
 86 & J0747+1429 & 07:47:14.29 & +14:29:54.3 & -26.5 & 1.59 & 1.93 & CL\\
 87 & J0245-0036 & 02:45:23.32 & -00:36:19.0 & -24.9 & 1.00 & 1.82 & CL\\
 88 & J0801+0534 & 08:01:30.16 & +05:34:37.0 & -25.7 & 1.15 & 2.07 & CL\\
 89 & J0840+0141 & 08:40:58.35 & +01:41:45.0 & -26.0 & 1.15 & 1.73 & CL\\
 90 & J1016+0833 & 10:16:25.72 & +08:33:09.8 & -25.2 & 0.97 & 2.50 & CL\\
 91 & J0159+0025 & 01:59:22.99 & +00:25:30.8 & -24.3 & 0.69 & 2.28 & PH\\
 92 & J0234-0044 & 02:34:41.17 & -00:44:43.7 & -24.4 & 1.00 & 2.10 & PH\\
 93 & J0243-0038 & 02:43:00.58 & -00:38:18.0 & -24.5 & 0.86 & 1.98 & PH\\
 94 & J0140-0202 & 01:40:20.33 & -02:02:43.0 & -25.5 & 0.78 & 1.94 & PH\\
 95 & J0252-0333 & 02:52:42.99 & -03:33:39.4 & -25.8 & 0.95 & 2.27 & PH\\
 96 & J0939+0451 & 09:39:29.36 & +04:51:47.4 & -25.1 & 0.74 & 2.13 & PH\\
 97 & J0208-0922 & 02:08:02.86 & -09:22:43.0 & -26.8 & 1.00 & 1.92 & PH\\
 98 & J0212-0602 & 02:12:48.13 & -06:02:16.9 & -25.7 & 1.00 & 2.15 & CL\\
 99 & J1205+1059 & 12:05:12.55 & +10:59:34.6 & -24.5 & 2.11 & 2.22 & CL\\
 100 & J1001-0007 & 10:01:22.20 & -00:07:52.7 & -25.3 & 0.80 & 2.41 & TN\\
 101 & J1145-0209 & 11:45:19.45 & -02:09:43.9 & -25.9 & 0.98 & 2.05 & CL\\
 102 & J1057+0804 & 10:57:37.23 & +08:04:11.2 & -26.1 & 0.79 & 2.20 & CL\\
 103 & J1244-0027 & 12:44:08.39 & -00:27:26.1 & -26.0 & 1.31 & 2.27 & CL\\
 104 & J1216+0454 & 12:16:29.25 & +04:54:34.8 & -25.7 & 1.67 & 1.93 & CL\\
 105 & J1048+0449 & 10:48:04.72 & +04:49:00.3 & -24.8 & 1.34 & 2.03 & CL\\
 106 & J0920-0048 & 09:20:56.50 & -00:48:07.0 & -25.8 & 0.86 & 1.65 & PH\\
 107 & J0223-0309 & 02:23:04.27 & -03:09:51.5 & -25.3 & 0.97 & 1.91 & PH\\
 108 & J1301-0020 & 13:01:12.26 & -00:20:48.9 & -26.5 & 1.00 & 2.10 & PH\\
 109 & J0823+0340 & 08:23:25.59 & +03:40:59.4 & -26.7 & 1.00 & 2.03 & CL\\
 110 & J0832+0450 & 08:32:20.15 & +04:50:28.3 & -25.5 & 0.95 & 2.00 & CL\\
 111 & J1022+0418 & 10:22:25.90 & +04:18:24.2 & -26.4 & 1.01 & 2.11 & PH\\
 112 & J1101+0314 & 11:01:05.12 & +03:14:03.9 & -26.3 & 1.67 & 2.09 & CL\\
 113 & J0207-0306 & 02:07:16.83 & -03:06:48.4 & -25.9 & 1.02 & 2.43 & CL\\
 114 & J1200+1528 & 12:00:26.16 & +15:28:16.1 & -25.4 & 0.96 & 1.87 & CL\\
 115 & J0252-0057 & 02:52:53.28 & -00:57:28.0 & -24.4 & 0.67 & 2.07 & PH\\
 116 & J0840+0636 & 08:40:26.81 & +06:36:32.8 & -26.3 & 1.06 & 2.08 & CL\\
 117 & J1244+0327 & 12:44:06.42 & +03:27:43.2 & -24.7 & 1.27 & 1.86 & CL\\
 118 & J1153-0141 & 11:53:02.24 & -01:41:13.2 & -26.4 & 1.11 & 1.94 & CL\\
 119 & J1019+0845 & 10:19:17.81 & +08:45:50.0 & -26.1 & 0.81 & 2.03 & CL\\
 120 & UM670-field & 01:17:22.24 & -08:41:43.5 & -23.6 & 0.80 & 2.22 & CL \B \\
 \hline
\end{tabular}
\tablefoot{
\tablefoottext{a}{Absolute $i-$band magnitude normalized to $z=2$ following \citet{Ross2013}.} 
\tablefoottext{b}{$1\sigma$ SB limit within 1\,arcsec$^2$ and a 30\,\AA\ narrow-band centered at the observed \lya wavelength in units of ${\rm 10^{-18}\,erg\,s^{-1}\,cm^{-2}\,arcsec^{-2} }$.}
\tablefoottext{c}{Sky conditions during the observations as described in the ESO observational log: PH-photometric; CL-clear; TN-thin cirrus.}
}
\end{table*}

The observations were carried out with the MUSE instrument on the VLT 8.2m telescope YEPUN (UT4) over roughly 6.5 years (2014-2021).
The data for the faint sample were taken as part of the European Southern Observatory (ESO) program 0106.A-0297(A) (PI: F. Arrigoni Battaia) in service mode on UT dates between 06-11-2020 and 10-03-2021 with good weather conditions (46.5\% with clear sky, 46.5\% with photometric sky, 7\% with thin clouds; details in Table~\ref{tab:Data})\footnote{From the acquired data under this ESO program we removed two unusable observations for this science case: the field corresponding to J0244-0059 was incorrectly observed and contains the 
acquistion star instead of the faint proposed quasar in the science datacube, and the field of J1103+0913 which shows fringes in all its datacubes.} 
The observations were taken using the Wide Field Mode of MUSE, which covers a field of view of $1'\times1'$ with a 0.2" pixel scale and a spectral range of 4750-9350\,\AA\ with a channel width of 1.25\,\AA\ and resolving power of $R\sim1750$ at 4984\,\AA\ (the expected \lya wavelength at the median redshift of the sample). The observational strategy is the same as in QSO MUSEUM I, 
and consists of three exposures of 900 seconds each per field with a dither of $<5"$ and 90 degree rotations with respect to each other.
The average seeing at the expected \lya wavelength of the 59 systems is 1.03". We summarize these observations in Table~\ref{tab:Data}.

Additionally, we include in this work the MUSE observations of the 61 quasars from QSO~MUSEUM~I (see their Table~2). 
These data were taken as part of the ESO programmes 094.A-0585(A), 095.A-0615(A/B), and 096.A-0937(B) (PI: F. Arrigoni Battaia). The 120 systems are reduced and analysed in this work using the exact same methods as described in Sections~\ref{sec:reduction}~and~\ref{sec:subtraction}.

\subsection{Data Reduction}\label{sec:reduction}

The data reduction for the whole sample comprised of 120 quasars was performed using the MUSE pipeline version 2.8.3 \citep{Weilbacher2012,Weilbacher2014,Weilbacher2020} following the method described in \citet{Farina2019} consisting in subtraction of bias and dark field, flat field correction, wavelength calibration, illumination correction and standard star flux calibration. We remove the sky emission in each datacube using the Zurich Atmospheric Purge \citep[ZAP;][]{Soto2016} software. 
The MUSE pipeline underestimates the variance due to the noise correlation between pixels \citep[][]{Bacon2015}, therefore we rescale each layer of the variance cubes to reflect the variance in the datacubes. The rescaled variance cubes are used to compute the surface brightness (SB) limits and errors in our estimations.

Following the method presented in \citet{GonzalezLobos2023}, the three\footnote{The fields of ID 8 and 37 had in total 12 and 6 exposures respectively (see Table~2 of \citealt{FAB2019}).} reduced exposures of each target are 
median combined after masking artifacts due to the separation of the MUSE IFUs (consisting of $\sim4\%$ of the field of view). The final products are a final science datacube and a variance datacube obtained by taking into account propagation of errors during the combination. This final variance datacube is once again checked against the data by rescaling each layer to the variance computed in the corresponding combined science layer. 

The average $2\sigma$ SB limit within 1\,arcsec$^2$ in a 30\,\AA\ narrow-band (NB) at the observed \lya wavelength of the sample is $2.2\times10^{-18}\,{\rm erg\,s^{-1}\,arcsec^{-2}\,cm^{-2}\,}$ (see Table~\ref{tab:Data} for the individual SB limits). Per channel (1.25~\AA) and within the same aperture and at the same wavelength, the data have an average $2\sigma$ SB limit of $4.3\times10^{-19}\,{\rm erg\,s^{-1}\,arcsec^{-2}\,cm^{-2}\,}$. These SB limits agree to those reported in \citet{FAB2019} despite the different analysis tools used in this work.


\subsection{Revealing extended emission around quasars}\label{sec:subtraction}

The unresolved bright emission from a quasar can easily outshine the fainter emission from the surrounding CGM \citep[e.g.,][]{Heckman1991, Heckman1991b,Moller2000}. 
Therefore, we need to subtract the point spread function (PSF) of the quasar, as it has been frequently described in the literature \citep[e.g.,][]{Borisova2016,Husemann2018,Farina2019,O'Sullivan2020,GonzalezLobos2023}, in order to characterize the extended CGM emission. 
In this work, we modify the Python routines developed and described in \citet{GonzalezLobos2023} to reveal extended emission around quasars with different luminosities. For the bright objects we use parameters similar to those in \citet{Borisova2016} and \citet{FAB2019}, while for the faint sample we use parameters similar to the ones used by 
\citet{Mackenzie2021}. 

Specifically, the PSF for faint quasars is constructed empirically using pseudo-NBs of 400 channels computed at each wavelength of the datacubes. This choice of pseudo-NB width is made to increase the signal to noise in the constructed PSF, similar to the 300 channels used in \citet{Mackenzie2021} and larger than the 150 channels used for brighter quasars (\citealt{Borisova2016,FAB2019}). Each pseudo-NB is normalized to the emission inside a region of 1\,arcsec$^{2}$ centered at the quasar coordinates, computed from the sigma clipped ($3\sigma$) average, and then subtracted from the datacube at the quasar position out to a radius of three times the seeing (see Table~\ref{tab:Data}).

Finally, we mask the 1\,arcsec$^2$ region located at the quasar position used for normalizing the pseudo-NB and exclude it from our analysis. Indeed this region is affected by PSF residuals (e.g., \citealt{Borisova2016}). We refer the reader to \citet{GonzalezLobos2023} for more details on the algorithm.

After the PSF-subtraction, any extended emission could be still contaminated by the presence of continuum sources. We remove them using a median-filtering approach (e.g., \citealt{Borisova2016}) using the \texttt{contsubfits} function (with default parameters) within the 
ZAP\footnote{\url{https://github.com/musevlt/zap}} \citep{Soto2016} software. Such a method has been already used in other works targeting extended emission around quasars (\citealt{FAB2019b,Herwig2024}). 


\section{Results}\label{sec:results}

\begin{figure*}
    \centering
    \includegraphics[width=0.85\linewidth]{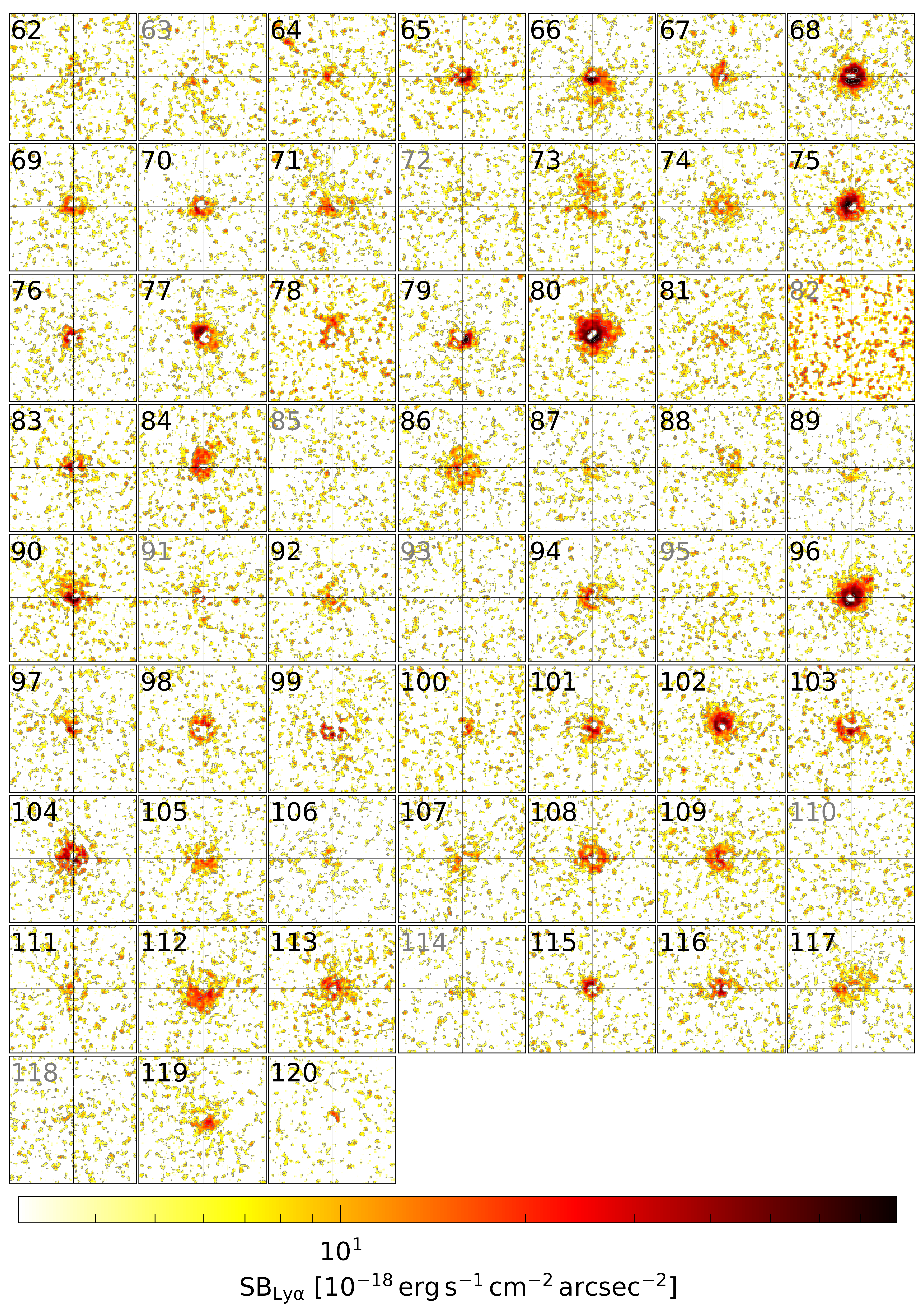}
    \caption{QSO MUSEUM III atlas of the faint quasars' \lya nebulae. \lya SB maps around the 59 faint quasars after PSF- and continuum-subtraction (see Section~\ref{sec:subtraction}), computed from 30\,\AA\ pseudo-NBs centered at the peak \lya wavelength of the nebula. 
    All images 
    show maps with projected sizes of  
    $20"\times20"$ 
    ($\sim150\,{\rm kpc\,}\times150$\,kpc at the median redshift of the sample). In each map, a black crosshair indicates the location of the quasar and their ID number is indicated in top left corner. For systems with no detected nebula (Section~\ref{sec:detection}), their ID number is indicated in color gray. The contours indicate levels of $[2, 4, 10, 20, 50]$ times the \lya SB limit within the pseudo-NB (Table~\ref{tab:Data}).}
    \label{fig:SBmaps}
\end{figure*}
\begin{figure*}
    \centering
    \includegraphics[width=0.9\linewidth]{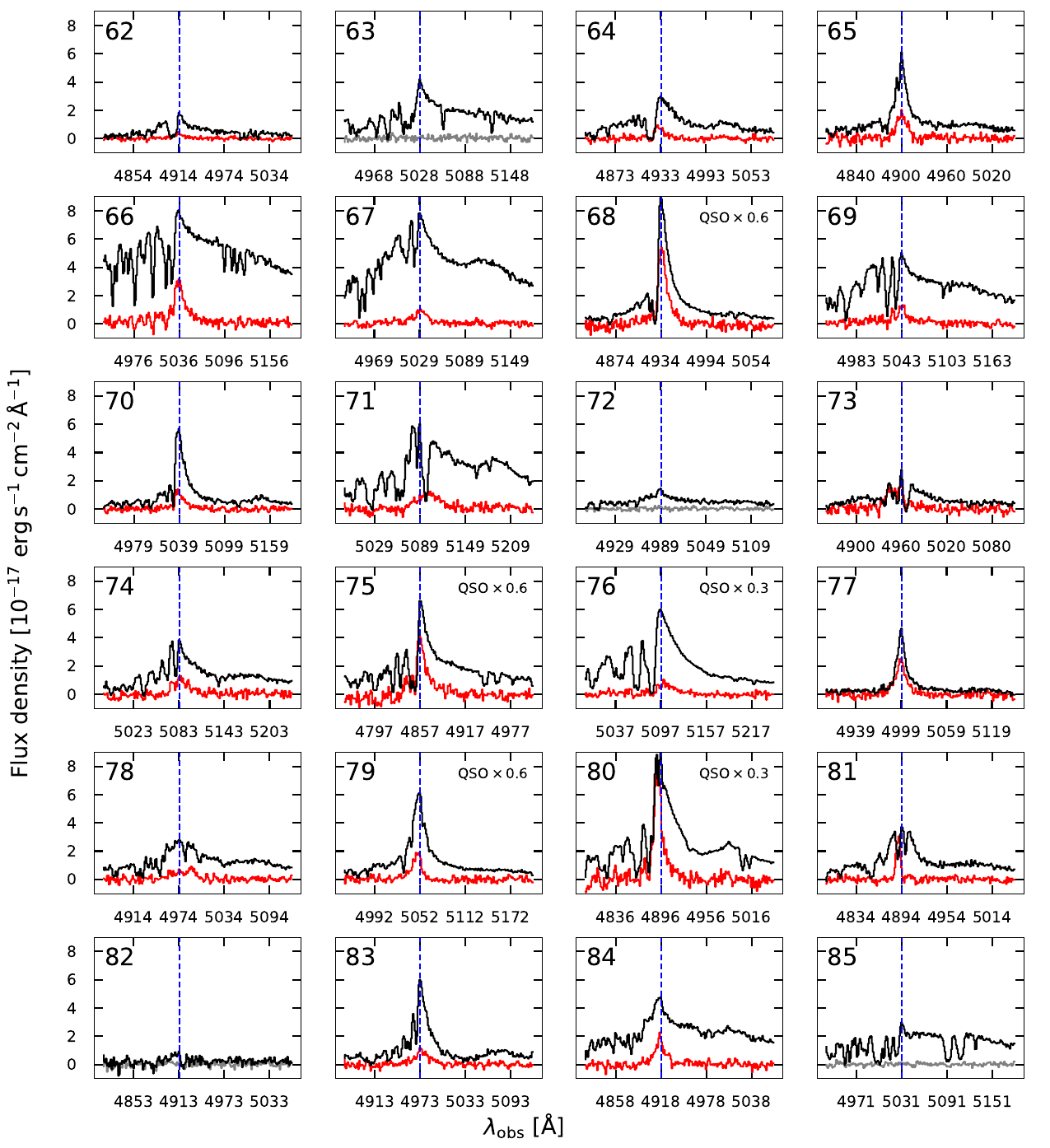}
    \caption{One-dimensional spectrum covering the 
    \lya line for ID 62-85 in the new faint quasars' sample. The ID of each quasar is shown in the top left corner of each panel. 
    Each panel shows the spectrum integrated from the MUSE datacube inside a 1.5'' radius aperture centered at the quasar location (black line) and the integrated spectrum of each detected nebula integrated from the PSF- and continuum-subtracted datacubes within the 2$\sigma$ isophotes from Figure~\ref{fig:SBmaps} (red line), respectively. 
    The wavelength of the peak of the \lya emission of each quasar is indicated with a blue vertical line. 
    The gray lines are the spectra integrated within a 1.5'' radius aperture from subtracted datacubes where we found no extended \lya emission. 
    We mask the 1''$\times$1'' PSF normalisation region when extracting 
    a spectrum using the PSF- and continuum-subtracted datacubes. 
    All the spectra are shown with  
    the same y-axis scale and we indicate with a label in the top right corner of the panel if the quasar spectra has been re-scaled by $0.6\times$ or $0.3\times$ for visualisation.
    }
    \label{fig:spec_atlas01}
\end{figure*}
\begin{figure*}
    \centering
    \includegraphics[width=0.9\linewidth]{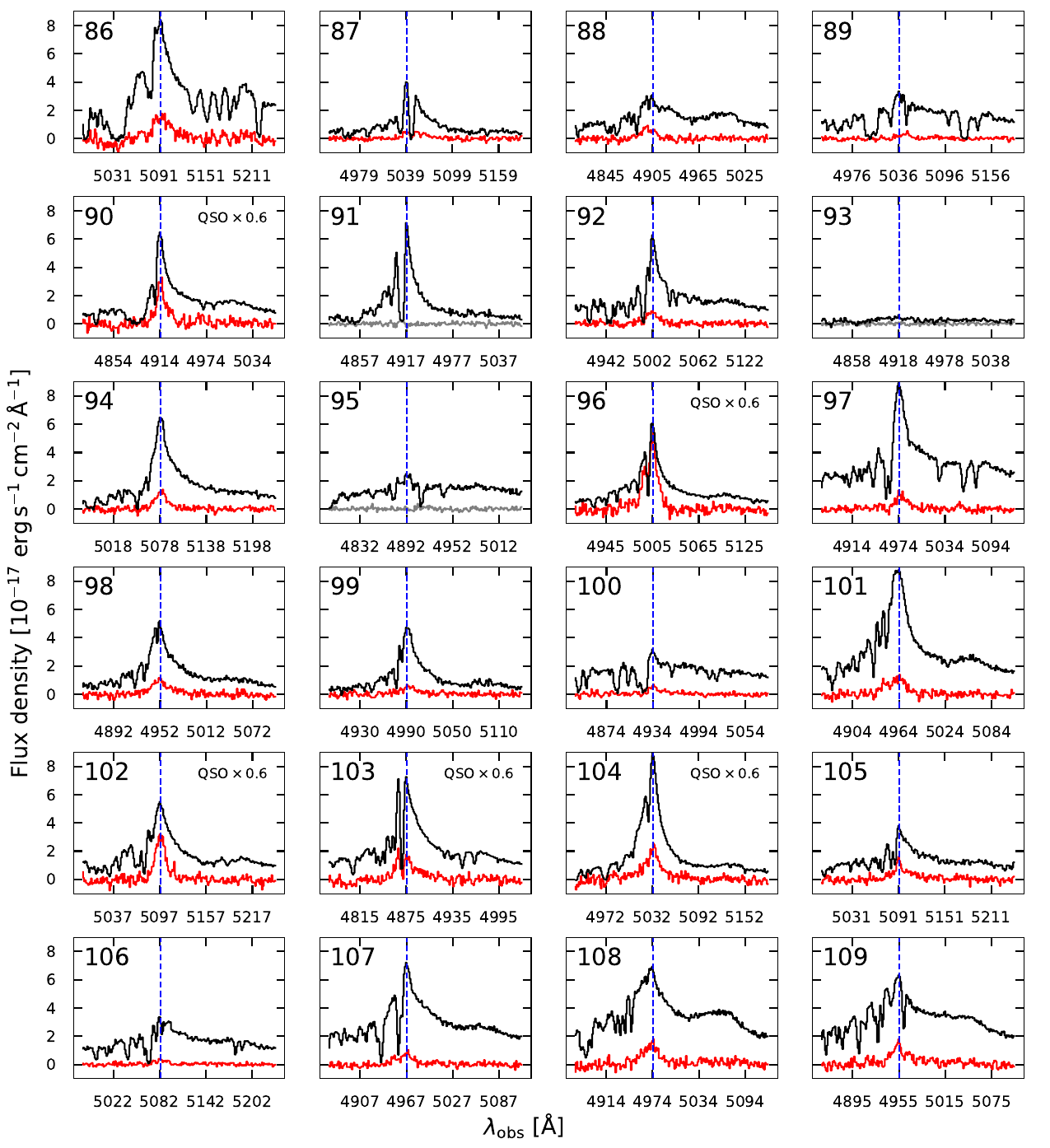}
    \caption{Same as Figure~\ref{fig:spec_atlas01} but for ID 86-109.}
    \label{fig:spec_atlas02}
\end{figure*}
\begin{figure*}
    \centering
    \includegraphics[width=0.9\linewidth]{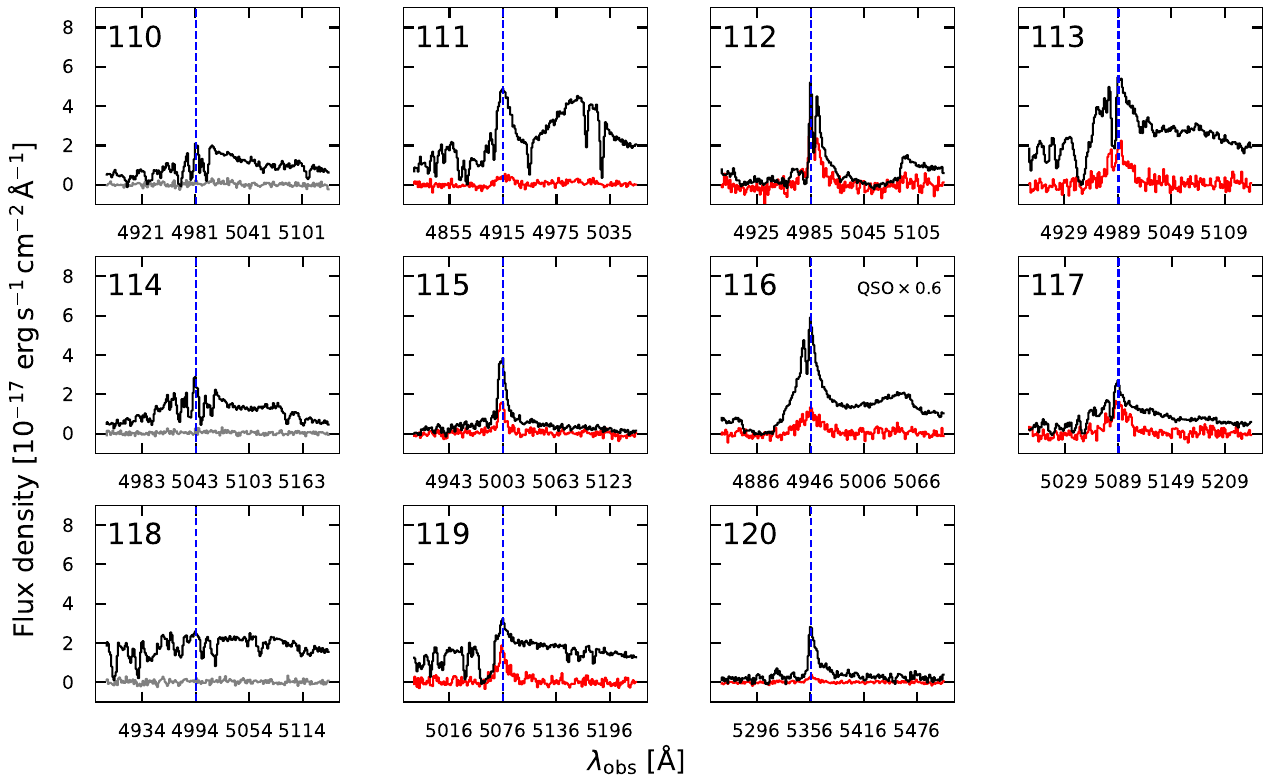}
    \caption{Same as Figure~\ref{fig:spec_atlas01} but for ID 110-120}
    \label{fig:spec_atlas03}
\end{figure*}

\begin{table*}[!htbp]
\footnotesize \centering
\caption{Properties of each quasar and extended Ly$\alpha$ emission.}
\label{tab:properties}
\begin{tabular}{|c|c|c|c|c|c|c|c|c|c|}
\hline
\T
ID & z$_{\rm Ly\alpha;peak}^{\rm QSO}$\tablefootmark{(a)} & z$_{\rm Ly\alpha;peak}^{\rm Neb}$\tablefootmark{(a)} & 
$L_{\rm Ly\alpha;peak}^{\rm QSO}$ & $L_{\rm Ly\alpha}^{\rm neb}$ 
& Area & ${\Delta v_{\rm gauss}}$ & ${\Delta v_{\rm m1}}$ & ${\rm FWHM_{gauss}}$\tablefootmark{(b)} & $\Delta \lambda_{\rm m2}$ \\
 &  &  & $10^{42}\,{ \rm erg\,s^{-1}\,\AA^{-1} }$ & $10^{42}\,{ \rm erg\,s^{-1} }$ & arcsec$^2$\ &
 ${\rm km\,s^{-1}}$ & ${\rm km\,s^{-1}}$ & ${\rm km\,s^{-1}}$ & ${\rm km\,s^{-1}}$ \B \\
\hline
\T

62 & 3.043 & 3.043 & 1.47$\pm$0.07 & 5.22$\pm$0.52 & 6$\pm$1 & 16$\pm$75 & 50$\pm$81 & 1421$\pm$176 & 587$\pm$63 \\
63 & 3.137 & \ldots & 3.63$\pm$0.09 & \ldots & \ldots & \ldots & \ldots & \ldots & \ldots \\
64 & 3.058 & 3.053 & 2.43$\pm$0.08 & 8.89$\pm$0.55 & 10$\pm$1 & -105$\pm$27 & -13$\pm$28 & 776$\pm$65 & 262$\pm$31 \\
65 & 3.031 & 3.031 & 4.85$\pm$0.07 & 23.52$\pm$0.9 & 22$\pm$2 & 24$\pm$30 & 26$\pm$22 & 1044$\pm$73 & 415$\pm$16 \\
66 & 3.143 & 3.145 & 6.96$\pm$0.09 & 44.31$\pm$1.13 & 45$\pm$5 & 32$\pm$24 & 85$\pm$16 & 1070$\pm$56 & 427$\pm$13 \\
67 & 3.138 & 3.136 & 6.75$\pm$0.09 & 17.99$\pm$0.93 & 18$\pm$2 & 51$\pm$43 & 31$\pm$42 & 1424$\pm$100 & 565$\pm$34 \\
68 & 3.059 & 3.059 & 12.05$\pm$0.07 & 58.92$\pm$0.96 & 42$\pm$4 & 147$\pm$11 & 175$\pm$8 & 828$\pm$28 & 344$\pm$6 \\
69 & 3.148 & 3.146 & 4.37$\pm$0.09 & 23.07$\pm$1.06 & 20$\pm$2 & -50$\pm$60 & -81$\pm$44 & 1671$\pm$142 & 749$\pm$29 \\
70 & 3.145 & 3.142 & 4.93$\pm$0.08 & 15.38$\pm$0.67 & 15$\pm$2 & 31$\pm$22 & 70$\pm$21 & 859$\pm$53 & 365$\pm$15 \\
71 & 3.186 & 3.201 & 5.38$\pm$0.09 & 34.79$\pm$1.45 & 30$\pm$3 & 721$\pm$58 & 675$\pm$46 & 1937$\pm$137 & 800$\pm$36 \\
72 & 3.105 & \ldots & 1.22$\pm$0.07 & \ldots & \ldots & \ldots & \ldots & \ldots & \ldots \\
73 & 3.081 & 3.07 & 2.29$\pm$0.07 & 32.89$\pm$1.36 & 35$\pm$4 & -510$\pm$45 & -470$\pm$39 & 1637$\pm$106 & 628$\pm$32 \\
74 & 3.182 & 3.184 & 3.41$\pm$0.08 & 23.87$\pm$1.13 & 26$\pm$3 & 156$\pm$46 & 213$\pm$40 & 1478$\pm$108 & 598$\pm$29 \\
75 & 2.996 & 2.994 & 8.47$\pm$0.08 & 44.71$\pm$1.05 & 30$\pm$3 & 95$\pm$23 & 129$\pm$13 & 963$\pm$54 & 401$\pm$9 \\
76 & 3.193 & 3.196 & 17.93$\pm$0.1 & 17.54$\pm$1.02 & 12$\pm$1 & 389$\pm$46 & 423$\pm$49 & 1588$\pm$116 & 655$\pm$36 \\
77 & 3.113 & 3.111 & 3.93$\pm$0.08 & 33.5$\pm$0.93 & 27$\pm$3 & -60$\pm$18 & -83$\pm$16 & 1031$\pm$42 & 437$\pm$12 \\
78 & 3.092 & 3.098 & 2.32$\pm$0.07 & 19.18$\pm$1.15 & 12$\pm$1 & 409$\pm$98 & 388$\pm$68 & 1984$\pm$232 & 858$\pm$46 \\
79 & 3.156 & 3.149 & 8.87$\pm$0.08 & 23.38$\pm$0.75 & 20$\pm$2 & -232$\pm$16 & -286$\pm$15 & 849$\pm$38 & 359$\pm$10 \\
80 & 3.028 & 3.023 & 24.28$\pm$0.1 & 74.65$\pm$1.29 & 49$\pm$5 & -235$\pm$11 & -210$\pm$7 & 765$\pm$27 & 306$\pm$5 \\
81 & 3.026 & 3.027 & 2.93$\pm$0.07 & 16.61$\pm$0.48 & 16$\pm$2 & -262$\pm$7 & -272$\pm$7 & 395$\pm$17 & 154$\pm$6 \\
82 & 3.042 & \ldots & 0.74$\pm$0.15 & \ldots & \ldots & \ldots & \ldots & \ldots & \ldots \\
83 & 3.091 & 3.091 & 4.99$\pm$0.08 & 19.04$\pm$0.97 & 15$\pm$2 & 177$\pm$41 & 185$\pm$42 & 1460$\pm$97 & 630$\pm$30 \\
84 & 3.046 & 3.045 & 3.83$\pm$0.08 & 23.46$\pm$0.9 & 22$\pm$2 & -64$\pm$26 & -62$\pm$23 & 1037$\pm$60 & 427$\pm$16 \\
85 & 3.139 & \ldots & 2.58$\pm$0.08 & \ldots & \ldots & \ldots & \ldots & \ldots & \ldots \\
86 & 3.189 & 3.191 & 7.46$\pm$0.09 & 38.1$\pm$1.46 & 37$\pm$4 & 154$\pm$52 & 152$\pm$35 & 1648$\pm$122 & 696$\pm$23 \\
87 & 3.146 & 3.159 & 3.47$\pm$0.07 & 12.18$\pm$0.78 & 12$\pm$1 & 742$\pm$92 & 793$\pm$70 & 1956$\pm$216 & 788$\pm$52 \\
88 & 3.035 & 3.029 & 2.46$\pm$0.07 & 12.85$\pm$0.77 & 16$\pm$2 & -374$\pm$52 & -416$\pm$44 & 1316$\pm$122 & 535$\pm$31 \\
89 & 3.143 & 3.152 & 2.9$\pm$0.08 & 5.63$\pm$0.51 & 8$\pm$1 & 467$\pm$54 & 493$\pm$58 & 1101$\pm$129 & 420$\pm$51 \\
90 & 3.043 & 3.044 & 8.64$\pm$0.09 & 34.51$\pm$1.18 & 33$\pm$3 & 105$\pm$21 & 125$\pm$19 & 946$\pm$49 & 418$\pm$12 \\
91 & 3.045 & \ldots & 5.73$\pm$0.08 & \ldots & \ldots & \ldots & \ldots & \ldots & \ldots \\
92 & 3.115 & 3.112 & 5.31$\pm$0.08 & 14.78$\pm$0.81 & 16$\pm$2 & -40$\pm$39 & -47$\pm$38 & 1246$\pm$93 & 556$\pm$25 \\
93 & 3.046 & \ldots & 0.51$\pm$0.06 & \ldots & \ldots & \ldots & \ldots & \ldots & \ldots \\
94 & 3.178 & 3.181 & 5.75$\pm$0.08 & 20.42$\pm$0.88 & 21$\pm$2 & 89$\pm$24 & 97$\pm$28 & 1113$\pm$57 & 449$\pm$20 \\
95 & 3.024 & \ldots & 1.96$\pm$0.08 & \ldots & \ldots & \ldots & \ldots & \ldots & \ldots \\
96 & 3.118 & 3.12 & 8.57$\pm$0.08 & 73.94$\pm$1.44 & 49$\pm$5 & -71$\pm$17 & -91$\pm$13 & 1203$\pm$41 & 477$\pm$10 \\
97 & 3.092 & 3.096 & 7.41$\pm$0.08 & 14.47$\pm$0.77 & 17$\pm$2 & 221$\pm$32 & 275$\pm$31 & 1074$\pm$78 & 394$\pm$27 \\
98 & 3.074 & 3.077 & 4.26$\pm$0.07 & 22.24$\pm$1.02 & 17$\pm$2 & 1$\pm$48 & 30$\pm$46 & 1762$\pm$113 & 749$\pm$30 \\
99 & 3.105 & 3.106 & 3.95$\pm$0.08 & 11.72$\pm$0.76 & 10$\pm$1 & 203$\pm$67 & 215$\pm$61 & 1631$\pm$156 & 720$\pm$40 \\
100 & 3.059 & 3.06 & 2.57$\pm$0.09 & 8.17$\pm$0.65 & 7$\pm$1 & 179$\pm$56 & 278$\pm$61 & 1310$\pm$132 & 507$\pm$55 \\
101 & 3.084 & 3.078 & 7.25$\pm$0.08 & 26.66$\pm$1.24 & 25$\pm$3 & -233$\pm$54 & -287$\pm$50 & 1873$\pm$129 & 648$\pm$47 \\
102 & 3.193 & 3.197 & 8.22$\pm$0.1 & 44.26$\pm$1.23 & 35$\pm$3 & 42$\pm$15 & 21$\pm$14 & 918$\pm$35 & 370$\pm$11 \\
103 & 3.01 & 3.002 & 9.43$\pm$0.08 & 32.57$\pm$1.28 & 22$\pm$2 & -23$\pm$73 & 97$\pm$46 & 2028$\pm$172 & 756$\pm$37 \\
104 & 3.14 & 3.143 & 12.58$\pm$0.09 & 43.01$\pm$1.27 & 28$\pm$3 & 36$\pm$29 & 3$\pm$23 & 1439$\pm$68 & 634$\pm$16 \\
105 & 3.188 & 3.188 & 3.4$\pm$0.08 & 22.51$\pm$1.02 & 19$\pm$2 & 138$\pm$46 & 223$\pm$39 & 1500$\pm$108 & 623$\pm$30 \\
106 & 3.181 & 3.18 & 2.99$\pm$0.07 & 5.38$\pm$0.47 & 8$\pm$1 & 111$\pm$57 & 145$\pm$57 & 1152$\pm$136 & 382$\pm$58 \\
107 & 3.086 & 3.083 & 5.97$\pm$0.08 & 15.08$\pm$0.88 & 18$\pm$2 & -109$\pm$54 & -158$\pm$54 & 1617$\pm$129 & 623$\pm$45 \\
108 & 3.092 & 3.092 & 5.77$\pm$0.09 & 30.23$\pm$1.23 & 26$\pm$3 & -141$\pm$43 & -157$\pm$37 & 1606$\pm$102 & 618$\pm$29 \\
109 & 3.076 & 3.077 & 5.19$\pm$0.08 & 29.45$\pm$1.22 & 27$\pm$3 & 60$\pm$54 & 180$\pm$44 & 1805$\pm$127 & 751$\pm$32 \\
110 & 3.098 & \ldots & 1.76$\pm$0.07 & \ldots & \ldots & \ldots & \ldots & \ldots & \ldots \\
111 & 3.043 & 3.045 & 3.91$\pm$0.07 & 6.47$\pm$0.55 & 9$\pm$1 & 171$\pm$62 & 154$\pm$56 & 1151$\pm$146 & 372$\pm$65 \\
112 & 3.101 & 3.103 & 4.35$\pm$0.08 & 38.03$\pm$1.2 & 41$\pm$4 & 212$\pm$24 & 217$\pm$19 & 1033$\pm$57 & 440$\pm$12 \\
113 & 3.105 & 3.108 & 4.58$\pm$0.1 & 39.69$\pm$1.66 & 34$\pm$3 & 48$\pm$42 & 33$\pm$41 & 1719$\pm$99 & 699$\pm$31 \\
114 & 3.148 & \ldots & 2.50$\pm$0.07 & \ldots & \ldots & \ldots & \ldots & \ldots & \ldots \\
115 & 3.116 & 3.116 & 3.25$\pm$0.07 & 12.71$\pm$0.49 & 12$\pm$1 & -83$\pm$12 & -97$\pm$12 & 567$\pm$28 & 240$\pm$9 \\
116 & 3.069 & 3.069 & 8.03$\pm$0.08 & 22.27$\pm$1.0 & 19$\pm$2 & 14$\pm$42 & -18$\pm$40 & 1591$\pm$100 & 650$\pm$30 \\
117 & 3.186 & 3.186 & 2.4$\pm$0.08 & 30.33$\pm$1.19 & 34$\pm$3 & 67$\pm$37 & 60$\pm$33 & 1444$\pm$86 & 575$\pm$25 \\
118 & 3.108 & \ldots & 2.18$\pm$0.08 & \ldots & \ldots & \ldots & \ldots & \ldots & \ldots \\
119 & 3.176 & 3.175 & 2.87$\pm$0.08 & 19.03$\pm$0.82 & 22$\pm$2 & -11$\pm$23 & 2$\pm$22 & 874$\pm$53 & 356$\pm$17 \\
120 & 3.407 & 3.406 & 2.81$\pm$0.08 & 3.56$\pm$0.31 & 4$\pm$0 & 94$\pm$26 & 113$\pm$32 & 648$\pm$63 & 255$\pm$27 \\
\hline
\end{tabular}
\tablefoot{
\tablefoottext{a}{Peak \lya redshift. The intrinsic uncertainty on this value is $\Delta z\sim0.001$.} 
\tablefoottext{b}{The reported values are not corrected for instrumental broadening.}
}
\end{table*}

\begin{figure*}
    \centering
    \includegraphics[width=0.9\linewidth]{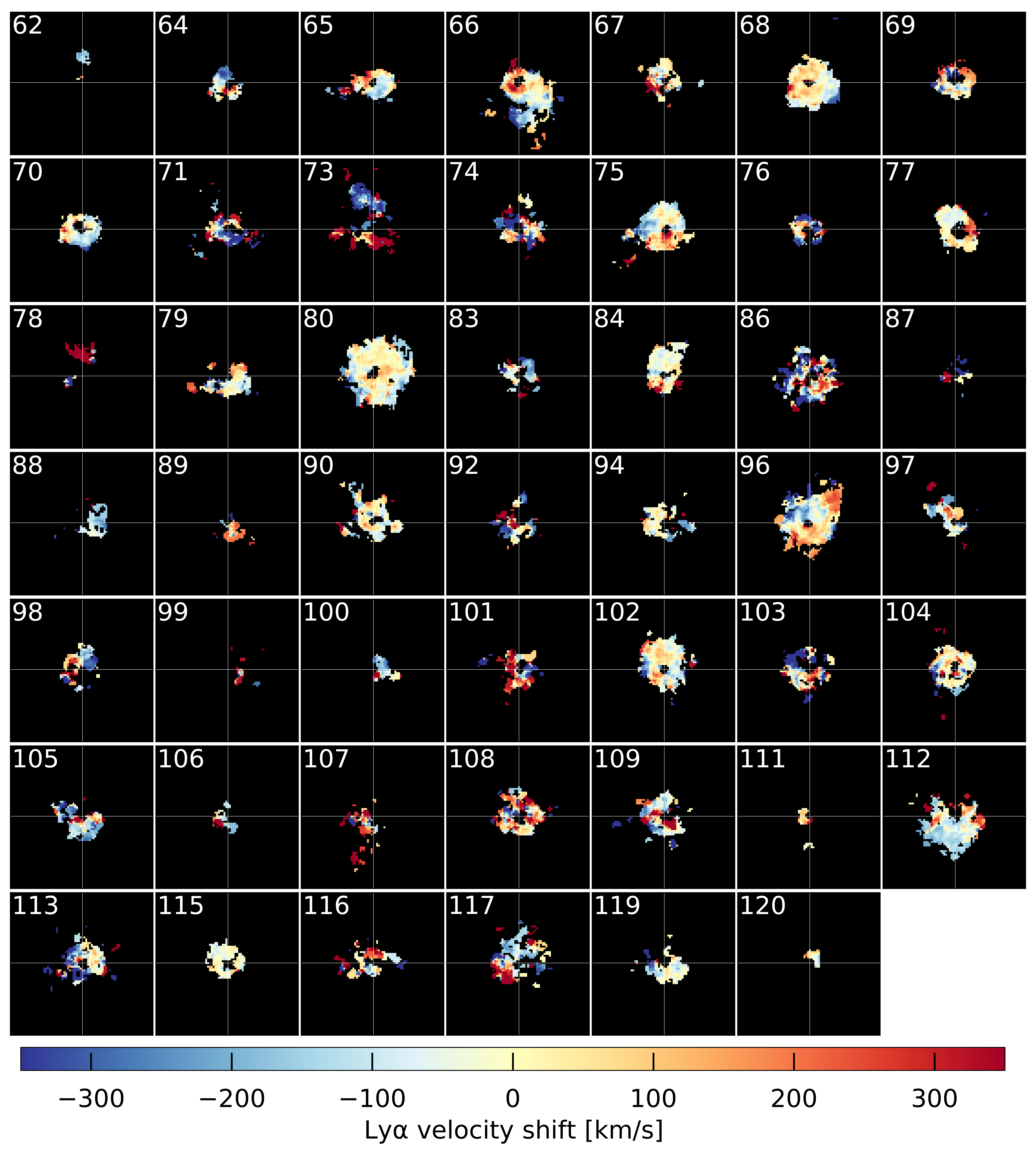}
    \caption{\lya velocity shift maps of the detected nebulae around $z\sim3$ faint quasars. The \lya velocity shift map is computed from the first moment of the PSF- and continuum-subtracted datacubes within a $\pm{\rm FWHM_{Ly\alpha}}$ with respect to the wavelength of the peak of the \lya emission of the nebula (Table~\ref{tab:properties}). 
    The panels are shown using the same projected scale as Figure~\ref{fig:SBmaps} ($\sim150\times150\,$kpc) and a white crosshair indicates the location of the quasar. The ID of each system is shown at the top left corner of each panel.}
    \label{fig:velocity-shift}
\end{figure*}
\begin{figure*}
    \centering
    \includegraphics[width=0.9\linewidth]{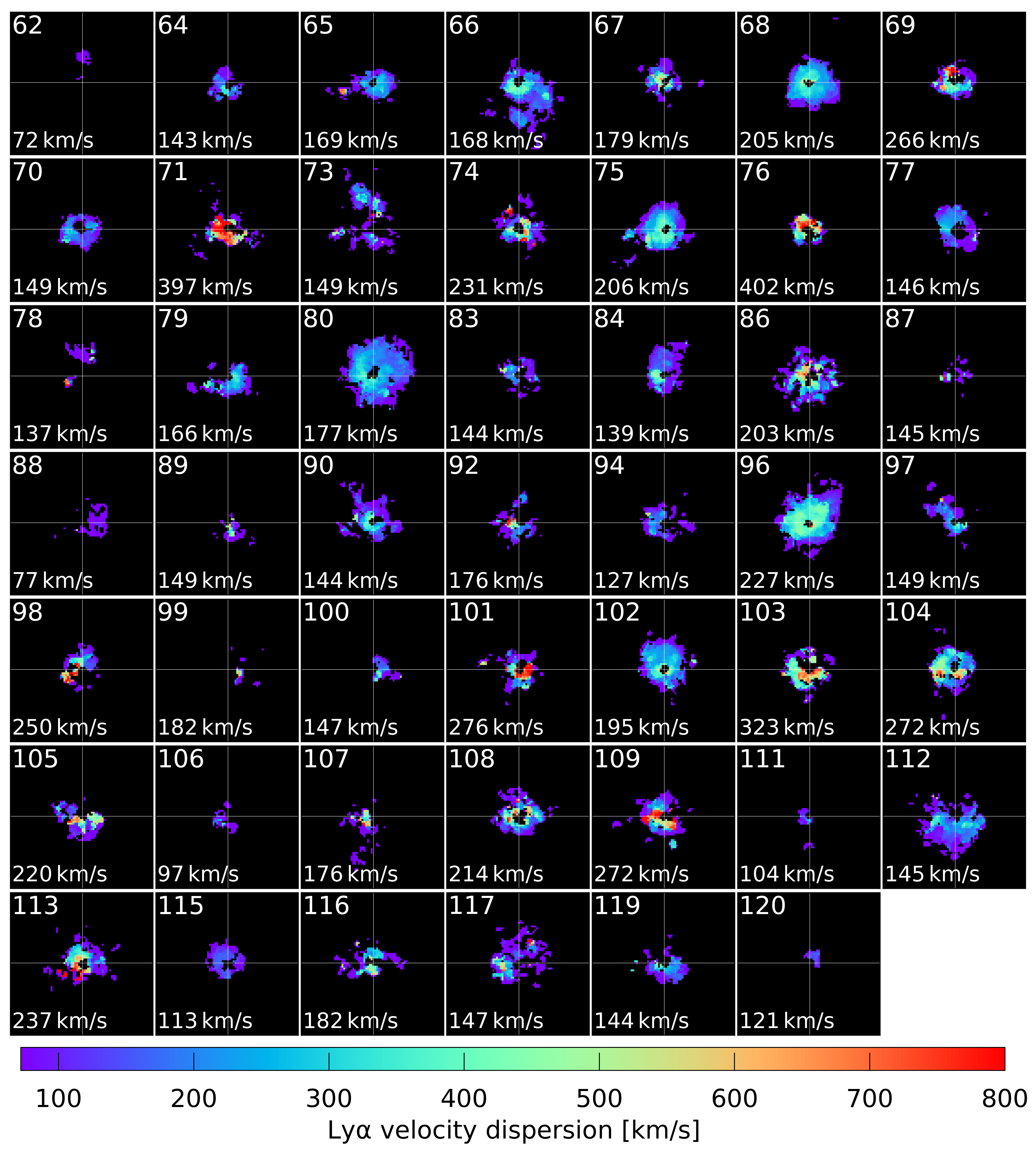}
    \caption{\lya velocity dispersion maps of the detected nebulae around  $z\sim3$ faint quasars. The \lya velocity dispersion map is computed from the second moment of the PSF- and continuum-subtracted datacubes within a $\pm{\rm FWHM_{Ly\alpha}}$ centered at the wavelength of the peak of the \lya emission of the nebula (Table~\ref{tab:properties}). 
    The panels are shown with the same projected scale as Figure~\ref{fig:SBmaps} ($\sim150\times150\,$kpc) and a white crosshair indicates the location of the quasar. The ID of each system and the average velocity dispersion within the mask are shown at the top and bottom left corner of each panel, respectively. The velocity dispersion of the faint quasars is on average lower than their bright counterparts (Figure~\ref{fig:velocity-dispersion_QSOMUSEUM}).}
    \label{fig:velocity-dispersion}
\end{figure*}

The final datacubes obtained in Section~\ref{sec:subtraction} contain only extended line emission, if detectable. In this study, we limit our investigation to the Ly$\alpha$  
transition, and we defer the analysis of additional emission lines (\ion{He}{ii}, \ion{C}{iv}) to subsequent works. We build \lya SB maps, integrated spectra, velocity shift and velocity dispersion maps using such PSF- and continuum-subtracted datacubes, and present them in Figures~\ref{fig:SBmaps},~\ref{fig:spec_atlas01},~\ref{fig:spec_atlas02},~\ref{fig:spec_atlas03},~\ref{fig:velocity-shift}~and~\ref{fig:velocity-dispersion}, respectively. 
In the following sections, we describe how nebulae are detected  and analyzed. 

\subsection{Nebula detection}\label{sec:detection}

Using the PSF- and continuum subtracted datacubes, we extract a spectrum inside a 1.5'' radius aperture centered at the quasar location, after masking the 1\,arcsec$^2$ normalisation region (see Section~\ref{sec:subtraction}). We use this spectrum to find \lya emission by setting a $3\sigma$ detection threshold based on the variance associated to the spectrum within the same aperture, then finding the peak emission above this threshold. This method is more sensitive to circular nebulae, however we expect the nebular emission to be rather uniform within this region (see Section~\ref{sec:properties}). 
We set the redshift of the detected nebulae using the wavelength of the peak \lya emission from this method and list them in Table~\ref{tab:properties}, where non-detections are marked with dashed lines. 
We report the detection of 110 nebulae out of the 120 targeted quasars. All the non-detections occur in the fainter quasar sample, 
first presented in this work. 
We further check both the detections and non-detections by building SB maps from 30~\AA\ pseudo-NB images and consider detected nebulae above 2$\sigma$ as usually done in the literature  (section~\ref{sec:SBmaps}). 

\subsection{Lyman-$\alpha$ SB maps and spectra}\label{sec:SBmaps}

We use the redshift of the nebulae identified in Section~\ref{sec:detection} to build \lya SB maps computed from $30\,\AA$ pseudo-NBs centered at the \lya line.
This wavelength range is chosen to allow comparison with previous studies. As in all former works (e.g., \citealt{Borisova2016}), we compute the residual background of each SB map after masking the location of continuum sources present in the original white image. 
We then subtract the background level from each SB map. 

The resulting SB maps for the 59 faint quasars and the 61 bright quasars are shown in Figures~\ref{fig:SBmaps}~and~\ref{fig:SBmaps_QSOMUSEUM}, respectively, after smoothing using a 2D Box kernel of 3 pixels. For non-detections, we build the SB maps in a $30\,\AA$ pseudo-NB centered at the peak Ly$\alpha$ emission of the quasar spectra. 
The SB maps of Figure~\ref{fig:SBmaps} correspond to $\sim150\,{\rm kpc\,}\times150$\,kpc at the median redshift of the sample and show that the \lya emission surrounding the fainter quasars appear dimmer and more compact than that observed around their bright quasar counterparts (Figure~\ref{fig:SBmaps_QSOMUSEUM}), consistent with the findings of 
\citet{Mackenzie2021}. 
As done in previous works, the detected nebulae are defined out to their $2\sigma$ isophote in the \lya SB maps. We use this isophote 
to characterize the nebulae physical properties in Section~\ref{sec:properties}.

For the 59 faint quasars, we integrate the spectra inside the $2\sigma$ isophotes and present the obtained one-dimensional spectra in Figures~\ref{fig:spec_atlas01},~\ref{fig:spec_atlas02}~and~\ref{fig:spec_atlas03} with red curves. For the non-detections, we show the spectrum integrated within a 1.5'' radius aperture in gray. 
The quasar spectra within a 1.5'' radius aperture are shown in black for comparison. 
The blue dashed line represents the wavelength of the peak \lya emission of the quasar. 
In most cases, the \lya emission of the detected nebulae resembles those of their associated quasars, even though the former is narrower, 
i.e, with similar absorption features and small shifts in wavelength compared to the \lya line of the quasar. 
Moreover, some systems display absorption features in the quasar and nebula spectra at the same wavelength (e.g, ID 87, 96, 103, 113), which could arise due to \lya radiative transfer effects, outflows or dense material in front of the system (e.g., \citealt{vanOjik1997,Gronke2015,Cai2018,FAB2019b}). However, the detailed modeling of these individual features is not the focus of this paper and we defer their interpretation for future research. 
Finally, there are cases (like ID 71, 73, 81, 92) in which the \lya emission occurs at wavelengths corresponding to absorption in the quasar spectrum. This result is similar to what has been found in the smaller sample studied in \citet{Mackenzie2021}. The resulting spectra for the 61 bright quasars from the QSO MUSEUM I sample 
are presented in Figures~\ref{fig:spec_atlas01_bright},~\ref{fig:spec_atlas02_bright}~and~\ref{fig:spec_atlas03_bright} of Appendix~\ref{sec:QSOMUSEUMI}, and shows that our analysis is consistent with that work.

\subsection{Nebulae morphologies, areas, integrated luminosities and kinematics}\label{sec:properties}

We characterize the extended \lya emission surrounding each quasar 
by 
integrating its \lya luminosity, area within the 2$\sigma$ isophote, 
morphology, 
velocity shift with respect to the quasar \lya line  
and linewidth, which we summarize in Table~\ref{tab:properties}. Additionally, we obtain the spatial kinematics of the nebulae using the first and second moment maps which are presented in Figures~\ref{fig:velocity-shift}~and~\ref{fig:velocity-dispersion}.

First, we characterize the \lya line spectra integrated inside the $2\sigma$ isophotes by fitting a gaussian function, from which we compute the centroid wavelength and full width at half maximum (FWHM). 
We estimate the total \lya luminosity of the nebulae by integrating the spectra within a wavelength range of $\pm$FWHM ($\pm2.355\sigma$) centered on the \lya line. This choice of wavelength range is selected so that we are considering most of the flux distribution into the calculation of the luminosity while excluding the noisiest part. 
The centroid of the gaussian fit is used to compute the velocity shift of the nebula with respect to the quasar's peak \lya emission. 
All these quantities are also listed in Table~\ref{tab:properties}. 
We set an upper limit of $14\,\AA$ for the standard deviation when performing the gaussian fit in some nebulae that show multiple components and/or absorption features in their integrated spectrum (see Table~\ref{tab:properties_QSOMUSEUM} and Figures~\ref{fig:spec_atlas01_bright}-\ref{fig:spec_atlas03_bright}), therefore focusing the fit on the brightest component. 

We also note that a gaussian function is, in general, not a good representation of the \lya line profile due to absorption features and possible radiative transfer effects affecting the propagation of \lya photons. Therefore, we compute, for comparison, the first and second moment of the \lya flux distribution to estimate the line centroid and linewidth within the $\pm$FWHM range obtained from the gaussian fit above. 
We find good agreement within uncertainties between the estimates from the gaussian fit and flux weighted moments of the \lya line flux distribution. We list in Table~\ref{tab:properties} the flux weighted first moment of the line in terms of the velocity shift with respect to the quasar peak \lya wavelength and the linewidth obtained from the second moment of the flux distribution.

In order to describe the kinematics of the observed nebulae, we compute the velocity shift and velocity dispersion maps shown in Figures~\ref{fig:velocity-shift}~and~\ref{fig:velocity-dispersion} for the faint quasars and Figures~\ref{fig:velocity-shift_bright}~and~\ref{fig:velocity-dispersion_QSOMUSEUM} for the bright quasars. These maps are computed using the first and second moment of datacubes 
constructed within the $\pm$FWHM range obtained from the gaussian fit and centered at the peak \lya wavelength of the nebulae.
The calculation of the moment maps is restricted only to regions with SNR>3 in the respective SB maps obtained using a 30\,\AA\ pseudo-NB from the aforementioned datacubes, after applying a spatial smoothing with a Gaussian kernel of $0.5\arcsec$. 
This spatial SNR$>3$ mask is applied to the PSF- and continuum- subtracted cubes before computing the first and second moment maps\footnote{The chosen conservative procedure leads to the exclusion of the nebula surrounding ID 81, as it has emission detected at a lower significance in the corresponding SB map.}. The velocities are computed relative to the wavelength of the peak of each extended Ly$\alpha$ emission spectra. 

The velocity shift maps 
of the faint sample (Figure~\ref{fig:velocity-shift}) are complex and diverse, with some nebulae presenting velocity components that roughly span the range $-200$~km~s$^{-1}$ to $+200$~km~s$^{-1}$ with respect to 
the peak \lya of the nebula (e.g., ID 66, 73, 96). Other nebulae appear to have more quiescent kinematics with velocity variations smaller than $100$~km~s$^{-1}$ from the \lya (e.g., ID 68, 80). 
On the other hand, the velocity shift maps of the bright sample (Figure~\ref{fig:velocity-shift_bright}) display similar variety of complex velocity structures, with both quiescent and disturbed features in their nebulae. In particular, the largest velocity gradients for the bright quasars span a larger range than the faint nebulae of up to $\sim600$~km~s$^{-1}$ across the peak \lya wavelength.

Most of the velocity dispersion maps of the faint quasars nebulae (Figure~\ref{fig:velocity-dispersion}) 
show quiescent kinematics with velocity dispersions around $100-200$~km~s$^{-1}$ and 
a tendency to increase up to $300-400$~km~s$^{-1}$ towards the location of the quasar. On the other hand, the bright quasar nebulae (Figure~\ref{fig:velocity-dispersion_QSOMUSEUM}) have on average larger velocity dispersions $\sim300$~km~s$^{-1}$ and a similar tendency to increase up to $500-600$~km~s$^{-1}$ towards the center. For both samples we see some cases where the velocity dispersion quickly increases to much higher values ($800-1000$~km~s$^{-1}$) at the center (radius smaller than $\sim25$~kpc). We further explore these findings in Section~\ref{sec:radial-profiles}.

\begin{figure}
    \centering
    \includegraphics[width=0.95\linewidth]{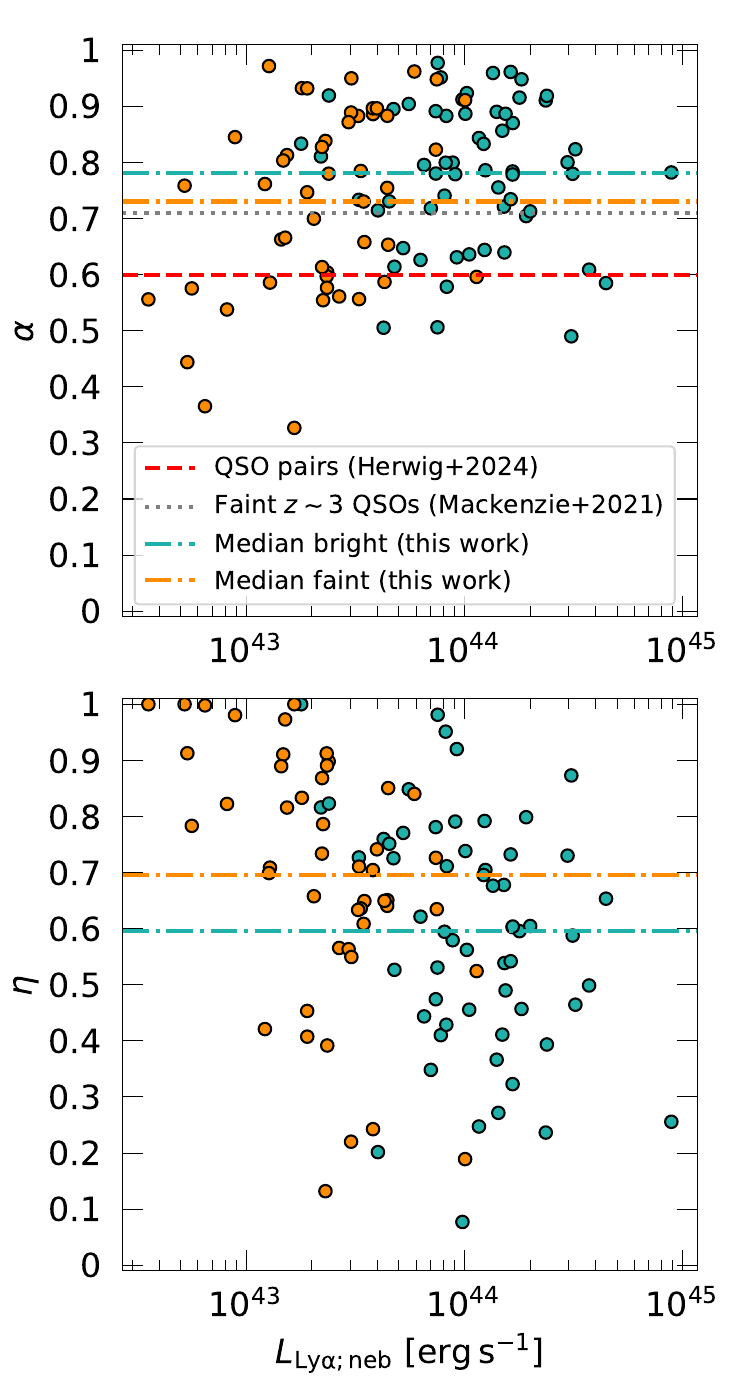}
    \caption{
    Level of elongation ($\alpha$) and lopsidedness ($\eta$) of the observed extended \lya emission around quasars. 
    In both panels, the orange and dark green points indicate the faint and bright sample of QSO MUSEUM III, respectively. The median values of the faint and bright samples are indicated with dot-dashed lines of the same colors. 
    \emph{Top:} Elongation $\alpha$ computed from the Stokes parameters (see Section~\ref{sec:properties}) within the $2\sigma$ isophote of the detected nebulae as a function of nebula \lya luminosity. Larger values of $\alpha$ indicate a rounder morphology. Additionally, we show the median values of $\alpha$ reported in the faint $z\sim3$ quasar sample from \citet{Mackenzie2021} and quasar pairs from QSO MUSEUM II with a gray dotted and red dashed line, respectively. 
    \emph{Bottom:} Nebula lopsidedness $\eta$ as presented in \citet{FAB2023} and described in the text. High values correspond to all emission within the $2\sigma$ isophote being on one side of the quasar. }
    \label{fig:asymmetry-luminosity}
\end{figure}

The \lya SB maps in Figures~\ref{fig:SBmaps}~and~\ref{fig:SBmaps_QSOMUSEUM} 
display a wide variety of morphologies, with some nebulae appearing centrally concentrated around the quasar position (e.g., ID 1, 39, 68, 80 among others), some seem to appear lopsided towards one direction from the quasar (e.g., ID 11, 36, 66, 109, 112) and finally some cases show large scale coherent structures (e.g., ID 13, 50, 56, 96). The strength of these effects could have an origin on the different powering mechanisms of extended \lya nebulae, such as photoionization due to anisotropic quasar radiation \citep[e.g,][]{Obreja2024} or \lya photons propagation 
\citep[e.g,][]{Costa2022}. Moreover, the origin could be related to the gas distribution around each system \citep[e.g.,][]{Cai2019} or viewing angle \citep[e.g,][]{Costa2022}. Finally, the environment or presence of companions could affect the observed \lya nebulae morphology \citep[see e.g.,][]{FAB2022,GonzalezLobos2023,Herwig2024}. We further discuss these scenarios in Section~\ref{sec:discussion} and here 
we attempt to quantify the observed nebulae morphologies using two different measurements. 

First, we quantify the asymmetry of the SB maps as previously done in the literature (e.g., \citealt{FAB2019}; \citealt{Herwig2024}), defined as the ratio between the semi-minor and semi-major axis within the 2$\sigma$ isophotes used to describe the area of the detected nebulae (Tables~\ref{tab:properties}~and~\ref{tab:properties_QSOMUSEUM}). 
The ratio can be obtained as: 
\begin{equation}
    \label{eq:asymmetry-stokes}
    \alpha = \dfrac{(1-\sqrt{Q^2 + U^2})}{1+\sqrt{Q^2 + U^2}}
\end{equation}
where $Q$ and $U$ 
are the stokes parameters computed from the flux weighted second order moments of the image following Equation~1 in \citet{FAB2019} within the $2\sigma$ isophote. 
The value of $\alpha$ corresponds to the aspect ratio of the isophote, with $\alpha=1$ representing a perfectly circular distribution. 
The top panel of Figure~\ref{fig:asymmetry-luminosity} shows the resulting $\alpha$ as a function of integrated \lya nebula luminosity (Tables~\ref{tab:properties}~and~\ref{tab:properties_QSOMUSEUM}) within the same isophote for the faint and bright quasars with orange and dark green circles, respectively. 
Most (90~\%) of the observed nebulae have values of $\alpha\gtrsim 0.6$, i.e., the nebulae tend to be rounder. 
Also, the median value of $\alpha$ estimated for the bright sample (green dot-dashed line) is consistent with the value presented in QSO MUSEUM I ($\alpha=0.77$), considering the different data reduction and analysis used in this work. 
Similarly, we show that the $\alpha$ reported in the faint quasar sample from  \citet{Mackenzie2021} (gray dotted line) appears to be consistent with the faint quasars presented in this work (orange dot-dashed line).
Additionally, 
the median $\alpha$ of the faint sample is smaller than that of the bright quasars, indicating that faint quasars present slightly more elongated nebulae. 
To further test this, we estimate the average error in the calculation of $\alpha$ of the sample following the method presented in \citet{Herwig2024}, who randomized the centroid coordinates in the flux weighted second moments in Equation~1 of \citet{FAB2019} and assumed the uncertainty is dominated by the seeing. 
This exercise results in a mean uncertainty of $\Delta\alpha=0.05$ which is comparable to the difference between the two median values of the bright and faint samples, therefore we cannot determine whether such a difference is significant.  
Further, we measure the Spearman's rank correlation coefficient between $\alpha$ and $L_{\rm Ly\alpha; neb}$ and obtain 0.19 with a $p$-value of 0.062, indicating a weak positive correlation that has low statistical significance. The same test between $\alpha$ and bolometric luminosities gives a similar result as expected due to the correlation between $L_{\rm bol}$ and $L_{\rm Ly\alpha; neb}$. Therefore, we infer that there is not a dependence between the elongation and luminosity of the nebulae or quasars. 
Also, 
the lower median $\alpha$ estimated for the faint sample is likely due to five of the dimmest nebulae ($L_{\rm neb}<10^{43}\,{\rm erg\,s^{-1}}$) studied here being associated with values representing more asymmetric or lopsided nebulae ($\alpha$<0.6). 
For comparison, we show the data-points of \lya nebulae around 
associated projected quasar pairs from QSO MUSEUM II (red dashed line).
Those systems present the lowest median $\alpha$, which can be explained by the elongated morphology tracing the direction connecting the pairs. 
Summarizing, we find that there is no clear indication of a trend between $\alpha$ and $L_{\rm Ly\alpha;neb}$, which is also reported in those three works we compared to. 

In addition, we use the same $2\sigma$ isophotes to compute the asymmetry as done in \citet{FAB2023}. 
In that work, the authors describe the asymmetry based on the ratio of areas $\eta=(A_{\rm max}^{\rm neb}-A_{\rm min}^{\rm neb})/A_{\rm max}^{\rm neb}$. 
Where $A_{\rm max}^{\rm neb}$ and $A_{\rm min}^{\rm neb}$ are computed as the maximum and minimum area on both sides of the direction that crosses the quasar position and which maximizes the difference between the two covered areas. 
In this case, $\eta$ represents the level of lopsidedness of the nebulae, i.e., a value of $\eta=1$ corresponds to all of the emission being on one side of the quasar. 
We plot $\eta$ as a function of nebula luminosity within the $2\sigma$ isophote in the bottom panel of Figure~\ref{fig:asymmetry-luminosity} using the same symbols as the left panel. 
This figure shows that the dimmest nebulae appear more lopsided with respect to brighter nebulae. 
This is similar to the results found in \citet{FAB2023}, where at a fixed SB threshold the smallest and dimmest nebulae appear more lopsided. Here, we do not use a fixed SB cut, but the systems studied here are found to lie close to the luminosity-area relation presented in that work for a common observed SB threshold corresponding to  $2.46\times 10^{-17}{\rm erg\,s^{-1}\,cm^{-2}\,arcsec^{-2}}$ at $z=2.0412$ (see their Table~1). 
Likewise, we compute the Spearman's rank correlation coefficient between $\eta$ and $L_{\rm Ly\alpha; neb}$ and find -0.29 with a $p$-value of 0.01 indicating a weak negative correlation.
Particularly, we report that the systems with $\alpha<0.5$ and $\eta\sim1$ correspond to ID 62, 81 and 99 which have very low signal to noise and their emission on the SB maps appear irregular. 
For these systems the constructed $2\sigma$ isophote lies on one side of the quasar, which is a consequence of their clumpy morphology due to low SB levels. 
On the other hand, 
some bright systems appear lopsided and this could be an
indication of larger scale structures. 
However, 
there is no clear trend between $\eta$ and the nebula \lya luminosity for nebulae around bright quasars.

By combining the information in both panels, we observe that most of the nebulae tend to have a circular morphology, but their emission tends to be preferentially towards one side of the quasar, i.e., the nebulae emission is asymmetric with respect to the quasar position. However, 
the values of $\eta$ span from centered to lopsided nebulae and there is no discernible strong trend in relation to nebula luminosity or $\alpha$. Finally, we note that the values of $\alpha$ and $\eta$ are sensitive to the SB limit of each observation. This effect is particularly important for fainter nebulae, for which it is possible that the morphology would change if their fainter outskirts could be detected. 
We briefly discuss the possible implications of these observations in Section~\ref{sec:pow_mech}.

\subsection{Surface brightness and velocity dispersion radial profiles}\label{sec:radial-profiles}

\begin{figure*}
    \centering
    \includegraphics[width=\linewidth]{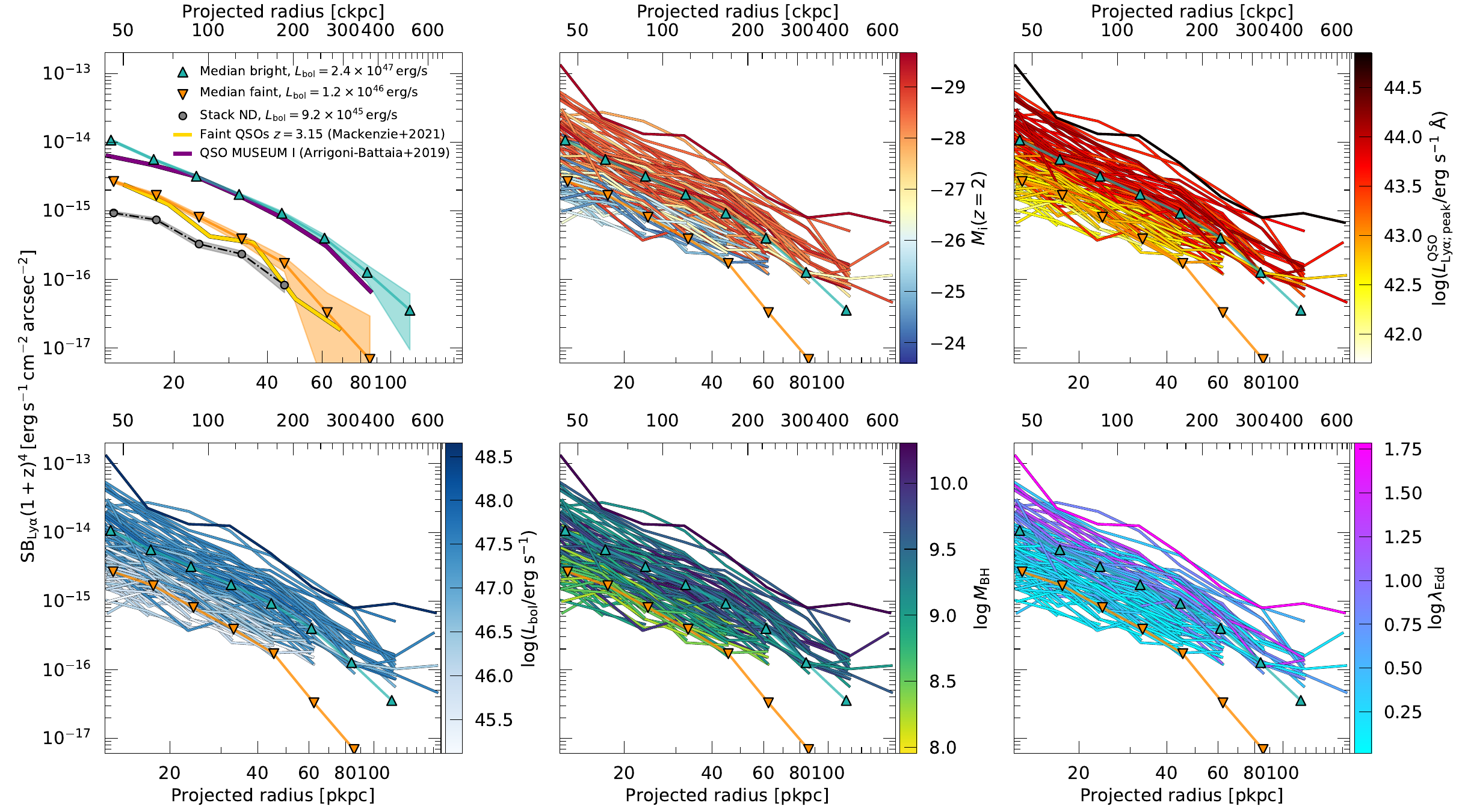}
	\caption[\lya SB radial profiles of the QSO MUSEUM III nebulae.]{\lya SB radial profiles of the QSO MUSEUM III nebulae. The profiles of the detected nebulae are computed by averaging the \lya SB maps of Figures~\ref{fig:SBmaps}~and~\ref{fig:SBmaps_QSOMUSEUM} inside annuli with logarithmically increasing radius centered at the quasar location, after masking the $1''\times1''$ normalization region. We cut the profiles when they reach values below the $2\sigma$ SB limit within the corresponding radial bin, 
 and only plot profiles that have at least two data-points satisfying this criteria (108/110 profiles). 
	The \lya SB 
    corrected by cosmological dimming 
    is shown as a function of physical (bottom axis) and comoving (top axis) projected distances. 
	\emph{Top left:} Median profiles of the faint and bright sample (orange and green triangles) and stacked non-detections (gray). Additionally, the median profiles from \citet{Mackenzie2021} and \citet{FAB2019} are shown with a yellow and purple line, respectively. 
	\emph{Top middle:} The profiles are color coded by the quasar's absolute $i$-band magnitude normalized to $z=2$. 
	\emph{Top right:} The profiles are color coded by the peak of the Ly$\alpha$ luminosity density of the quasar. 
	\emph{Bottom left:} The profiles are color coded by the bolometric luminosity of the quasar, computed from the monochromatic luminosity at 1350\AA\ (Appendix~\ref{app:QSOfit}). 
	\emph{Bottom middle:} The profiles are color coded by their quasar's black-hole mass (Appendix~\ref{app:QSOfit}). 
	\emph{Bottom right:} The profiles are color coded by their quasar's Eddington ratio (Appendix~\ref{app:QSOfit}).}    
    \label{fig:Rprofiles}
\end{figure*}
\begin{figure*}
    \centering
    \includegraphics[width=0.8\linewidth]{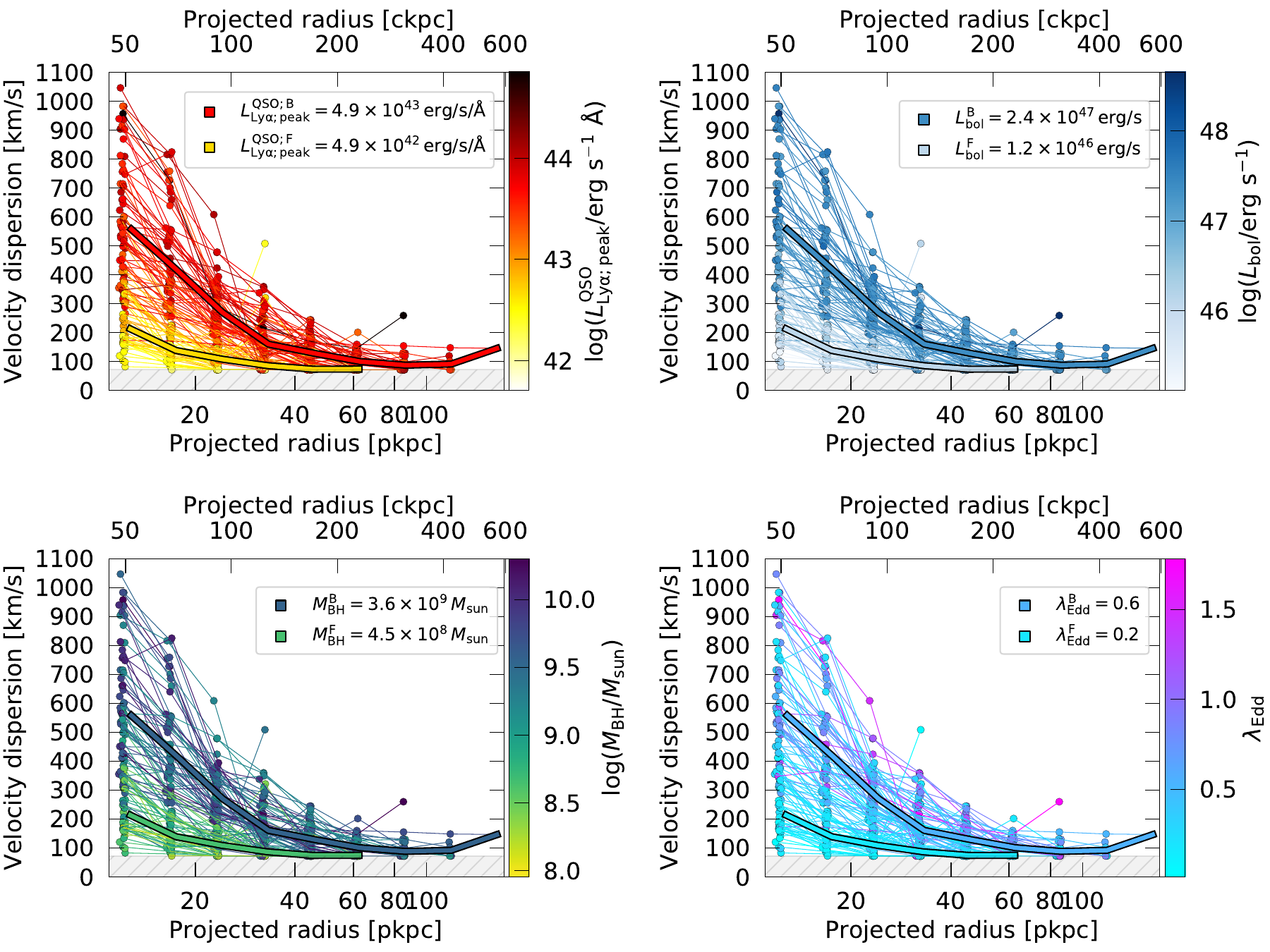}
    \caption{Individual \lya velocity dispersion radial profiles of the QSO MUSEUM III nebulae. Each profile is computed by averaging the \lya velocity dispersion maps from Figures~\ref{fig:velocity-dispersion}~and~\ref{fig:velocity-dispersion_QSOMUSEUM} inside the 
    SNR~$>3$ mask, using the same annuli as the profiles from Figure~\ref{fig:Rprofiles}. \emph{Top left:} The profiles are color coded by the peak \lya luminosity density of the quasar. \emph{Top right:} The profiles are color coded by the bolometric luminosity of the quasar. \emph{Bottom left:} The profiles are color coded by the black hole mass of the quasar. \emph{Bottom right:} The profiles are color coded by the Eddington ratio of the quasar. Additionally, we show in each panel the median velocity dispersion profile of the faint and bright samples, color coded by each property with the median value of each bin indicated at the top right corner of the panel. 
    In each panel, values below the MUSE spectral resolution limit ($\sigma=72$\,km~s$^{-1}$) are shown with a dashed gray area.}
    \label{fig:sigma-profiles}
\end{figure*}

We have shown that the nebulae uncovered in this work are characterized by a diversity of morphologies, luminosities and kinematics. In this section, we quantify those differences by building radial profiles using the SB maps of Figures~\ref{fig:SBmaps}~and~\ref{fig:SBmaps_QSOMUSEUM} and the velocity dispersion maps of Figures~\ref{fig:velocity-dispersion}~and~\ref{fig:velocity-dispersion_QSOMUSEUM}. 
The \lya SB radial profiles are computed by averaging the maps inside annuli with logarithmically increasing radii centered at the quasar position, then corrected by cosmological dimming, as usually done in the literature \citep{FAB2019,GonzalezLobos2023}. We extract the SB radial profiles for all systems and compute the median profile for the faint and bright sample and show them in the top left panel of Figure~\ref{fig:Rprofiles} with orange and blue triangles, respectively. These profiles are plotted only out to the last radius where they exceed the $2\sigma$ detection limit of the stack, and the shaded areas represent their mean uncertainty. 
The profile shapes for the faint and bright samples are strikingly similar, however the SB normalisation factor of the bright sample is about 8 times larger than the faint. The median values of each AGN property within the faint and bright sample bins are shown in the legend of Figure~\ref{fig:sigma-profiles}.
For comparison, we show the median profile of 12 similarly faint $z\sim3$ quasars from \citet{Mackenzie2021} (yellow line), and the analysis of the bright sample in QSO MUSEUM I (purple line). 
The agreement between our analysis and those in previous studies further substantiates the validity of our approach.

Additionally, we construct a stacked SB map for the ten non-detections by median combining 30~\AA\ pseudo-NB images centered at the wavelength of the peak of the quasar \lya emission. This decision is based on the frequent observation that \lya nebulae peak emission occurs at a similar wavelength of the quasar \lya peak emission (\citealt{FAB2019,Cai2019} and Section~\ref{sec:pow_mech}). We similarly propagate the associated variances to obtain errors. The stacked map shows a clear detection of which we show its SB profile in the top left panel of Figure~\ref{fig:Rprofiles} with gray circles and shaded area as the mean uncertainty. This SB profile is at the low end of all the observed individual nebulae and bolometric luminosities, confirming that these sources are just below our detection limit for individual systems.

Further, the remaining panels in Figure~\ref{fig:Rprofiles} illustrate the individual \lya SB profiles, which are color-coded according to their quasar properties: 
the absolute $i$-band magnitude normalized at $z=2$ (top middle), the quasar's peak Ly$\alpha$ luminosity density (top right), the quasar's bolometric luminosity (bottom left), the quasar's black hole mass (bottom middle) and Eddington ratio (bottom right). Each profile is cut when it reaches a value below the $2\sigma$ SB limit within a radial bin.  
Also, we only show profiles with more than two data-points, resulting in a total of 108 profiles. 
The profiles span almost 2 orders of magnitude in their SB level and their shapes are similar across the whole sample. We further study the shape of the SB profiles by fitting a power and exponential function in Section~\ref{sec:stacked-profiles}. 
Figure~\ref{fig:Rprofiles} confirms -- with $4\times$ more systems -- the correlation between quasar \lya and UV luminosities and \lya SB of the nebulae, also found in \citet{Mackenzie2021}.
In general, the profiles of nebulae around the faint sample have \lya SB about one order of magnitude lower than the nebulae around bright quasars. Also, the extended \lya around the bright quasars is detected out to projected distances of $\sim100$~pkpc, while for the faint quasars the \lya is detected mostly out to 60~pkpc. Additionally, the SB profile normalisation factor increases as the peak \lya luminosity and bolometric luminosity of the quasar increase.
Such trend is not as clearly apparent with the absolute magnitude, as the profiles present larger \lya SB scatter at a fixed quasar magnitude. Similarly, the black hole masses show scatter across the range of \lya SB.
Finally, there is a tendency for the brighter nebulae to also present the more extreme Eddington ratios, suggesting a potential link between nebula SB and AGN accretion rate. 

In addition, we present in Figure~\ref{fig:sigma-profiles} radial profiles that were constructed similarly to the SB profiles, but using the velocity dispersion maps of Figures~\ref{fig:velocity-dispersion}~and~\ref{fig:velocity-dispersion_QSOMUSEUM}. 
These profiles are computed by averaging inside each annuli only where there is velocity dispersion data (within the SNR$>3$ mask of Figures~\ref{fig:velocity-dispersion}~and~\ref{fig:velocity-dispersion_QSOMUSEUM}). We overlay in the figure the spectral resolution limit of MUSE at the median \lya wavelength ($\sigma=72\,$km~s$^{-1}$)  with a shaded gray region. Note that the profiles approach this limit as the distance to the center increases, indicating that we need higher spectral resolution to resolve the lines 
at such distances. However, the velocity dispersion profiles tend to increase towards the quasar position. We color code the profiles by their quasar properties such as the peak \lya luminosity density of the quasar (top left), the quasar's bolometric luminosity (top right), the black hole mass (bottom left) and Eddington ratio (bottom right). We see a wide range of velocity dispersions, ranging from $\sim100$~km~s$^{-1}$ up to $\sim1000$~km~s$^{-1}$ within the innermost regions of the CGM of the targeted quasars ($\sim12$~kpc). This scatter decreases to about $300$~km~s$^{-1}$ as the projected distance increases ($\sim40$~kpc), due to the steep decrease on the velocity dispersion of the nebulae around the brightest quasars. 
Indeed, 
the median velocity dispersion profile for the faint and bright quasar sample shown in Figure~\ref{fig:sigma-profiles} with thicker lines indicates that the velocity dispersion of bright quasar nebulae are about three times larger than fainter quasars at $\sim12$~kpc and only two times larger at $\sim20$~kpc. 
If we compare the velocity dispersion radial profiles to properties such as black hole masses and Eddington ratios we observe a similar behavior but 
less 
significant, probably because of the larger scatter of these properties across the range of luminosities probed with our sample.


\section{Discussion}\label{sec:discussion}

\begin{figure*}
    \centering
    \includegraphics[width=0.5\linewidth]{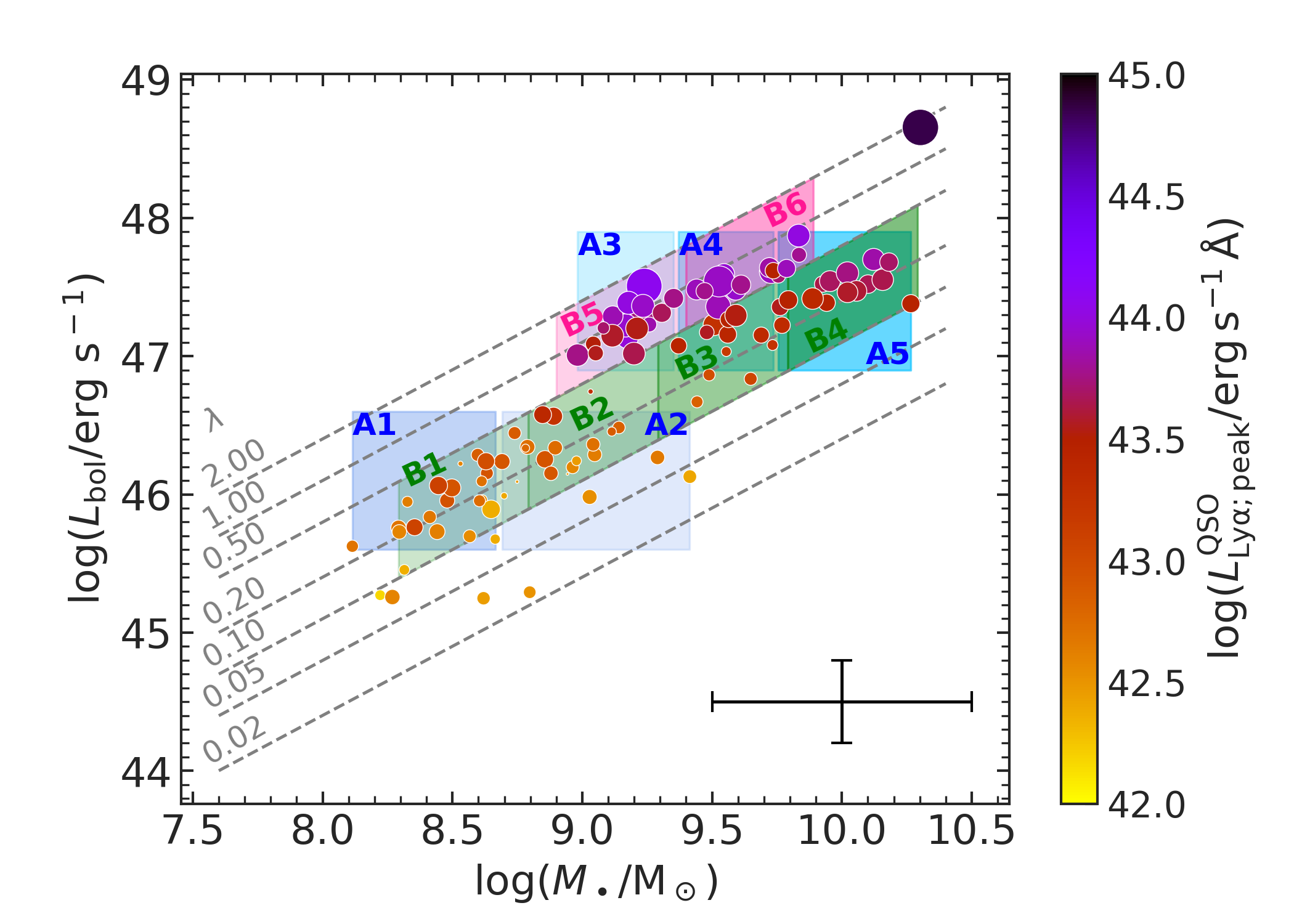}\includegraphics[width=0.5\linewidth]{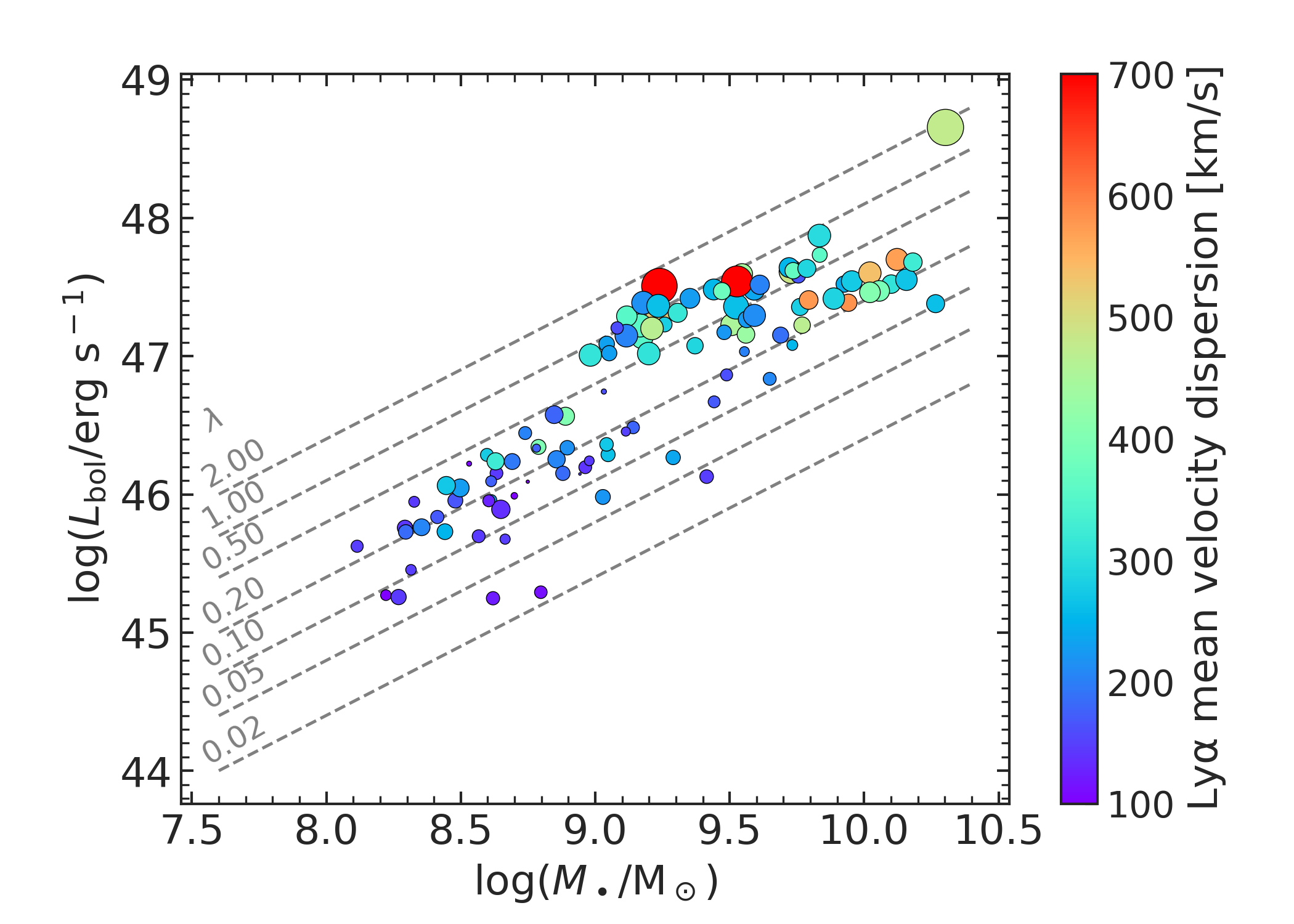}
    \caption{Comparison of observed quasar properties with nebulae properties. We show for each detected nebulae, their quasar bolometric luminosity as a function of black hole mass. The size of each point is proportional to the integrated \lya luminosity of the nebula divided by the total nebula area in kpc$^2$, with larger symbols having larger ratios. Different values of constant Eddington ratios ($\lambda_{\rm Edd}$) are shown with gray dashed lines. \emph{Left:} The points are color coded by the peak \lya luminosity density of each quasar (Table~\ref{tab:properties_QSOMUSEUM}). The black errorbars in the lower right corner indicate the intrinsic uncertainty of the $M_{\rm BH}$ and $L_{\rm bol}$ estimators (see Section~\ref{sec:qso_phys_prop}). 
    \emph{Right:} The points are color coded by the mean velocity dispersion shown in the maps of Figures~\ref{fig:velocity-dispersion}~and~\ref{fig:velocity-dispersion_QSOMUSEUM}. Additionally, we overlay in the left panel the bins covering different quasar properties described in Section~\ref{sec:stacked-profiles}.}
    \label{fig:parameter_space}
\end{figure*}

By targeting 59 additional faint ($-27<M_{\rm i}(z=2)-24$) $z\sim3$ quasars, this work extends the current catalog of known \lya nebulae by 49 additional detections, covering two orders of magnitude fainter systems than the bolometric luminosities studied in the QSO MUSEUM I survey (see Figure~\ref{fig:LbolLpeak}), 
and representing a factor of $4\times$ more nebulae than are currently available at these quasar luminosities. 
This extended dynamic range 
allows us, for the first time with sufficient statistics, to study the CGM of quasars 
in emission 
across the faint and bright ends of the known $z\sim3$ quasar population (Figure~\ref{fig:Lpeak}), and link it to their AGN properties in addition to the physical mechanisms driving the emission.

\subsection{Stacked surface brightness and velocity dispersion radial profiles}\label{sec:stacked-profiles}

We design figures that summarize the main results on how the properties of
quasar nebulae scale with quasar properties. 
The left panel of Figure~\ref{fig:parameter_space} 
shows the bolometric luminosity as a function of black hole mass of the 110 quasars with detections in a log-log plot, color coded by the peak \lya luminosity density of the quasars. 
In this plot, quasars with similar Eddington ratios follow gray dashed diagonal lines. In the lower right corner of the same panel, we indicate with black errorbars the intrinsic uncertainty of the $\log(M_{\rm BH}/M_\odot)$ and $\log(L_{\rm bol}/L_\odot)$ estimators (Section~\ref{sec:qso_phys_prop}), which correspond to 0.5 and 0.3 dex, respectively. We discuss the effect of these uncertainties further in Section~\ref{sec:stacking-bins}. 
Additionally in this figure, the size of the symbols is proportional to the ratio between the total luminosity of the \lya nebula and its area in kpc$^2$ within their $2\sigma$ isophote, basically tracing the average \lya SB of the nebula. 
This plot illustrates both the correlation between the quasar \lya luminosity and their bolometric luminosity (see also Figure~\ref{fig:LbolLpeak}), and the increase in \lya SB for brighter systems also 
seen  
in the \lya SB profiles of Figure~\ref{fig:Rprofiles}. 
Additionally, we show the same properties in the right panel of Figure~\ref{fig:parameter_space}, 
but color coded by the mean velocity dispersion measured in the second moment maps of Figures~\ref{fig:velocity-dispersion}~and~\ref{fig:velocity-dispersion_QSOMUSEUM}. 
The nebulae have a tendency to present larger average velocity dispersions with increasing bolometric luminosity (and hence also peak \lya quasar luminosity), which was also observed in the velocity dispersion radial profiles of Figure~\ref{fig:sigma-profiles}. 
Therefore, 
the nebulae become brighter and potentially more turbulent with increasing bolometric luminosity at fixed black hole mass, or in other words with increasing Eddington ratios. 
In this section we explore the main drivers of these tendencies and the implications of these observations.

\begin{figure*}
    \centering 
    \includegraphics[width=0.95\linewidth]{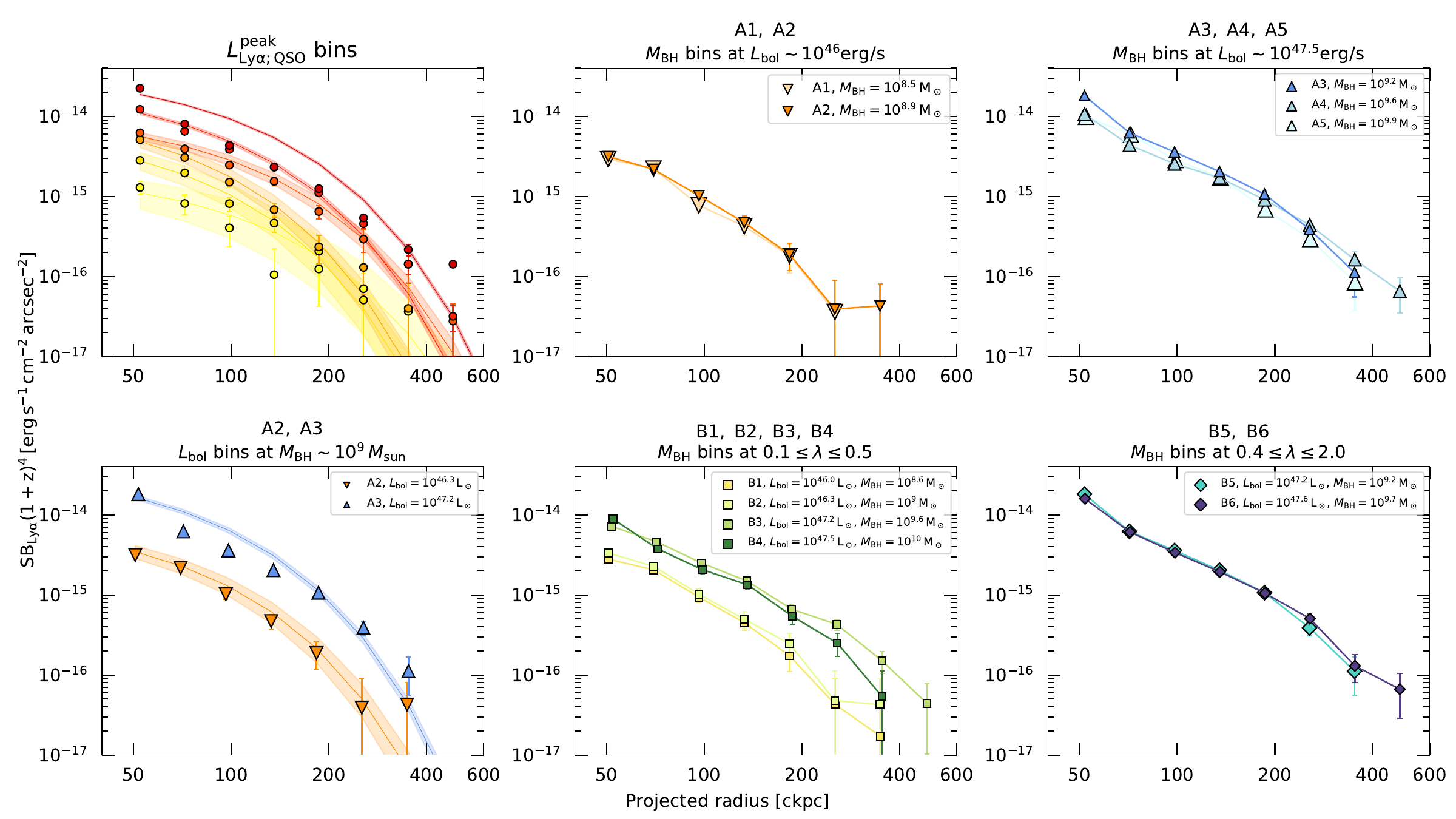}
    \caption{Median \lya SB radial profiles 
    for bins of 
    different quasar properties. The panels show the median \lya SB corrected by cosmological dimming as a function of projected comoving distances within different bins, the errorbars represent the mean $1\sigma$ uncertainty of the stacked datapoints. 
    \emph{Top left:} Median profiles within equally spaced bins of peak \lya luminosity density of the quasars (see Section~\ref{sec:stacked-profiles}), color-coded by the median $L_{\rm Ly\alpha;peak}^{\rm QSO}$ of the bin (same color as used for the individual radial profiles as Figure~\ref{fig:Rprofiles}). 
    The exponential fits (see Section~\ref{sec:MCMC-fits}) of the profiles are shown by the associated shaded areas. 
    \emph{Top middle:} Median profiles within bins A1 and A2 of Figure~\ref{fig:parameter_space}. 
    \emph{Top right:} Median profiles within bins A3, A4 and A5. 
    \emph{Bottom left:} Median profiles within bins A2 and A3. 
    \emph{Bottom middle:} Median profiles within bins B1, B2, B3 and B4. 
    \emph{Bottom right:} Median profiles within bins B5 and B6. 
    The median bolometric luminosity and black hole mass of bins A1-5 and B1-6 are indicated in the legend of each panel.     
    Additionally, we show in the leftmost column the resulting fits to an exponential function of the form ${\rm SB_{Ly\alpha}}(1+z)^4={\rm SB_0}\times e^{(r/R_{h})}$ to the median profiles with a solid curve and shaded regions indicating their 1$\sigma$ uncertainties (see text for details), using the same colors as the datapoints.}
    \label{fig:binned_SBprofiles}
\end{figure*}
\begin{figure*}
    \centering 
    \includegraphics[width=0.9\linewidth]{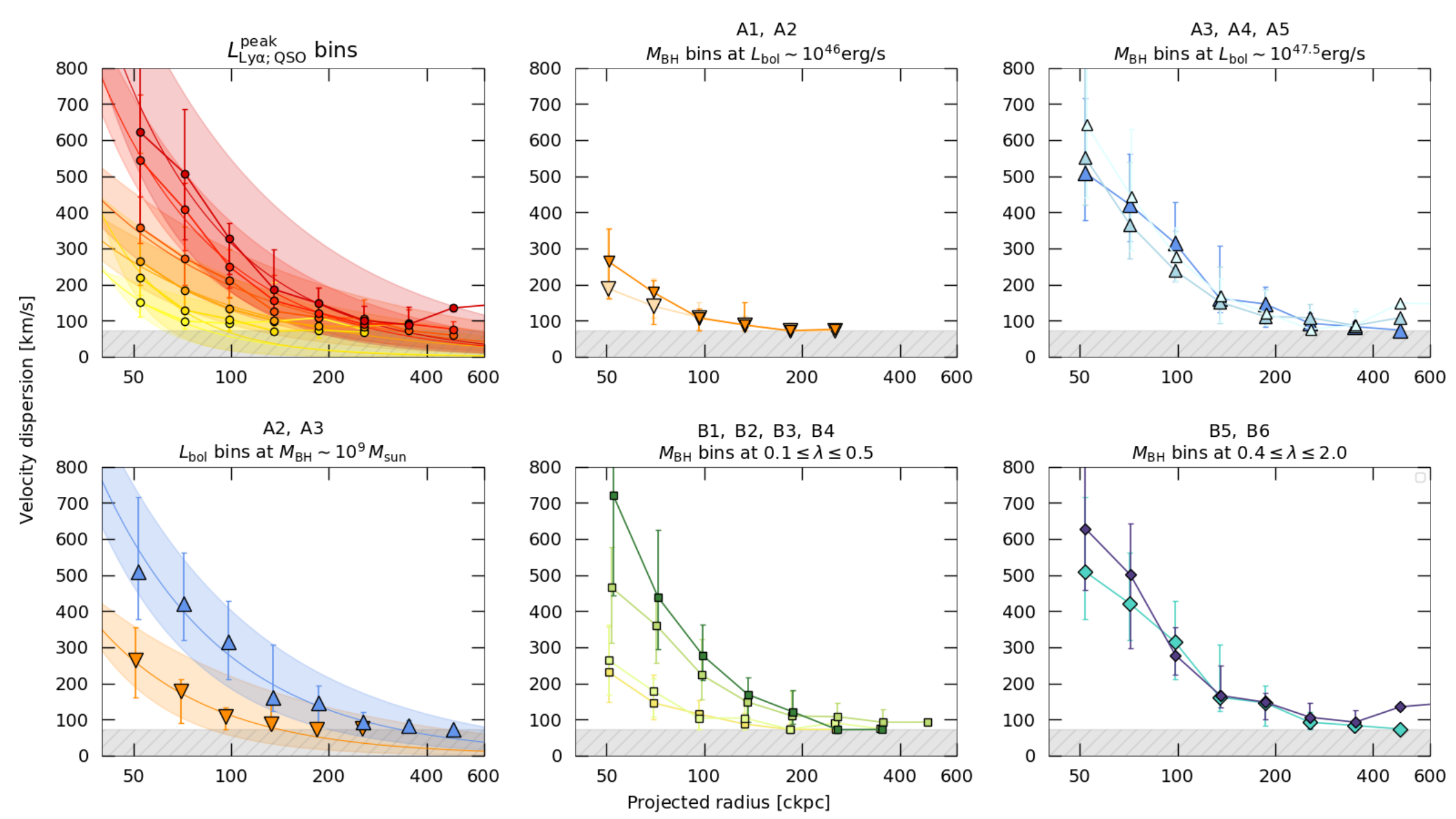}
    \caption{Median \lya velocity dispersion radial profiles 
    for bins of different quasar properties. 
    The panels show the median velocity dispersion of the \lya nebular line as a function of projected comoving distances within the same bins, layout and colors as Figure~\ref{fig:binned_SBprofiles}. The errorbars represent the uncertainty obtained from the 16 to 84 percetile of the stacked datapoints. 
    In each panel, 
    the shaded gray regions mark the values below the spectral resolution limit of the observations. 
    Additionally, we show in the leftmost column the resulting power law fits of the form $\sigma=\sigma_{50}(r/50\,{\rm ckpc})^{-\beta}$ to the median profiles with a solid curve and shaded regions indicating their 1$\sigma$ uncertainties (see text for details). 
    }
    \label{fig:binned_Sigma-profiles}
\end{figure*}
\begin{figure}
    \centering
    \includegraphics[width=\linewidth]{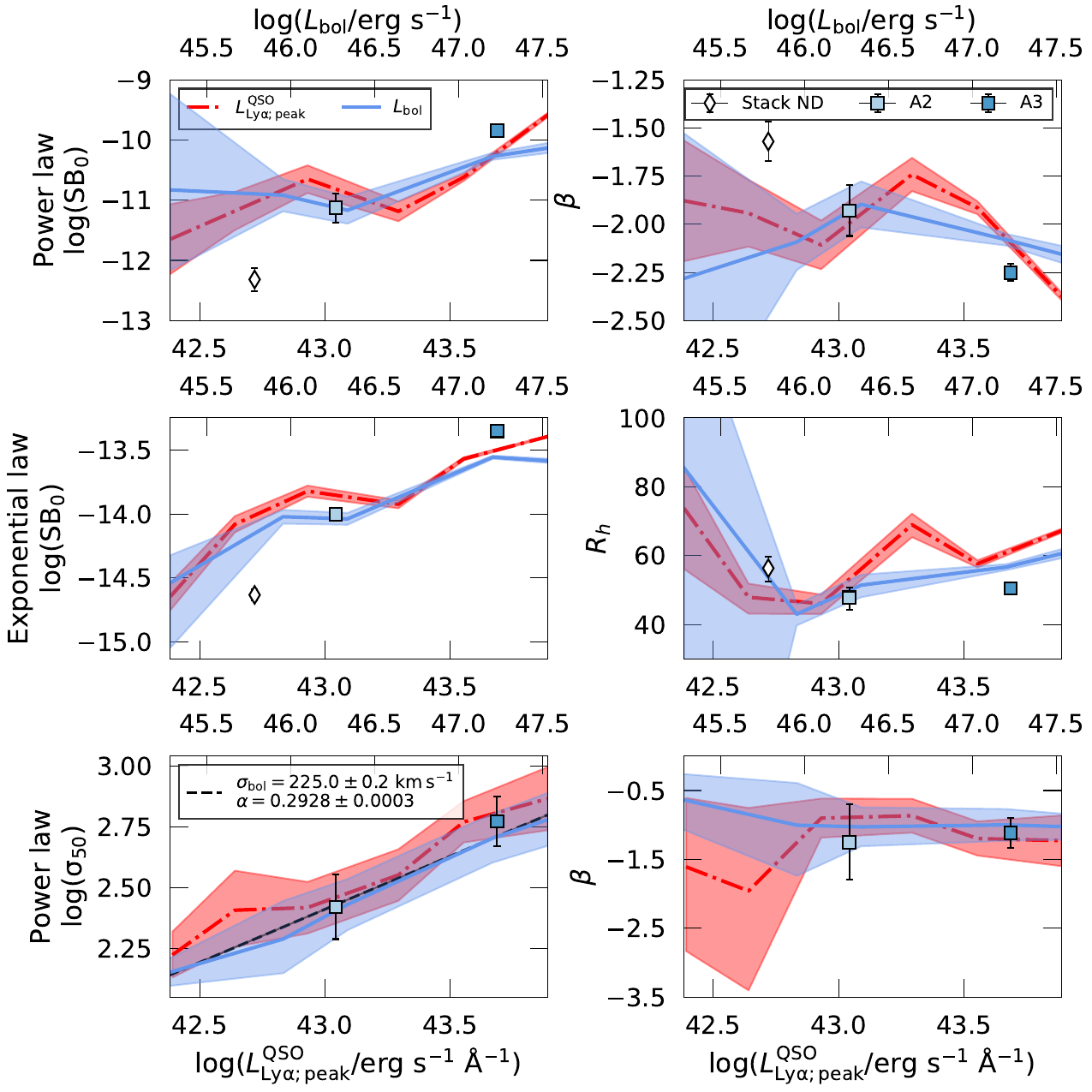}
    \caption{Fitted parameters from a power and exponential law to the stacked SB and velocity dispersion profiles as 
    functions of their median quasar luminosities. 
	The first row shows the fitted $\log({\rm SB_0})$ (left) and $\beta$ (right) parameters of a power law versus the median peak Ly$\alpha$ luminosity density (top axis, red dot dashed line) and bolometric luminosity (bottom axis, blue line) of the quasar, with their respective uncertainties as shaded areas. 
	The second row shows the $\log({\rm SB_0})$ and $R_{\rm h}$ parameters of an exponential law fit as a function of their median quasar luminosities. 
	The bottom row shows the fitted $\log(\sigma_{50})$ and $\beta$ of a power law fit to the median velocity dispersion profiles as a function of their quasar luminosity. We also show in the left panel a power law fit to the normalization factor as a function of bolometric luminosity of the form $\sigma_{50}=\sigma_{\rm bol}(L_{\rm bol}/{[10^{46}~\rm erg\, s^{-1}]})^\alpha$ and parameters shown in the legend with a black dashed line and $1\sigma$ uncertainty shaded area. 
	Finally, the fitted parameters for the profile of the stacked non-detections are shown with a white diamond and errorbars representing the $1\sigma$ uncertainty, while the fitted parameters of bins A2 ans A3 are shown with light blue and dark blue squares, respectively.}
    \label{fig:emcee_plots}
\end{figure}

\subsubsection{Stacking data by AGN properties}\label{sec:stacking-bins}

We explore the 
principal factors 
responsible for the emergence of these trends by computing the median \lya SB profiles using several bins encompassing different quasar properties. 
Stacking is done by first re-binning the profiles and their uncertainties into a common grid of projected comoving distances using a cubic spline interpolation, then we compute the median value at each radial bin. 
Since the stacking procedure decreases the noise potentially revealing fainter emission than in individual profiles, 
we also include in the stack the datapoints falling below the 2$\sigma$ \lya SB limit of each individual profile. 
We compute the mean uncertainty of each stacked radial bin by propagating their variances. 
Similarly, we re-scale the 2$\sigma$ \lya SB limit in the stacked bins and use it to cut the stacked profile at the radial distance at which it falls below the 2$\sigma$ detection limit.

First, we build 
equally spaced bins of the peak Ly$\alpha$ luminosity density of the quasar and show their median \lya SB profiles in the top left panel of Figure~\ref{fig:binned_SBprofiles} and color code them by the median peak \lya luminosity of the quasars within the bin using the same color scale as the radial profiles shown in the middle panel of Figure~\ref{fig:Rprofiles}. 
The bins range from 
$\log(L_{\rm Ly\alpha;peak}^{\rm QSO}/{\rm [erg\,s^{-1}\,\AA^{-1}]})=42.5$
to 
$44$ with sizes of 0.3 dex, resulting in 6 bins with 20, 23, 16, 9, 23 and 29 stacked profiles each. 
Within this range of peak \lya luminosity densities, the SB level of the stacked profiles has a tendency to increase from ${\rm SB}_{\rm Ly\alpha}(1+z)^4\sim10^{-15}$ to $10^{-14}$~erg~s$^{-1}$~cm$^{-2}$~arcsec$^{-2}$ in their inner regions ($\sim50$~ckpc or $\sim12.5$~pkpc) and maintains a similar order of magnitude difference as the projected distance increases. 
This suggest that the profiles all have similar shape. 

Additionally, we stack the profiles in the same way but using equally spaced bolometric luminosity bins starting with $\log(L_{\rm bol}/{\rm [erg\,s^{-1}]})=45$ to $48.7$ and sizes of 0.6 dex, resulting in 5 bins with 8, 31, 21, 31 and 28 profiles each. We do not show these profiles, 
but we note that a similar trend is observed between the SB level and the bolometric luminosity as expected from the relationship between the bolometric luminosity and the peak \lya luminosity density of the quasar 
(Figure~\ref{fig:LbolLpeak}).

It is clear from these observations that the \lya SB level of the nebulae is correlated with the quasar luminosities. 
We test whether other AGN properties could also be responsible for the changes in SB level. 
For this, we construct additional bins at different black hole masses and Eddington ratios that span the range of properties of the targeted quasars. These bins are overlayed and labeled in the left panel of Figure~\ref{fig:parameter_space}. Bins A1 and A2 are constructed to cover the low end of bolometric luminosities ($L_{\rm bol}\sim10^{46}\,{\rm erg\,s^{-1}}$) and black hole masses between $M_{\rm BH}\sim10^{8.1-8.6}$ and $\sim10^{8.6-9.5}\,{\rm M_{\odot}}$, respectively. Bins A3, A4 and A5 are constructed to cover the higher end of bolometric luminosities ($L_{\rm bol}\sim10^{47.5}\,{\rm erg\,s^{-1}}$) and black hole masses of $M_{\rm BH}\sim10^{9.0-9.4}$, $\sim10^{9.4-9.7}$ and $\sim10^{9.7-10.2}\,{\rm M_{\odot}}$, respectively. Bins B1, B2, B3 and B4 are constructed to cover the lower end of Eddington rations ($\lambda_{\rm Edd}\sim0.2$) and black hole mass bins of $\log(M_{\rm BH}/{\rm M_{\odot}})=8.3$ to $10.3$ and sizes of 0.5 dex.  
Finally, bins B5 and B6 are constructed to cover the higher end of Eddington ratios ($\lambda_{\rm Edd}\sim1.0$) and black hole mass bins of $\log(M_{\rm BH}/{\rm M_{\odot}})=8.9-9.4$ to $9.4-9.9$. We stress once again that the uncertainties in the estimation of the black hole masses and bolometric luminosities are of the order of 0.5 and 0.3 dex, respectively (Section~\ref{sec:qso_phys_prop}). This is why we conservatively adopt  
these large 
bin sizes. 

We stack the \lya SB profiles inside these bins using the same aforementioned method 
and plot the resulting profiles in Figure~\ref{fig:binned_SBprofiles}, where each panel corresponds to a set of bins indicated in their legend. 
The top middle panel of this figure shows that the increasing black hole mass at low bolometric luminosity (bins A1 and A2) does not have an impact in the median SB profile, indicating that the \lya radiation around faint quasars is independent of the black hole mass. 
Similarly, at high bolometric luminosity but different black hole mass (bins A3, A4, A5) there is no apparent evolution of the profiles. 
However, the bottom left panel of Figure~\ref{fig:binned_SBprofiles} shows that at fixed black hole mass ($M_{\rm BH}\sim10^9\,{\rm M_{{\odot}}}$), the lower end of bolometric luminosities (bin A2) has around 7 times dimmer nebulae than the higher end of bolometric luminosities (bin A3), confirming the trend observed between individual radial profiles and bolometric luminosity of Figure~\ref{fig:Rprofiles}. 
On the other hand, if we compare the bins with low Eddington ratios and different black hole masses (bins B1, B2, B3 and B4) we see a slight increase in the SB level of the stacked profiles, more likely originating from the increase in bolometric luminosity than the increase in black hole mass, as we do not see changes in profiles due to black hole mass in other plots. 
Finally the two bins at high Eddington ratio (B5 and B6) show very similar profiles despite the different black hole mass and slightly different bolometric luminosity. 

Additionally, we use the same bins to compute median \lya velocity dispersion radial profiles which are shown in Figure~\ref{fig:binned_Sigma-profiles} using the same colors as in Figure~\ref{fig:binned_SBprofiles}. The stack is carried out by computing the median value at each projected distance bin and masking out values that fall below the spectral resolution limit ($\sigma=72$~km~s$^{-1}$ at the median redshift for \lya). These plots show that all profiles seem to approach the resolution limit at projected distances between 100 and 200~ckpc. Additionally, we see that the velocity dispersion profiles follow similar trends as the SB radial profiles as a function of bolometric luminosity. Indeed, the velocity dispersions become higher towards the center with increasing peak \lya luminosity density of the quasar and bolometric luminosity. 
This effect is clearly seen in the bottom left panel where the velocity dispersion increases from 300~km/s to 500~km/s in the center for bins A2 and A3, respectively. 
This trend can be understood as an effect of AGN activity on the CGM kinematics. 
Similarly to Figure~\ref{fig:binned_SBprofiles}, the profiles do not seem to be strongly affected by the change in black hole mass. 

Finally, given the large intrinsic uncertainties in the quasar parameters, 
we test how the stacked profiles change by varying the positions of each system in the $\log(M_{\rm BH}/{\rm M_\odot})$--$\log(L_{\rm bol}/{\rm L_\odot})$ plane (Figure~\ref{fig:parameter_space}). Specifically, we verify this for the bins where we see most of the variation: bins A2 and A3. We generate A2 and A3 samples by randomly drawing the individual values from Gaussian distributions defined by the intrinsic uncertainties in the estimators and centered on the datapoints in Figure~\ref{fig:parameter_space}. For each random realization of the plot in Figure~\ref{fig:parameter_space} we stack the profiles of sources within the A2 and A3 ranges. This process is repeated 2000 times. We find that the median profiles and uncertainties obtained from this exercise are consistent with the ones presented in Figures~\ref{fig:binned_SBprofiles}~and~\ref{fig:binned_Sigma-profiles}.

\subsubsection{Exponential and power law fits to median profiles}\label{sec:MCMC-fits}

With these tests we confirm that the increase in SB level and velocity dispersion of the nebulae is correlated to the bolometric luminosity, therefore we focus in quantifying these differences by fitting with a power or exponential law the median profiles inside the quasar luminosity bins described above. We perform the fits using a Markov Chain Monte Carlo (MCMC) approach, which is implemented in the \texttt{emcee}\footnote{\url{https://emcee.readthedocs.io/en/stable/index.html}} module for Python \citep{Foreman-Mackey2013}. First, we fit the data using the \texttt{curve\_fit} function of \texttt{scipy} \citep{Virtanen2020} and use these to initialize the positions of the MCMC walkers. We generate 50 samples of each parameter and run the MCMC process on 20000 steps. 
For reference, we show in the top left panels of Figures~\ref{fig:binned_SBprofiles}~and~\ref{fig:binned_Sigma-profiles} the resulting exponential and power law fits, respectively, with a solid line and shaded regions representing their $1\sigma$ uncertainty and using the same colors as the datapoints of each corresponding median profile. 
Additionally, we fit the median profiles of the non-detections, bin A2 and bin A3 with the same method. 
For the \lya SB corrected by cosmological dimming, we fit the profiles (in comoving units) using both a power law of the form ${\rm SB_{Ly\alpha}}(1+z)^4={\rm SB_0}\times r^{\beta}$ and exponential function of the form ${\rm SB}_{\rm Ly\alpha}(r)={\rm SB}_0\times e^{-(r/R_h)}$. 
On the other hand, for the median velocity dispersion profiles within the same bins, we fit only a power law of the form $\sigma=\sigma_{50}(r/50\,{\rm ckpc})^{\beta}$. 

We summarize in Table~\ref{tab:MCMC-fit} the resulting fitted parameters to the radial profiles stacked in the different bins described in this section. Additionally, Figure~\ref{fig:emcee_plots} shows the results of the MCMC fit to these bins. In that figure, each panel shows the fits inside $L_{\rm Ly\alpha;peak}^{\rm QSO}$ and $L_{\rm bol}$ bins with a dot-dashed red line and solid blue line, respectively. 
The top row shows the power law fit parameters to the stacked SB profiles, where we see that the normalization factor ${\rm SB_0}$ of the median profiles tends to increase up to an order of magnitude with increasing quasar luminosity and the power $\beta$ ranges between -2.25 to -1.75. 
Similarly, the middle row shows the fitted parameters for the exponential function, which shows a similar behavior in the normalization factor. The characteristic radius $R_{\rm h}$ of the exponential fit appears to be constant at $\sim50$\,ckpc. 
The fitted parameters of the stacked non-detections are shown with a white diamond, while the fits of bins A2 and A3 are shown with light blue and dark blue squares. The fitted values for these three bins follow similar tendencies as the luminosity bins. 
Interestingly, the profile obtained by stacking the non-detections has a lower normalization factor than the lowest luminosity bin (which has comparable median luminosity). 
This could indicate that at these low luminosities there is potentially a mechanism that can suppress the 
extended \lya emission, such as host galaxy orientation \citep[e.g.,][]{Costa2022}, small opening angles \citep{Obreja2024} and/or molecular and dust content of the host galaxy \citep{Nahir2022,GonzalezLobos2023}, which we further discuss in Section~\ref{sec:pow_mech}. 

Additionally, we assess the goodness of the power law versus exponential fits to each stacked profile by calculating the root-mean-square-deviation (RMSD) between the profiles and their fitted functions and weighted by the data uncertainty $\sigma$ as:
\begin{equation}
	{\rm RMSD_{SB}}=\sqrt{\dfrac{1}{N}\sum_{i=1}^{N}\dfrac{({\rm SB}_{{\rm Ly\alpha; }i}-{\rm fit}_i)^ 2}{\sigma_i^2}},
\end{equation}
The results of the RMSD calculation to each fit are presented in Table~\ref{tab:MCMC-fit}. 
We find that the exponential function is a better fit to the profile of stacked non-detections than a power law (RMSD of 2.9 and 4.2, respectively). Similarly, bin A2 is also better described with an exponential than power law (RMSD of 1.6 and 3.3, respectively). 
Opposite to this, bin A3 is better described with a power law rather than an exponential law (RMSD of 3.6 and 13.04, respectively). This is a consequence of the increased SB towards the center in comparison to the points at larger distance (see Figure~\ref{fig:binned_SBprofiles}), which biases the fit towards larger normalization factors. 
If this first point is not considered in the fit, an exponential law can be a good description of the profile in bin A3. 
Finally, the bottom row of Figure~\ref{fig:emcee_plots} shows the fitted parameters to the median velocity dispersion profiles for a power law of the form $\sigma = \sigma_{50}(r/50\,{\rm ckpc})^{\beta}$. From the bottom right panel, we see that the power remains basically constant at $\beta\sim-1.0$. 
On the other hand, 
the normalization factor $\sigma_{50}$ is related to the bolometric luminosity through a power law. Therefore, we further fit a power law of the form $\sigma_{50}=\sigma_{\rm bol}(L_{\rm bol}/[{\rm erg\,s^{-1}}])^\alpha$ and show the result in that panel with a dashed black line, together with a list of the parameters values in the legend ($\alpha=0.2928\pm0.0003$, $\sigma_{\rm bol}=225.0\pm0.2$~km~s$^{-1}$). This indicates that all of the observed inner velocity dispersions have a clear scaling 
with the quasar bolometric luminosities, suggesting that the \lya emission could be tracing 
the impact from AGN radiation and/or winds/outflows at these scales. This finding could be used in future works to test AGN feedback models and their energy coupling with the gas.

To understand whether the inner velocity dispersions are 
a possible sign of outflowing gas, we compare them to the average one-dimensional root mean square (rms) velocity of a massive $M_{\rm DM}\sim10^{12.5}\,{\rm M_\odot}$ halo expected to host these quasars. 
A velocity dispersion comparable or smaller than this value should indicate gravitational motions, while larger values candidate outflowing gas. 
Given that the maximum of the average 1D rms velocity $\sigma_{\rm rms-1D}$ is related to the maximum circular velocity within a dark matter halo by $\sigma_{\rm rms-1D}=V_{\rm circ}^{\rm max}/\sqrt(2)$ \citep{Tormen1997}, assuming such a massive halo at $z\sim3$ characterized by an NFW profile \citep{Navarro1997} with a concentration parameter of $c=3.7$ \citep{Dutton:2014}, the maximum circular velocity is $V_{\rm circ}^{\rm max}=360$~km~s$^{-1}$, giving 
$\sigma_{\rm rms-1D}=250\,{\rm km\,s^{-1}}$ (similar calculation in \citealt{FAB2019}). 
This value corresponds to $\log L_{\rm bol}/{\rm L_\odot}=46.15$ in the obtained power-law relationship. This bolometric luminosity is similar to the median bolometric luminosity of the QSO MUSEUM faint sample ($\log L_{\rm bol}^{\rm F}/{\rm L_\odot}=46.08$), for which we observe lower inner velocity dispersions ($\sigma\sim300\,{\rm km\,s^{-1}}$), suggesting that these systems could be less affected by AGN feedback. Conversely, the more luminous quasars in the sample may have imprinted the signature of AGN feedback in their associated extended emission, which is distinguished by elevated inner velocity dispersions ($\sigma\sim600\,{\rm km\,s^{-1}}$; Figs.~\ref{fig:parameter_space} right and \ref{fig:binned_Sigma-profiles}). This finding has to be tested with non-resonant lines 
as the velocity dispersion is likely increased by resonant scattering effects (e.g., Figure B1 in \citealt{Costa2022}). 
Robustly, the \lya emitting gas at projected distances larger than 30~kpc seems to be dominated by gravitational motions ($\sigma<250\,{\rm km\,s^{-1}}$). 

\begin{table*}
\begin{center}
\caption{Results of the MCMC fits to the \lya SB (after correcting for cosmological dimming) and velocity dispersion profiles in comoving projected distances.}
\label{tab:MCMC-fit}

\resizebox{\linewidth}{!}{%
\begin{tabular}{ccccccccccc}
 \multicolumn{11}{c}{\textbf{Peak Ly$\boldsymbol{\alpha}$ luminosity density bins}} 
 \B \\
 \hline \T 
  \multirow{2}*{Median} &
  \multirow{2}*{\# Stacked} & 
 \multicolumn{3}{c}{SB power law} & 
 \multicolumn{3}{c}{SB exponential} & 
 \multicolumn{3}{c}{Velocity dispersion power law}
 \\
\cmidrule(lr){3-5} \cmidrule(lr){6-8} \cmidrule(l){9-11}
$\log(L_{\rm Ly\alpha;peak}^{\rm QSO}/{\rm L_\odot})$ & profiles & 
$\log{\rm SB_0}$ & $\alpha$ & RMSD & 
$\log{\rm SB_0}$ & $R_{\rm h}$ & RMSD &
$\log{\rm \sigma_{50}}$ & $\beta$ & RMSD 
\B \\

\hline
\T 

42.38 & 20 & 
$-11.65\pm0.58$ & $-1.88\pm0.31$ & 0.39 &
$-14.65\pm$0.11 & 73.78$\pm$14.80 & 0.99 &
2.22$\pm0.09$ & $-1.61\pm1.11$ & 1.83 
\\  

42.64 & 23 & 
$-11.18\pm0.32$ & $-1.94\pm0.17$ & 1.03 &
$-14.08\pm$0.06 & 47.89$\pm$4.38 & 0.73 &
1.34$\pm0.16$ & $-1.96\pm1.33$ & 1.75 
\\ 

42.93 & 16 & 
$-10.65\pm0.23$ & $-2.11\pm0.13$ & 1.02 &
$-13.82\pm$0.04 &46.02$\pm$2.83 & 0.89 &
2.42$\pm$0.11 & $-0.90\pm$0.28 & 0.76 
\\

43.29 & 9 & 
$-11.19\pm0.16$ & $-1.74\pm0.09$ & 1.19 &
$-13.92\pm$0.03 & 68.96$\pm$3.42 & 1.21 &
2.55$\pm$0.11 & $-0.87\pm$0.25 & 0.48 
\\

43.55 & 23 & 
$-10.62\pm0.06$ & $-1.91\pm0.03$ & 4.99 &
$-13.57\pm$0.01 & 57.61$\pm$1.03  & 4.12 &
2.77$\pm$0.09 & $-1.2\pm$0.25 & 1.11 
\\

43.89 & 29 & 
$-9.58\pm0.04$ & $-2.37\pm0.02$ & 4.73 &
$-13.39\pm$0.01 & 67.19$\pm$0.61 & 22.93 &
2.87$\pm$0.13 & $1.23\pm$0.37 & 0.31 

\\ 
\hline \\
\B\\
\end{tabular}
}

\resizebox{\linewidth}{!}{%
\begin{tabular}{ccccccccccc}
 \multicolumn{11}{c}{\textbf{Bolometric luminosity bins}} 
 \\
 \hline \T 
  \multirow{2}*{Median} &
  \multirow{2}*{\# Stacked} & 
 \multicolumn{3}{c}{SB power law} & 
 \multicolumn{3}{c}{SB exponential} & 
 \multicolumn{3}{c}{Velocity dispersion power law}
 \\
\cmidrule(lr){3-5} \cmidrule(lr){6-8} \cmidrule(l){9-11}
$\log(L_{\rm bol}/{\rm L_\odot})$ & profiles & 
$\log{\rm SB_0}$ & $\alpha$ & RMSD & 
$\log{\rm SB_0}$ & $R_{\rm h}$ & RMSD &
$\log{\rm \sigma_{50}}$ & $\beta$ & RMSD 
\B \\

\hline
\T 
45.28 & 8 &  
$-10.83\pm1.46$ & $-2.28\pm0.83$ & 0.20 &
$-14.54\pm$0.36 &85.48$\pm$62.75 & 1.43 &
2.15$\pm0.06$ & $-0.63\pm$0.4 & 0.24 
\\

45.95 & 31 & 
$-10.92\pm0.27$ & $-2.09\pm0.15$ & 1.12 &
$-14.02\pm$0.05 & 42.95$\pm$3.06 & 0.68 &
2.29$\pm1.15$ & $-1.0\pm$0.67 & 1.14
\\

46.34 & 21 & 
$-11.17\pm0.22$ & $-1.897\pm0.12$ & 1.26 &
$-14.04\pm$0.05 & 51.4$\pm$3.29 & 0.70 &
2.43$\pm$0.11 & $-1.03\pm$0.28 & 0.40
\\

47.2 & 31 &
$-10.27\pm0.06$ & $-2.08\pm0.03$ & 2.27 &
$-13.55\pm$0.01 & 56.66$\pm$0.86 & 5.48 &
2.71$\pm$0.1 & $-1.0\pm$0.23 & 0.28
\\

47.53 & 28 &
$-10.13\pm0.08$ & $-2.15\pm0.04$ & 1.22 &
$-13.58\pm$0.02 & 60.58$\pm$1.34 & 5.29 &
2.78$\pm$0.11 & $-1.02\pm$0.19 & 3.01
\\ 
\hline \\
\B\\
\end{tabular}
}

\resizebox{\linewidth}{!}{%
\begin{tabular}{ccccccccccc}
 \multicolumn{11}{c}{\textbf{Other bins}} \\
 \hline \T 
 &
 \multirow{2}*{\# Stacked} &
 \multicolumn{3}{c}{SB power law} & 
 \multicolumn{3}{c}{SB exponential} & 
 \multicolumn{3}{c}{Velocity dispersion power law}
 \\
\cmidrule(lr){3-5} \cmidrule(lr){6-8} \cmidrule(l){9-11}

Bin & profiles & 
$\log{\rm SB_0}$ & $\alpha$ & RMSD & 
$\log{\rm SB_0}$ & $R_{\rm h}$ & RMSD &
$\log{\rm \sigma_{50}}$ & $\beta$ & RMSD 
\B \\   
\hline
\T 

Stack ND\tablefootmark{(a)} & 10 &
$-12.32\pm$0.19 & $-1.57\pm$0.1 & 1.9 & 
$-14.63\pm$0.04 & 56.33$\pm$3.65 & 1.3 & 
-- & -- & -- \\

A2 & 21 & 
$-11.13\pm$0.24 & $-1.93\pm$0.13  &  1.2 & 
$-14.00\pm$0.05 & 47.78$\pm$3.24 & 0.6 & 
2.42$\pm$0.13 & $-1.25\pm$0.54 &  0.73 
\\

A3 & 19 & 
$-9.84\pm$0.08 & $-2.25\pm$0.04 &  1.3 & 
$-13.35\pm$0.02 & 50.47$\pm$1.04 & 4.6 & 
2.77$\pm$0.1 & $-1.12\pm$0.22 & 0.29
\\ 
\hline \\
\B

\end{tabular}}
\tablefoot{
\tablefoottext{a}{ND = Non-detections}
}
\end{center}

\end{table*}

\begin{figure*}
    \centering
    \includegraphics[width=\linewidth]{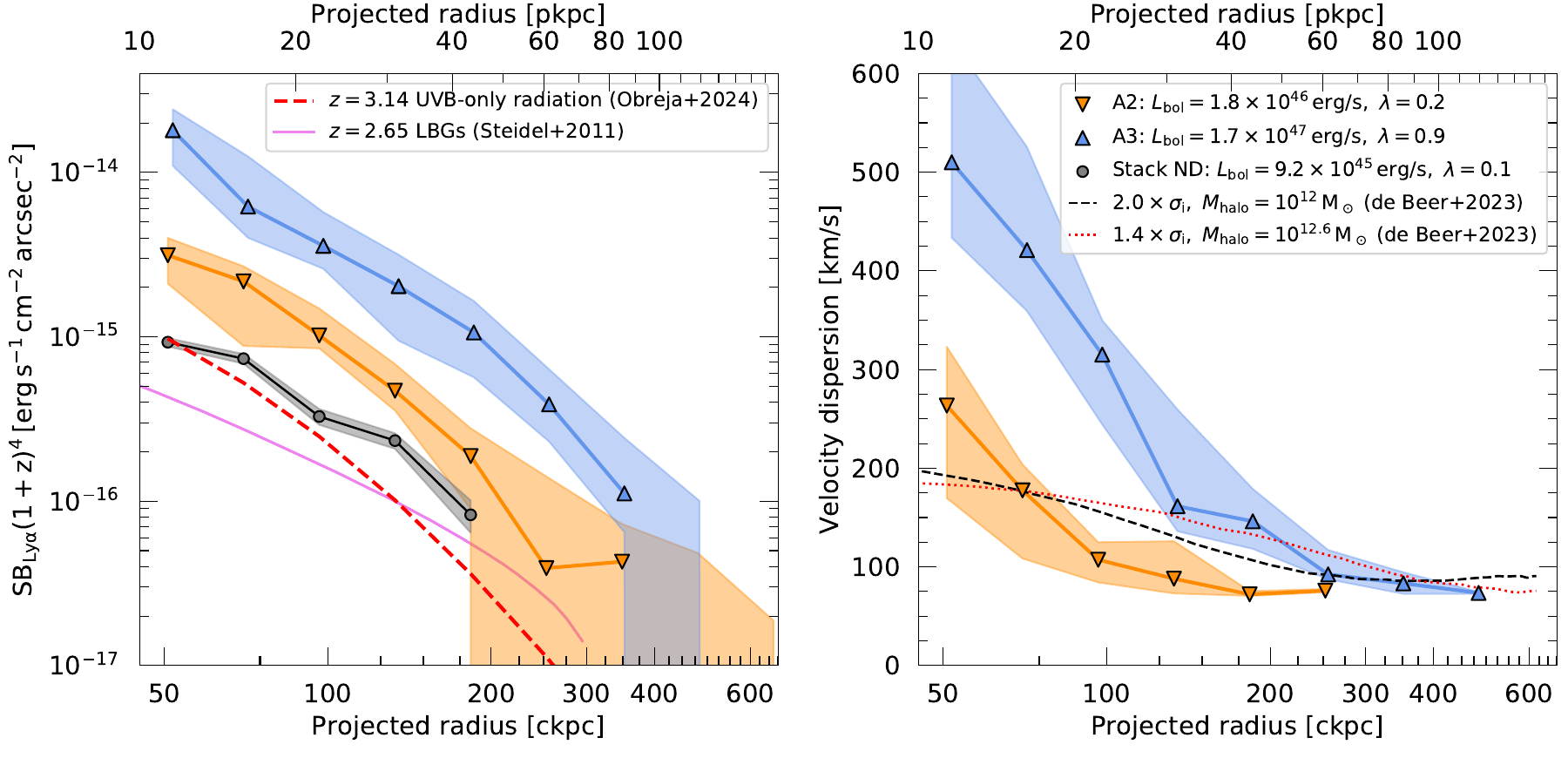}
    \caption{Median \lya SB and velocity dispersion of stacked radial profiles. 
		\emph{Left:} \lya SB corrected by cosmological dimming as a function of projected comoving distance for the stacked non-detections in gray dots and bins A2 and A3 with orange and blue triangles, respectively. The simulated \lya profile obtained from the illumination of a 
  $10^{12.45}$~M$_{\odot}$ halo by only the UVB 
  (\citealt{Obreja2024}) is shown with a red dashed line. The profile around LBGs \citep{Steidel2011} is shown with a magenta line.
		\emph{Right:} Velocity dispersion as a function of projected comoving distances of bins A2 and A3 with orange and blue triangles, respectively, and their shaded region representing the 25$^{\rm th}$ to 75$^{\rm th}$ percentiles. 
		Additionally, we show with a black dashed line and a red dot-dashed line the intrinsic \lya velocity dispersion ($\sigma_{\rm int}$) from \citet{deBeer2023} using cosmological simulations of haloes at $z=3.017$ with $M_{\rm halo}=10^{12}$ and $10^{12.6}\,{\rm M_\odot}$, respectively, after rescaling to match the velocity dispersion at 17\,pkpc of bin A2 (see Section~\ref{sec:pow_mech}). 
		The median bolometric luminosity and Eddington ratio for the sources in bins A2 and A3 are indicated in the legend.
  }
    \label{fig:SB+sigma-profiles}
\end{figure*}

\subsection{Evidence for instantaneous AGN feedback on CGM scales}\label{sec:inst_feedback}

AGN feedback is a key ingredient of galaxy formation models, especially for reproducing the massive end of the galaxy population (e.g., \citealt{DiMatteo2005}). 
Nevertheless, there is still considerable debate regarding the 
impact of AGN on their host galaxies and the surrounding medium (e.g., \citealt{Harrison2024}). 
One 
key issue 
is the lack of conclusive observations of the instantaneous, rather than cumulative, effects of AGNs on large scales. Such observations are essential to determine how the energy couples with the gas, and ultimately how AGN feedback works. In this section, we discuss the evidence for likely instantaneous AGN feedback and its impact on CGM scales as traced by the \lya emission, considering that all of our targets are quasars expected to be in the radiatively-efficient regime (e.g., \citealt{Harrison2024}).

We discussed that the observed \lya and bolometric luminosity of the quasars have the strongest relation with 
the SB and velocity dispersion of the extended \lya emission (Section~\ref{sec:stacked-profiles}).  In other words, brighter (more accreting) quasars, expected to generate stronger feedback, are associated with brighter and more turbulent extended Ly$\alpha$ emission. These results were not obvious a priori. Indeed, the fact that these relations are in place implies that the mechanisms responsible for the \lya emission act 
on timescales similar to the known variability of quasar activity as otherwise any relation with the observed current quasar activity would be washed out. Given the rather short quasar 
variability 
estimates (see section~\ref{sec:intro}), the observed relations between quasar luminosity and CGM emission seems an indication of instantaneous AGN feedback.

For example, if we assume that the nebulae are predominantly powered by photoionization followed by recombination, then a drop in luminosity due to ''shut-down'' at the end of their current quasar activity could cause small \lya nebulae, or parts of them, to fade. 
However, we do not see this effect. Indeed, in order for this to happen the recombination time needs to be much shorter than the period of 
``shut-down'', which has been inferred to be of the order of $10^{4-5}$\,years by some observations \citep[e.g.,][]{Lintott2009, Davies2015}. 
The timescale for case B recombination is $t_{\rm rec}=1/\alpha_{\rm B}(T) n_{\rm H}$ where $n_{\rm H}$ is the gas number density and $\alpha_{\rm B}(T) = 2.6\times10^{-13}(T/10^4\,{\rm K})^{-0.7}\,{\rm cm^{-3}\,s^{-1}}$ is the recombination coefficient \citep{Dijkstra2019}. Then, for this phenomenon to occur in $2\times10^{4}$\,K gas, the \lya would be tracing volume densities of $n_{\rm H}>>2$\,cm$^{-3}$. 
The fact that we do not find non-detections or donut-like shaped \lya nebulae (i.e. the inner portion of the CGM would be the first to decrease in SB after the turn off of a quasar) among most quasars seem to confirm that the quasar ``shut-down'' phases, if any, are very short and much shorter than the recombination time in the \lya emitting gas. This means that the densities of the \lya emitting gas are not as high.

Figure~\ref{fig:SB+sigma-profiles} summarizes the AGN electromagnetic impact on the \lya SB and velocity dispersion median radial profiles by focusing on the bins A2 and A3, which were constructed at roughly fixed black hole mass ($M_{\rm BH}\sim 10^9\,{\rm M_\odot}$) and increasing bolometric luminosity, hence Eddington ratio $\lambda=0.2$ and $\lambda=0.9$, respectively. 
The left panel 
shows the median \lya SB corrected by cosmological dimming as a function of projected distance, with the shaded areas representing the 25 and 75 percentiles of the stacked profiles. 
Additionally, we show the median \lya SB profile of the stacked non-detections with gray dots and shaded region indicating its $1\sigma$ uncertainty. 
For comparison, we show in the same panel the simulated profile obtained in \citet{Obreja2024} by assuming that the cool CGM of a high-resolution zoom-in $10^{12.45}\,{\rm M_\odot}$ halo is only illuminated by the ultraviolet background (UVB) (red dashed curve). The simulation has been run only with stellar feedback down to $z\sim3$, but it is in agreement with the halo mass -- stellar mass relation (\citealt{Moster2018}). This profile has a similar scaling to the stacked profile of non-detections, but decays more steeply with increasing radius. Suggesting that even 
faint quasars can affect the surrounding cool CGM emission.

Lastly, as reference, 
the \lya SB profile around Lyman break galaxies (LBGs; \citealt{Steidel2011}) is shown in the same panel with a magenta solid line. LBGs are expected to be hosted in less massive ($M_{\rm DM}\sim10^{12}\,{\rm M_\odot}$) and less active halos than those hosting quasars.

The right panel of Figure~\ref{fig:SB+sigma-profiles} shows the median velocity dispersion profiles within the same bins, with shaded areas representing their 25 and 75 percentiles. 
The velocity dispersion in bin A2 is on average a factor of $2\times$ lower than the velocity dispersion of bin A3. 
Both profiles drop quickly from the inner regions and seem to plateau at larger radii, with the latter effect likely due to the instrument spectral resolution. 
Given that bins A2 and A3 span a similar range
of black hole masses, the difference in the central velocity dispersion ($<$40\,pkpc) could indicate that the 
inner CGM turbulence is modulated by the intensity of AGN feedback in these systems, and that the \lya emission better traces the violent AGN impact on these inner CGM scales. Indeed, the quasar lifetimes quoted in Section~\ref{sec:intro} ($\sim10^7$~years) are compatible with the time that an outflow with $1000-5000\,{\rm km\,s^{-1}}$ requires to reach the observed CGM scales ($10-50\,{\rm kpc}$). 
Therefore,  the found power law relating the velocity dispersion to the bolometric luminosity (section~\ref{sec:stacked-profiles} and Figure~\ref{fig:emcee_plots}) could be used in future works to test AGN feedback models and their energy coupling with the cool gas.

It is important to note that the velocity dispersion of extended \lya emission around quasars has been frequently 
associated with 
gravitational motions within massive halos of $M_{\rm halo}\sim10^{12.5}\,{\rm M_\odot}$, 
both using averages over the full detected nebulae \citep{FAB2019,Farina2019} and 
the ratio of the median velocity dispersion in two specifically selected radial annuli \citep{deBeer2023}. 
The latter work assumed that the intrinsic \lya velocity dispersion profiles obtained from their cosmological hydrodynamic simulations 
under the assumption of only quasar photoionization followed by recombination 
can be simply scaled in normalization to match observations, and therefore estimate the halo mass. 
The right panel of Figure~\ref{fig:SB+sigma-profiles} shows the simulated intrinsic \lya velocity dispersion ($\sigma_{\rm int}$) from \citet{deBeer2023} using mock observations of cosmological simulations of halos of $M=10^{12}$ and $10^{12.6}\,{\rm M_\odot}$ at $z=3.017$ with a black dashed line and red dot-dashed line, respectively, after rescaling them to match the observed velocity dispersion at an average projected distance of 17\,pkpc (or $\sim70$\,ckpc) of bin A2. This normalization 
considers 
that \citet{deBeer2023} used as first radial annuli for their ratio calculation the range of projected distances of 40-100\,ckpc, 
corresponding 
to 9.76-24.4\,pkpc at the median redshift of our sample. Their simulated profile 
is much broader than the one observed here, indicating that their working assumptions cannot be entirely valid. 

Additionally, it has been long ago proposed a correlation between the SMBH mass and the mass of the host dark matter halo (\citealt{Ferrarese2002}), and this relation should be tighter at earlier cosmic times (\citealt{Volonteri2011}). As proposed by \citet{Farina2022} one could obtain a first estimate of the halo circular velocity under the
assumption that the extended \lya emission traces on average gas in 
gravitational motion within the dark matter halo. 
We attempted such a calculation for our sample and found 
no strong correlation and a similarly large scatter as in \citet{Farina2022} (see their Figure 13). This fact could be due to the large uncertainties in SMBH mass values and on the use of an average velocity dispersion which could be dominated by more violent kinematics around highly accreting sources. Observations targeting lower SMBH masses are needed to test the presence of this relation and its evolution (if any) at $z\sim3$. 

Given the lack of agreement with gravitational effects, we propose that other mechanisms, like the AGN activity earlier discussed, could be contributing to the observed inner shape of the velocity dispersion profile. 
Particularly, the regions more sensitive to AGN impact ($R<40$\,pkpc) seem to be 
where the nebulae appear more circular, while larger scales are tracing more quiescent material. In the following section, we discuss the different possible roles of this AGN feedback in powering extended \lya emission.

\subsection{Insights on the extended \lya powering mechanisms}  
\label{sec:pow_mech}

As mentioned in Section~\ref{sec:intro}, there are several physical mechanisms that can collectively power the observed extended \lya emission, such as photoionization due to the quasar (and 
nearby sources) radiation followed by recombination, resonant 
scattering of \lya photons originating from the BLR 
of the quasar, shocks due to galactic/AGN winds and outflows, 
and gravitational cooling. In Section~\ref{sec:stacked-profiles}, we proposed that the observed  \lya SB and velocity dispersion are modulated by the central AGN due to their dependence with the bolometric luminosities of the quasars at fixed black hole masses. 
Given this dependence, we exclude a gravitational cooling scenario, which would instead depend mainly on the halo mass of the system studied (e.g., \citealt{Dijkstra&Loeb2008, Faucher-Giguere2010}). 
Next, we 
discuss the interplay that can happen between AGN luminosities 
and the remaining powering mechanisms, and ultimately 
their possible contribution to the observed extended emission.

\subsubsection{Photoionization due to AGN radiation followed by recombination}\label{sec:pow-photoionization}

In this scenario, gas illuminated by any quasar considered in this work is expected to be almost fully ionized \citep{Mackenzie2021}, and optically thin to ionizing radiation. This results in a \lya SB level that depends on the gas properties such as hydrogen volume and column densities (${\rm SB_{Ly\alpha}}\propto n_{\rm H}N_{\rm H}$; Equation 10 in \citealt{Hennawi2013}), 
translating to a dependence in $n_{\rm H}$ and mass reservoir in the cool phase $M_{\rm H}$. 
If we assume a fixed $M_{\rm H}$ or similarly the median value of column density obtained in absorption studies of quasar CGM $N_{\rm H}=10^{20.5}\,{\rm cm^{-2}}$ \citep{Lau2016}, we find that the median \lya SB profiles of the faint and bright quasar nebulae imply gas densities of $n_{\rm H}\sim0.5$ and $2\,{\rm cm^{-3}}$, respectively, assuming a fixed projected distance of $R\sim20$\,kpc, the median redshift of our sample, and a covering fraction of $f_{\rm C}=1$ \citep{FAB2015b}. 
As expected, these values are consistent with the densities estimated in previous works (e.g., \citealt{Hennawi2015, FAB2019b}). Thus, if the recombination scenario in optically thin gas were the dominant mechanism powering the nebulae, then the observed \lya SB may imply that the bright quasars reside in denser gas reservoirs. 
Similarly, if we re-do the same calculation but fix 
$n_{\rm H}$, the change in SB could be due to a more (4$\times$) massive reservoir of cool gas around bright quasars.
However, both faint and bright quasars at $z\sim3$ seem to reside in similarly massive halos as estimated by clustering measurements
(\citealt{White2012,Timlin2018}). 
Also, at low redshift the black hole mass scales with the halo X-ray temperature, hence halo mass (e.g., \citealt{Gaspari2019}). 
If this relation holds also out to $z\sim3$, we should see a trend of increasing velocity dispersion as a function of black hole mass if the velocity dispersion of \lya nebulae is a proxy of
quasar halo mass (\citealt{FAB2019,Farina2019,deBeer2023}). 
Here, we do not see strong evidence for such a trend (see Figure~\ref{fig:binned_Sigma-profiles}). Hence, we do not expect drastic changes in $n_{\rm H}$ and $M_{\rm H}$ in the targeted quasars, unless the effect of quasar feedback is considered. Brighter quasars might increase densities in the inner halo because of entrained material in stronger winds/outflows (e.g., \citealt{Costa2014}), but can simultaneously reduce the reservoir of cool gas due to stronger ionizing radiation (e.g., \citealt{Obreja2024}). 
Alternatively, other mechanisms, such as stripped satellite galaxies, could supply additional gas mass to power the observed nebulae. However, this effect would need to be more pronounced around brighter quasars, which requires further observational tests (e.g., \citealt{Chen2021, Bischetti2021}). 
Another way to vary the cool gas mass as a function of quasar luminosity is to assume a change in the fraction of gas illuminated by the AGN \citep{Obreja2024}, e.g., by requiring that the ionization cone angle increases with bolometric luminosity. 
We further discuss this in Section~\ref{sec:simple_model}.

\subsubsection{Resonant scattering of quasar BLR \lya photons}\label{sec:pow-scattering}

In this scenario, \lya photons originating from the BLR of the quasar scatter through the surrounding medium until they escape the CGM. 
Therefore, the observed flux distribution contains information on the \lya source as well as the location and kinematics of the last scattering event (see e.g., \citealt{Blaizot2023}). 
In this case, we expect that the \lya luminosity of the nebulae is correlated to the amount of \lya photons originating from the quasar (and consequently $L_{\rm bol}$, Figure~\ref{fig:LbolLpeak}).

\begin{figure}
    \centering
    \includegraphics[width=\linewidth]{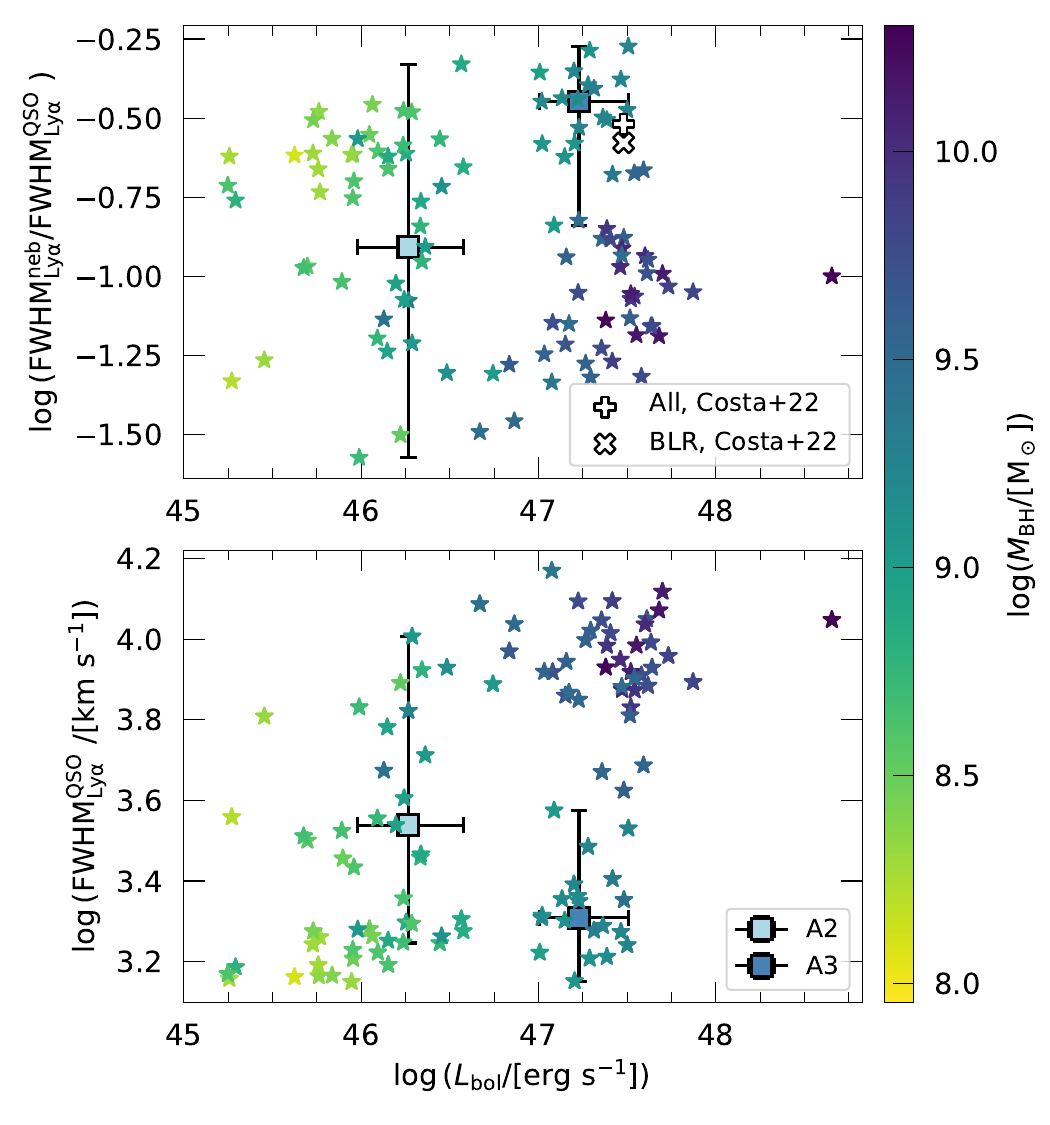}
    \caption{\lya FWHM versus bolometric luminosity. 
            In both panels each star represents a system with nebula detected and colors indicating the black hole mass. Additionally, we show in both panels the median values of bins A2 and A3, with their errorbars representing the full range of values in the bin.
            \emph{Top:} Ratio between the \lya  FWHM of the nebula, computed from the mean velocity dispersion of the second moment maps (Figures~\ref{fig:velocity-dispersion}~and~\ref{fig:velocity-dispersion_QSOMUSEUM}), and the \lya FWHM of the quasar as a function of bolometric luminosity. Datapoints obtained from the $z\sim6$ cosmological simulations in \citet{Costa2022} considering all powering mechanisms (white plus) and scattering from the BLR only (white cross).
            \emph{Bottom:} FWHM of the \lya line of the quasar (Tables~\ref{tab:QSOfit-bright}~and~\ref{tab:QSOfit-faint}) as a function of the bolometric luminosity of the quasar (see Appendix~\ref{app:QSOfit}).}
    \label{fig:FWHM-vs-Lbol}
\end{figure}

To test this scenario, 
we compare with a high-resolution cosmological, radiation-hydrodynamic simulation of a quasar host halo at $z>6$ post-processed with a \lya radiative transfer code \citet{Costa2022}. Their predictions were able to match the \lya emission out to scales of $\sim100$\,kpc in $z\sim6$ quasar host halos. These simulations indicated that BLR photons, even alone, 
can power \lya nebulae. 
Additionally, they proposed that AGN outflows are essential for the formation of extended \lya emission, by clearing out the gas within the galactic nucleus and allowing \lya photons to scatter through the CGM. 
This could be consistent with our observations, which show that an increase in the luminosity of the quasar or its Eddington ratio (hence AGN feedback) corresponds to brighter nebulae with larger velocity dispersions at their center, which could arise in the case of AGN winds/outflows injecting turbulence into the ISM and CGM cool gas. 
The simulations from 
\citet{Costa2022} predicted that only quasar's photons with $|\Delta v|\lesssim 1000$~km~s$^{-1}$ scatter and create \lya nebulae, while photons in the broad wings usually freely stream through the CGM. 
We test this by comparing in the top panel of Figure~\ref{fig:FWHM-vs-Lbol} the ratio between the \lya FWHM of the nebulae (computed from the average of the second moment maps) and the \lya FWHM of the quasars (see Appendix~\ref{app:QSOfit} for an explanation of its calculation) as a function of $L_{\rm bol}$, where the datapoints are colored by their black hole mass.
In the same panel, the median values of bins A2 and A3 are shown using blue squares and errorbars representing the full range of values within the bin. 
The plot shows that the \lya extended emission always has a width narrower than that of the quasar's line, ranging from about 50\% to 2\%, 
in agreement with the expectations from this scenario. 
Additionally, at fixed black hole mass (bins A2 and A3), a higher Eddington ratio corresponds to an increase of the ratio between FWHMs. 
Once taking into account the drop in FWHM$^{\rm QSO}_{\rm Ly\alpha}$ for these bins, there is still a residual increasing trend suggesting that the FWHM$_{\rm neb}$ around more accreting systems encodes information on the impact from AGN feedback on the CGM kinematics as we have discussed earlier in Section~\ref{sec:inst_feedback}. 
Alternatively, it could also be possible that larger FWHM$_{\rm neb}$ appear around brighter quasars because of the increased amount of BLR \lya photons available for scattering and/or higher amounts of gas (and optical depth) in the vicinity of these AGN. 

The bottom panel of Figure~\ref{fig:FWHM-vs-Lbol} shows that faint and bright quasars span the whole range of measured FWHM$^{\rm QSO}_{\rm Ly\alpha}$ and there is no clear indication of a correlation with their bolometric luminosity. However, we see that quasars displaying larger \lya FWHM tend to also have relatively larger black hole masses at fixed bolometric luminosity, as expected from BH virial mass estimators. 
For comparison, the datapoints from the $z\sim6$ cosmological simulation of \citet{Costa2022} (Table~2 in that work) considering all powering mechanisms (recombination, collisions, broad-line region scattering) and only broad-line-region scattering of \lya photons are shown in the top panel with a white plus sign and white cross, respectively. 
These simulations assumed in input for the radiative transfer calculation that ${\rm FWHM^{QSO}_{Ly\alpha}}=2400\,{\rm km\,s^{-1}}$, $M_{\rm BH}\sim10^9\,{\rm M_\odot}$ and $L_{\rm bol}=3\times10^{47}\,{\rm erg\,s^{-1}}$, which are comparable to the values of the high accreting bin A3. The linewidth ratios from the simulation are comparable to the measured values, but from this observation it is not possible to disentangle the mechanisms. This is even more true considering the variations in the FWHM when observing the simulated system 
adopting different line of sights (the standard deviation of the FWHM values along the six line of sights studied in \citet{Costa2022} is of about 150~km/s). 

\begin{figure}[h]
    \centering
    \includegraphics[width=\linewidth]{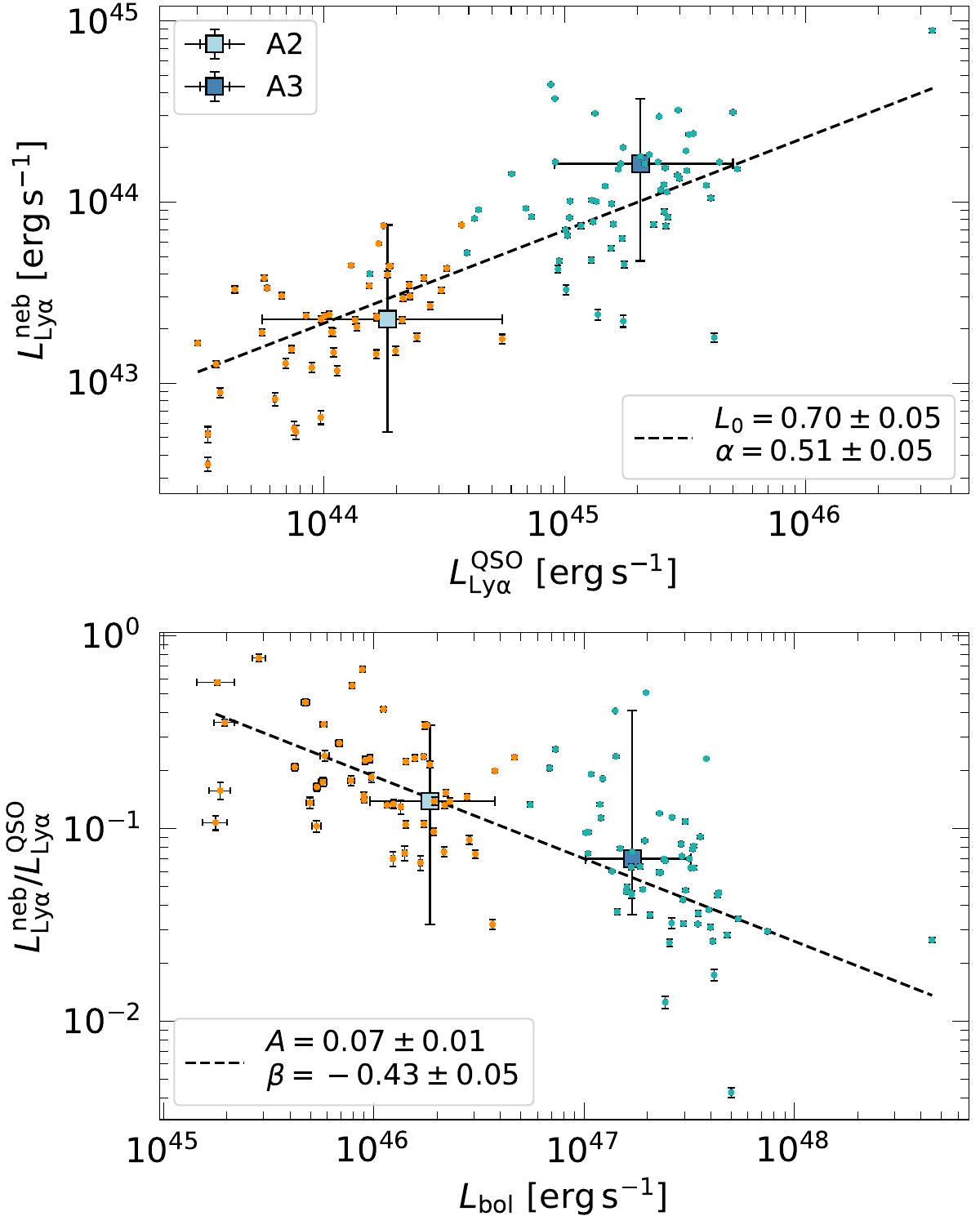}
    \caption{Comparison of nebular luminosities as a function of quasar luminosities. 
    \emph{Top:} \lya luminosity of the detected nebulae versus \lya luminosity of the quasar, integrated within the $\pm$FWHM range of the nebular \lya line (see Section~\ref{sec:properties} and Tables~{\ref{tab:properties}~and~\ref{tab:properties_QSOMUSEUM}}) and errorbars representing the $1\sigma$ uncertainty of the measurements. The median luminosities of bins A2 and A3 are shown with blue squares and their errorbars represent the full range of values within each bin. Additionally, we plot with a dashed black line a power law fit of the form $L_{\rm Ly\alpha}^{\rm neb}/[10^{44}\,{\rm erg\,s^{-1}}]=L_0(L_{\rm bol}/[10^{45}\,{\rm erg\,s^{-1}}])^\alpha$ and list the fitted parameters in the legend. 
    \emph{Bottom:} Ratio of \lya luminosities between the nebula and quasar as a function of quasar's bolometric luminosity. Similarly, the median values and full range within bins A2 and A3 are shown with blue squares and errorbars, respectively. Also, we plot with a dashed black line a power law fit of the form $(L_{\rm Ly\alpha}^{\rm neb}/L_{Ly\alpha}^{\rm QSO})=A(L_{\rm bol}/[10^{47}\,{\rm erg\,s^{-1}}])^\beta$ and show the fitted parameters in the legend.} 
    \label{fig:qsoL-vs-nebL}
\end{figure}
\begin{figure}
    \centering
    \includegraphics[width=\linewidth]{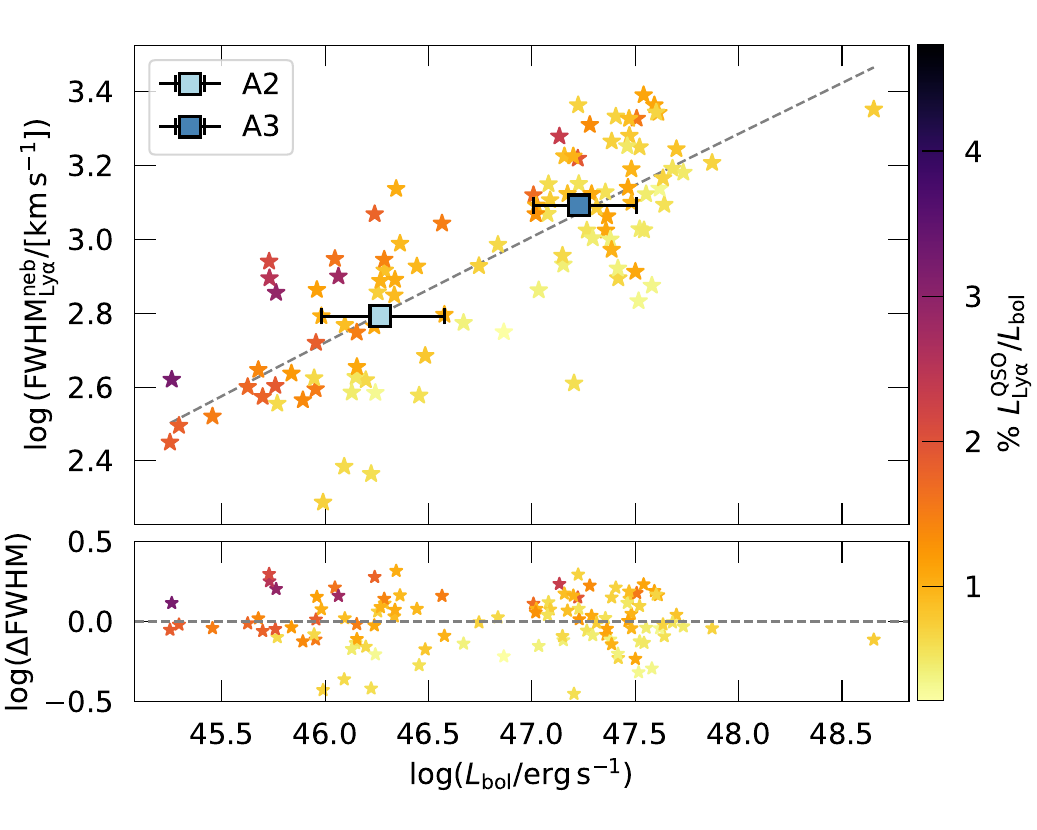}
    \caption{\lya FWHM as a function of quasar luminosities. 
    In both panels each star represents a system with nebula detected and colors correspond to the percentage of quasar's \lya luminosity (integrated within the $\pm$FWHM of the nebula) with respect to the bolometric luminosity.  \emph{Top:} \lya FWHM at the center of the nebula, computed from the first point in the velocity dispersion radial profiles (Figure~\ref{fig:sigma-profiles}) as a function of bolometric luminosity of the quasar. The median values of bins A2 and A3 are indicated with blue squares and error-bars indicating the range of $L_{\rm bol}$ within each bin. The gray dashed line represents a power law fit to the datapoints. \emph{Bottom:} Distance from the datapoints to the power law fit shown in the top panel.}
    \label{fig:FWHM-vs-Lratio}
\end{figure}

Having demonstrated that the FWHM ratio is comparable with predictions for a  broad-line region photon scattering scenario, we further test this case by using the 
same approach from \citet{GonzalezLobos2023} and compute in first approximation the number of BLR \lya photons available for scattering $L_{\rm Ly\alpha}^{\rm QSO}$. For this, we integrate the quasar spectra within $\pm$FWHM of the nebular \lya line and compare it with the integrated \lya luminosity of the nebula within the same velocity range (Tables~\ref{tab:properties}~and~\ref{tab:properties_QSOMUSEUM}). 
The top panel of Figure~\ref{fig:qsoL-vs-nebL} shows that these two quantities are positively correlated, consistent with \citet{GonzalezLobos2023} and our results on the SB profiles. 
Moreover, the Spearman's rank correlation coefficient between $L_{\rm Ly\alpha}^{\rm neb}$ and $L_{\rm Ly\alpha}^{\rm QSO}$ is 0.8, indicating a very strong positive correlation. 
The relationship is fitted with a power law of the form $L_{\rm Ly\alpha}^{\rm neb}/[10^{44}\,{\rm erg\,s^{-1}}]=L_0(L_{\rm bol}/[10^{45}\,{\rm erg\,s^{-1}})^\alpha$, where $L_0=0.70\pm0.05$ and $\alpha=0.51\pm0.05$, and shown with a black dashed line. Additionally, the values of bins A2 and A3 are shown with blue squares and errorbars indicating the full range of values within each bin. 
The found relation could be a hint that broad-line region photons scattering is in place. However, 
because quasars with stronger ionizing radiation can reduce the amount of neutral hydrogen (\citealt{Obreja2024}) available for scattering, 
the ratio between $L_{\rm Ly\alpha}^{\rm neb}$ and $L_{\rm Ly\alpha}^{\rm QSO}$ should decrease with increasing ionizing radiation. 
The bottom panel of Figure~\ref{fig:qsoL-vs-nebL} shows this behavior, where the \lya luminosity ratio (fraction of scattered quasar photons) tends to decrease with increasing $L_{\rm bol}$. 
For this case we also compute the Spearman's rank correlation coefficient and find a value of -0.73, indicating a strong negative correlation. A power law fit of the form $(L_{\rm Ly\alpha}^{\rm neb}/L_{\rm Ly\alpha}^{\rm QSO})=A(L_{\rm bol}/[10^{47}\,{\rm erg\,s^{-1}}])^\beta$ with values $A=0.07\pm0.01$ and $\beta=-0.43\pm0.05$ is shown with a black dashed line.
Together, these relations are suggestive of a contribution from resonant scattering of quasar's \lya photons to the overall extended \lya emission.

Finally, as a last test, we consider that resonant scattering of \lya should produce an increase of the nebular FWHM with respect to a recombination-only scenario (see e.g., \citealt{Costa2022}), and this increase should be stronger when more \lya photons are available. The top panel of Figure~\ref{fig:FWHM-vs-Lratio} shows the \lya FWHM of each nebula as a function of the bolometric luminosity of the quasar, color coded by the percentage of quasar's \lya luminosity (integrated within the $\pm$FWHM$_{\rm Ly\alpha}^{\rm neb}$, Section~\ref{sec:properties}) with respect to the bolometric luminosity (Section~\ref{sec:qso_phys_prop}). 
The median values of bins A2 and A3 are shown with blue squares and error-bars representing the uncertainty of $L_{\rm bol}$. At fixed $L_{\rm bol}$ there is, on average, an increase of the ratio of luminosities. This indicates that quasars with more \lya photons available for scattering present broader nebulae. To further quantify this, we fit a power law relationship between the nebular FWHM and $L_{\rm bol}$ (gray dashed line) and compute the distance in FWHM to it. The result is shown in the bottom panel of Figure~\ref{fig:FWHM-vs-Lratio}, where darker points (larger luminosity ratio) have on average positive differences with respect to the dashed line. However, the scatter is large. Overall, also this test is suggestive of resonant scattering being important in shaping the extended \lya emission around quasars.

\subsubsection{Shocks due to galactic/AGN outflows}\label{sec:pow-outflows}

\begin{figure}[t]
    \centering
    \includegraphics[width=\linewidth]{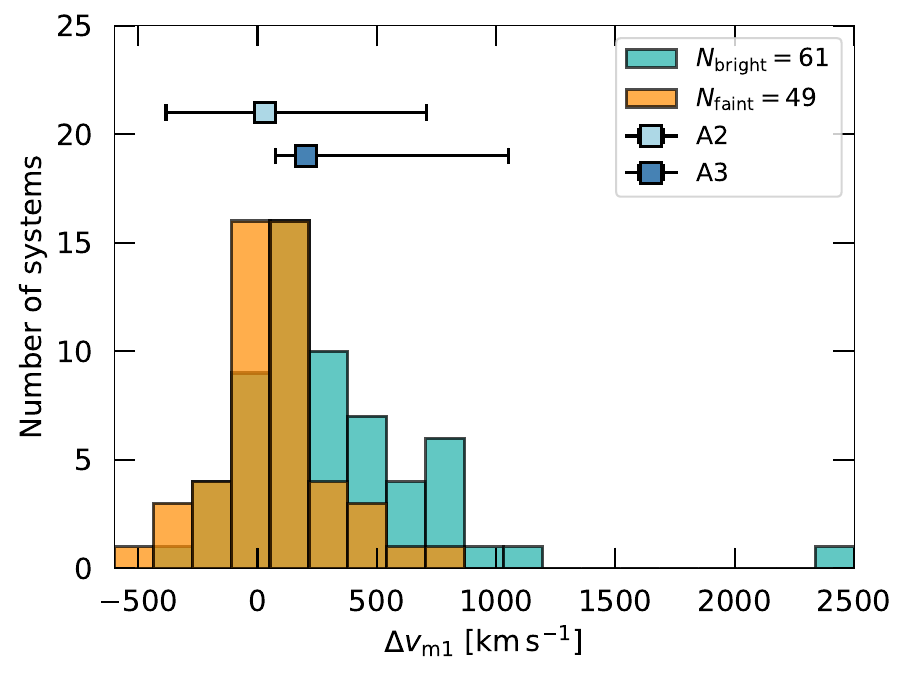}
    \caption{Distribution of \lya nebula velocity shift with respect to the quasar \lya emission peak. The plot shows histograms of the velocity shifts ($v_{\rm m1}$) computed from the first moment of the flux distribution (Tables~\ref{tab:properties}~and~\ref{tab:properties_QSOMUSEUM}) for the extended \lya emission around the bright (green) and faint (orange) QSO MUSEUM samples. The size of each histogram bin corresponds to $163\,{\rm km\,s^{-1}}$. 
    We also show the median $v_{\rm m1}$ for bins A2 and A3 (Figure~\ref{fig:parameter_space}) at the respective number of systems. Their errorbars indicate the full range of velocity shifts within A2 and A3. 
    }
    \label{fig:vshift-histogram}
\end{figure}

Fast shocks are expected to provide additional ionizing photons and boost collisions. 
In such a case, the \lya flux is expected to scale as $F_{\rm Ly\alpha}\propto n_{\rm H}v^3_{\rm shock}$, where $v_{\rm shock}$ is the shock speed \citep{Allen2008}. Given the presence of increased \lya SB around more accreting and hence active quasars, we cannot exclude 
shocks and collisions 
in powering the observed \lya nebulae. 
If only shock speed would be responsible for the difference in \lya SB in the A2 and A3 bins, bright quasars would push $2\times$ faster outflows at fixed $n_{\rm H}$. 
This is not far from the debated scaling invoked by \citet{Fiore2017} between $L_{\rm bol}$ and the outflow velocity $v_{\rm out}$, with $v_{\rm out}\propto L_{\rm bol}^{1/5}$. 
While this relation, if valid, is computed on smaller scales, it could be that shock velocities may follow similar relations and therefore contribute to the powering of the \lya nebulae. 
The exquisite spatial resolution of JWST \citep{Gardner2006} may help us in uncovering morphological and velocity signatures of shocks in the \lya nebulae surrounding quasars as traced by rest-frame optical emission lines (e.g., [OIII], \citealt{Bo2024}).

Further, 
winds/outflows from an AGN 
can cause the peak of the observed \lya flux distribution to appear redshifted with respect to the intrinsic emission due to \lya resonant scattering (see e.g., \citealt{Chang2023, Chang2024}). 
In Figure~\ref{fig:vshift-histogram}, we show the distribution of velocity shifts ($\Delta v_{\rm m1}$) of the nebulae computed from the first moment of their \lya flux distribution with respect to the peak \lya wavelength of the quasar listed in Tables~\ref{tab:properties}~and~\ref{tab:properties_QSOMUSEUM} of the faint and bright quasars with green and orange bins, respectively. 
Additionally, the number of nebulae in bins A2 and A3 as a function of their median $\Delta v_{\rm m1}$ with errorbars representing the velocity range covered by each bin are shown with blue squares. 
Most of the nebulae have redshifts similar to the quasar's \lya, however the bright quasar sample shows more systems with velocity shifts $>200\,{\rm km\,s^{-1}}$. This would be consistent with the proposed scenario in which the most luminous quasars are producing stronger feedback, resulting in a more redshifted \lya emission with respect to fainter quasars. 
Additionally, we also studied the distribution of velocity shifts derived from the gaussian fits ($\Delta v_{\rm gauss}$ in Tables~\ref{tab:properties}~and~\ref{tab:properties_QSOMUSEUM}) and found consistent results. More precise information on the systemic redshifts of the quasars is needed to draw further conclusions.

\subsubsection{A plausible simple scenario: a quasar opening angle increase with bolometric luminosity}\label{sec:simple_model}

The previous subsections discussed evidence in favor of all the \lya powering mechanisms related to AGN feedback, whether it pertains to electromagnetic (photoionization, BLR \lya photon scattering) or mechanical (shocks) processes. In this subsection we assume that all these processes could act together and present a simple calculation to explain the uncovered observational trends.

The calculation is based on the idea proposed by \citet{Costa2022}, which indicates that the observation of a \lya nebula in the vicinity of a quasar requires the presence of an AGN outflow. 
Therefore, we compute whether the targeted quasars are capable of clearing out part of their host galaxy ISM. Specifically, \citet{Costa2018} estimated a 
bolometric luminosity threshold over which an AGN is able to power momentum-driven mechanical outflows, which can be computed as the luminosity at which the outbound radiation pressure acceleration balances the inward gravitational acceleration on the entirety of ISM gas:
\begin{equation}
    \frac{L_{\rm ths}}{c} = 
    \frac{G M_{\rm 200}^{\rm 2}}{R_{\rm ISM}^2} \frac{M_{\rm ISM}}{M_{\rm 200}} \frac{M_{\rm ISM}+M_{\rm\bigstar}+M_{\rm DM}(<R_{\rm ISM})}{M_{\rm 200}},
\end{equation}
where $G$ is the gravitational constant, $c$ the speed of light, $R_{\rm ISM}\approx0.1R_{200}$ is the outer boundary of the ISM, and $M_{\rm ISM}$, $M_{\rm\bigstar}$ and $M_{\rm DM}(<R_{\rm ISM})$ are the gas, star and dark matter masses within the ISM region. 
A typical $z\sim3$ quasar host dark matter halo has $M_{\rm 200}$=10$^{\rm 12.5}$M$_{\rm\odot}$ and $R_{\rm 200}\approx$100~kpc. 
If we assume a Navarro-Frenk-White dark matter profile \citep{Navarro1997} and use the concentration--halo mass relation for $z=3$ of \citet{Dutton:2014} $M_{\rm DM}(<R_{\rm ISM})/M_{\rm 200}\approx0.056$, 
we can estimate the ISM mass fraction $M_{\rm ISM}/M_{\rm 200}$ using the EAGLE simulation \citep[the dashed green curve in figure 2 of ][for $z=2$]{Mitchell:2022} to be 8\% for our target halo mass, where we assumed the cosmic baryon fraction value of the Planck cosmology $f_b=\Omega_b/\Omega_M=0.157$. 
For the stellar mass fraction, we can use the $z$-appropriate relation of \citet{Moster2018} to get $M_{\rm\bigstar}/M_{\rm 200}\approx 0.027$. 
Plugging all these numbers in the equation above, an AGN at $z=3$ has to have a bolometric luminosity of at least $L_{\rm ths}=$1.08$\times$10$^{\rm 48}$~erg/s for radiation pressure to push away all of the ISM. 
This limit luminosity is computed assuming a spherical symmetric ISM distribution. However, quasars are typically associated with star forming galaxies, which have their ISM organized in a disk. 
Therefore, the amount of ISM that needs to be removed for the AGN radiation to escape in free channels 
is lower. 
For example, if we assume the AGN needs to remove only the ISM within a typical ionization bi-cone of angle $\alpha=60^\circ$ \citep[e.g.][]{Obreja2024}, the luminosity limit becomes [1-cos($\alpha$/2)]$\times$1.08$\times$10$^{\rm 48}$erg/s=1.4$\times$10$^{\rm 47}$erg/s. 
Interestingly, this new $L_{\rm bol}$ limit coincides with the minimum luminosity of the QSO MUSEUM I (bright quasar) sample, which has larger average \lya SB than the QSO MUSEUM faint sample as given e.g., by the points' sizes in Figure~\ref{fig:parameter_space}. 
The observed \lya SB of the bright nebulae could then be produced because the AGN has been able to push enough ISM material. In turn this would make more efficient all the powering mechanisms related to AGN feedback: there will be (i) more ionizing photons escaping the galaxy and ionizing a larger mass of CGM gas (subsection~\ref{sec:pow-photoionization}), (ii) larger channels of least resistance for the propagation of \lya photons (subsection~\ref{sec:pow-scattering}), and (iii) faster shocks (subsection~\ref{sec:pow-outflows}).

Using the above reasoning we could estimate how the opening angle of the targeted quasars changes as a function of their $L_{\rm bol}$ by requiring:
\begin{equation}\label{eq:Lbol-opening-anlge}
	 L_{\rm bol}=L_{\rm ths}\times[1-{\rm cos}(\alpha/2)].
\end{equation}
\begin{figure}
    \centering
    \includegraphics[width=\linewidth]{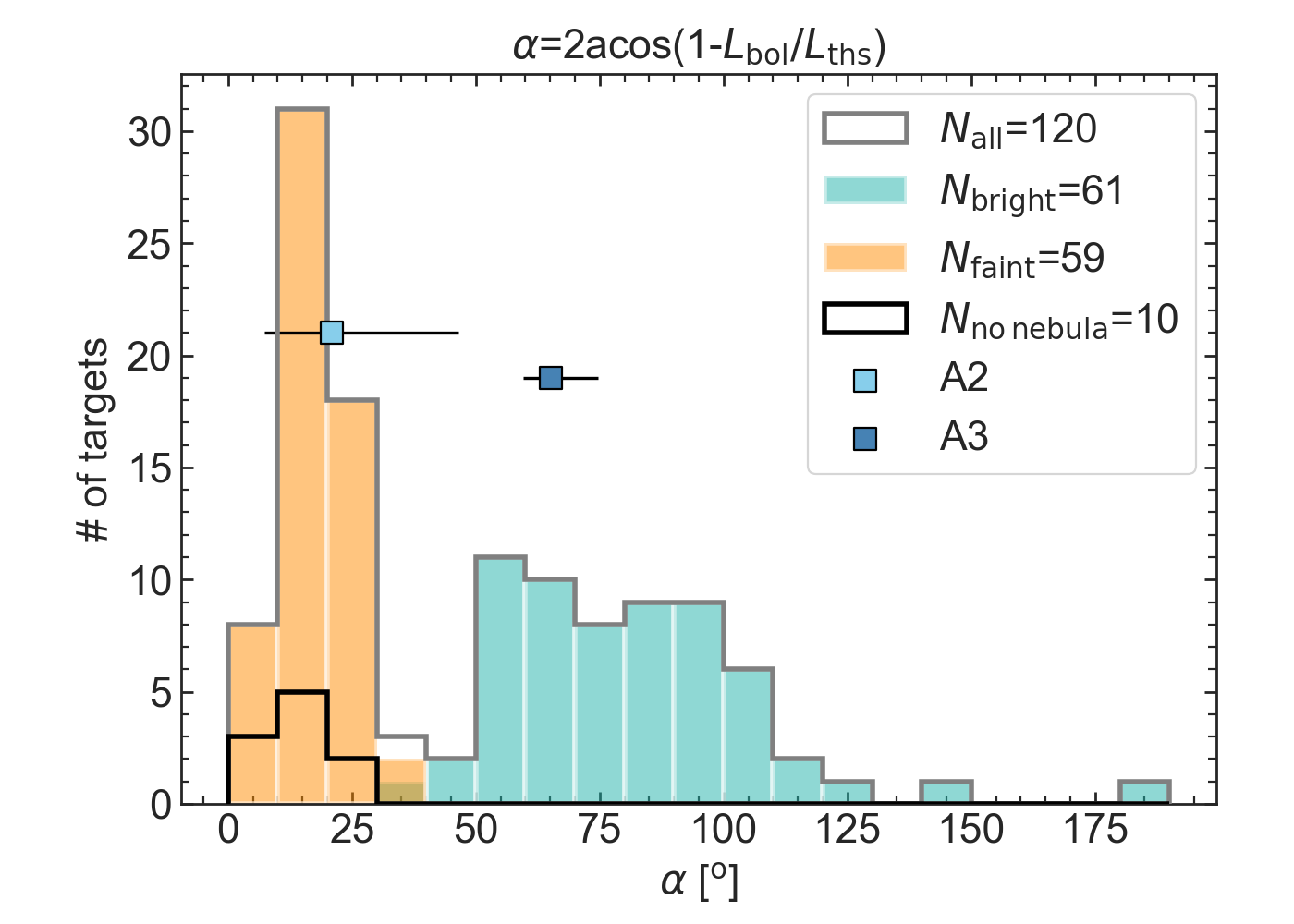}
    \caption{Distribution of the opening angle estimated for the targeted quasars using Equation~\ref{eq:Lbol-opening-anlge}. 
    The histogram shows the number of QSO MUSEUM faint (orange) and bright (dark green) quasars with their estimated opening angle following our discussion in Section~\ref{sec:pow_mech}. 
    The values of bins A2 and A3 are shown with blue squares. The thick black histogram represents the non-detections.}
    \label{fig:openingAngle}
\end{figure}
Figure~\ref{fig:openingAngle} shows the result of solving for $\alpha$ in Equation~\ref{eq:Lbol-opening-anlge}, with $L_{\rm ths}~=~1.08\times10^{\rm 48}\,{\rm erg\,s^{-1}}$, for the full sample. Brighter quasars would have larger opening angles. 
The same figure shows that the median opening angles of the subsamples A2 (faint) and A3 (bright) should be $21^\circ$ and $65^\circ$, respectively. This means that the ratio of illuminated volumes is $\sim9$, which is very similar to the ratio of their SB levels ($\sim7$), indicating the plausibility of such scenario. 
The median angle we find for the brigther sample is in agreement with the work by \citet{Obreja2024}, which studied the impact, in post-processing, that a varying AGN opening angle has in the fraction of cool gas within a simulated halo, using a combination of black hole masses of $M_{\rm BH}=10^8$~and~$10^9\,{\rm M_{\odot}}$ and Eddington ratios of $\lambda=0.1$~and~$1.0$. They concluded that a combination of $M_{\rm BH}=10^9\,{\rm M_\odot}$, $\lambda=0.1$ and opening angle of $60^\circ$ was able to reasonably reproduce the average SB radial profile in different rest-frame UV transitions obtained in the stacking analysis presented in \citet{Fossati2021}. 
Furthermore, quasars associated with non-detections are at the low end of the distribution of opening angles, which is consistent with their stacked SB profile being above the predictions for UVB-only illumination\footnote{In the case of very faint AGN, we expect star formation and cooling radiation to become more important in powering the \lya emission and add to the UVB contribution.} (\citealt{Obreja2024}, left panel of Figure~\ref{fig:SB+sigma-profiles}). 
It is important to note that the value estimated for $L_{\rm ths}$ is valid only in an outflow scenario due to radiation pressure, where the outward force is produced if every photon scatters only once. However, in very compact and dusty galaxies it is possible that re-processed infrared photons are scattered multiple times by dust grains \citep{Costa2018}, which lowers the luminosity threshold needed to push out the galaxy's ISM. Additionally, in the presence of energy-driven outflows \citep{King2003,Ward2024} it is possible to have an extra contribution to the outwards force due to the shocked gas, which can lower the luminosity threshold by a factor of $\sim10$, as shown in simulations of $z>6$ quasars \citep{Costa2014,Costa2020}. Adopting this factor into Equation \ref{eq:Lbol-opening-anlge} results in much larger opening angles. The faint quasars bolometric luminosities would require opening angles of $10^\circ$ to $\sim100^\circ$, while most bright quasars would exceed $180^\circ$ and therefore illuminate all the gas around them. These large opening angles for bright quasars are however ruled out by several previous studies (e.g., \citealt{Hennawi2013, Borisova2016, Obreja2024}. Consequently, our discussion is conducted within the first approximation. However, it is acknowledged that the actual scenario may exhibit intermediate characteristics between these two regimes.

It has been argued that so different quasar opening angles 
might result in observable difference in the asymmetries of the 
nebulae \citep{Mackenzie2021}. However, our observations do not support the presence of such variations, which is consistent with earlier research. We hypothesize that the absence of observable differences can be attributed to the fact that at small opening angles, the level of emission is comparable to the level of \lya emission resulting from other mechanisms, such as host galaxy star formation, cooling radiation, and UVB contribution, hence washing out any geometrical difference due to quasar illumination alone.
Also, small opening angles for faint quasars can explain the more frequent presence of narrow \lya nebulae with very similar spectrum to some portions of the \lya emission of the quasar itself (see Figures~\ref{fig:spec_atlas01}-\ref{fig:spec_atlas03}). Such emission is likely due to ISM and inner CGM and potentially not illuminated by the quasar (see also \citealt{Mackenzie2021}). 
We stress that a luminosity dependent opening angle means that the fraction of obscured or type-II AGN $f_{\rm N}$ would vary as a function of $L_{\rm bol}$. Specifically, following geometrical arguments $f_{\rm N}={\rm cos(\alpha/2)}$, resulting in $f_N^{\rm faint}=0.93^{+0.06}_{-0.24}$ and $f_N^{\rm bright}=0.42^{+0.09}_{-0.15}$ in our case. These values are in overall agreement with those obtained using infrared luminosities and X-ray surveys (see e.g., Figure 4 in \citealt{Treister2008}, \citealt{Marconi2004,Merloni2014,HickoxAlexander2018}), also considering that in the quoted errors we do not include the large uncertainties in $L_{\rm bol}$. Even though these fractions are still matter of debate, we argue that statistical studies of quasar CGM could help in assessing this problem further.


\section{Summary}\label{sec:summary}

We have presented the largest effort to date to map the CGM around quasars in \lya emission, namely 120 $z\sim3$ quasars (QSO~MUSEUM~III). This effort builds upon our first sample around bright quasars (QSO~MUSEUM~I, \citealt{FAB2019}) and adds 59 new faint quasar fields, resulting in a global sample with a median redshift of $z=3.13$, and bolometric luminosities, black hole masses and Eddington ratios in the ranges $45.1 < {\rm log}(L_{\rm bol}/[{\rm erg\ s^{-1}]}) < 48.7$, $7.9 < {\rm log}(M_{\rm BH}/[{\rm M_{\odot}]}) < 10.3 $ and $0.01< \lambda_{\rm Edd} <1.8$, respectively. We were able to detect 110 \lya nebulae around the 120 targets using an homogeneous strategy consisting of snapshot VLT/MUSE observations (45 minutes/source), totalling 120 hours of MUSE time when considering observation overheads. All the non-detections are associated with the new faint sample. We characterized the \lya SB level, kinematics, morphology (area and asymmetry) of the extended emission, and looked for trends with quasar properties (luminosities, black hole mass, Eddington ratio). Our observational results can be summarized as follows:
\begin{itemize}
\item The surface brightness of the CGM \lya emission depends on the luminosities of the central quasar and therefore decreases around fainter quasars (Section~\ref{sec:radial-profiles},  Figure~\ref{fig:Rprofiles}) while keeping the same radial profile shape. 
More precisely, a factor of ten decrease in quasar bolometric luminosity (corresponding to a factor of 5 in peak \lya luminosity) results in about seven times dimmer surface brightness radial profiles.
\item The morphology of the extended \lya emission characterized using measures of elongation and lopsidedness 
shows that the emission is overall circular but it tends to be preferentially displaced towards one side of the quasar. There are no particularly evident trends with quasar luminosity, apart a tentative increase of more lopsided and asymmetric systems at low luminosity (Figure~\ref{fig:asymmetry-luminosity}).
\item A stack of the 10 non-detections reveals extended \lya emission just below the individual SB limits (Section~\ref{sec:radial-profiles}), thereby confirming that the apparent lack of \lya emission around these systems is due to the faintness of their associated quasars, together with possible line-of-sight effects. 
\item The \lya velocity dispersion profiles show a similar behavior as the surface brightness profiles: the velocity dispersion around brighter quasars is higher than around faint quasars (Section~\ref{sec:radial-profiles},  Figure~\ref{fig:sigma-profiles}). Specifically, a factor of ten decrease in quasar bolometric luminosity (corresponding to a factor of 5 in peak \lya luminosity) results in at least two times smaller \lya velocity dispersion in the inner CGM ($<40$~kpc).
\item The large sample size allowed us to apply different binnings in quasar properties to the data, and show that the quasar bolometric luminosity is the key parameter governing the changes in surface brightness levels and velocity dispersion. When considering a fixed black-hole mass bin, such a change corresponds to a different accretion regime or Eddington ratio. For $M_{\rm BH}\sim10^9$~M$_{\odot}$, a change in the accretion from $\lambda_{\rm Edd}=0.2$ to $0.9$ results in 7 times brighter \lya surface brightness profiles and in at least two times larger \lya velocity dispersion (Figure~\ref{fig:SB+sigma-profiles}). 
\item The binned \lya radial velocity dispersion profiles can be well fit by a power law of the form $\sigma = \sigma_{50}(r/50\,{\rm ckpc})^{\beta}$, with $\beta$ found to be $\sim-1$ throughout the full range of luminosities considered. Interestingly, the normalization factor $\sigma_{50}$ clearly scales with the bolometric luminosity following a power law $\sigma_{50}\propto L_{\rm bol}^{\alpha}$, with $\alpha\sim0.29$ (Section~\ref{sec:stacked-profiles}). 

\end{itemize}

Based on these observational findings, we propose that: 

\begin{itemize}
\item The relations between the ``current'' quasar bolometric luminosity and observed surface brightness and velocity dispersion radial profiles are likely a probe of 
instantaneous AGN feedback (Section~\ref{sec:inst_feedback}). Indeed, the relation with the observed current quasar activity implies that the \lya powering mechanisms act on timescales similar to their variability.
\item The dependence of the CGM \lya emission with the quasar luminosity is due to different ionization cone opening angles at different quasars luminosity, with brighter quasars having larger opening angles (Section~\ref{sec:simple_model}). Larger ionization cone opening angles can boost the \lya signal by increasing the amount of ionized gas (photoionization) or increasing the fraction of illuminated gas (resonant scattering of BLR \lya photons).
\item A combination of quasar anisotropic photoionization followed by recombination, quasar's \lya photon resonant scattering, CGM \lya photons resonant scattering, and shocks seem to be needed to explain the observation of extended \lya emission around quasars (Section~\ref{sec:pow_mech}).
\end{itemize}

In conclusion, the findings of this study highlight the key importance of using large samples of quasars for targeted investigations into their CGM emission properties. To improve and test our findings, future observational efforts should focus on obtaining large samples targeting additional line emissions beside \lya (e.g., H$\alpha$, \ion{He}{ii},  \ion{C}{iv}, [\ion{O}{III}]; e.g. \citealt{Bo2024}) to assess the balance between \lya powering mechanisms and firmly constrain the CGM physical properties (e.g., densities, metal enrichment, ionization state). Such a work targeting \ion{C}{iv} and \ion{He}{ii} is ongoing and will be presented in a future accompanying article.

Given the complexity of these systems, the comparison to cosmological simulations will be invaluable. In particular, QSO MUSEUM has already enough statistics to provide tests for AGN feedback implementations. High-resolution simulations (e.g., \citealt{Costa2022,Obreja2024}) are however required to be run down to the probed redshifts.

\begin{acknowledgements}
Based on observations collected at the European Organisation for Astronomical Research in the Southern Hemisphere under ESO programmes 094.A-0585(A), 095.A-0615(A), 095.A-0615(B), 096.A-0937(B), 0160.A-0297(A).
A.O. is funded by the Carl-Zeiss-Stiftung through the NEXUS programm. 
E.P.F. is supported by the international Gemini Observatory, a program of NSF NOIRLab, which is managed by the Association of Universities for Research in Astronomy (AURA) under a cooperative agreement with the U.S. National Science Foundation, on behalf of the Gemini partnership of Argentina, Brazil, Canada, Chile, the Republic of Korea, and the United States of America. 
\end{acknowledgements}





\bibliographystyle{aa}
\bibliography{aanda}

\begin{appendix}

\section{Estimating quasar properties from their 1-D spectra}
\label{app:QSOfit}

\FloatBarrier

\begin{table*}
\footnotesize \centering
\caption{Quasar properties for the bright sample (QSO MUSEUM I).}\label{tab:QSOfit-bright}
\begin{tabular}{|c|c|c|c|c|c|c|c|}
\hline
\T
ID & z\tablefootmark{(a)} & $L_{\rm Ly\alpha;peak}^{\rm QSO}$ & $\lambda L_\lambda(1350\,\AA)$ & $L_{\rm bol}$ & FWHM(\ion{C}{iv}) & $M_{\rm BH}$ & $\lambda_{\rm Edd}$ \\
   &  & [${ 10^{43}\, \rm erg\,s^{-1}\,\AA^{-1} }$] & [${ \rm 10^{46}\, erg\,s^{-1} }$] & [${ \rm 10^{47}\, erg\,s^{-1} }$] & [${\rm km\,s^{-1}}$] & [${\rm 10^{9}\, M_{sun}}$] &  \B \\
\hline
\T
1  & 3.166 & $5.4\pm0.3 $ & $5.00\pm0.02$ & $1.90\pm0.01$ & $3724\pm9$ & $1.71\pm0.01$ & $0.885\pm0.006$ \\
2  & 3.133 & $6.1\pm0.3 $ & $14.17\pm0.03$ & $5.40\pm0.01$ & $5653\pm50$ & $6.84\pm0.12$ & $0.627\pm0.011$ \\
3  & 3.110 & $9.1\pm0.5 $ & $10.27\pm0.03$ & $3.91\pm0.01$ & $4415\pm18$ & $3.52\pm0.03$ & $0.883\pm0.008$ \\
4  & 3.219 & $4.3\pm0.2 $ & $8.71\pm0.02$ & $3.32\pm0.01$ & $7142\pm38$ & $8.43\pm0.09$ & $0.312\pm0.003$ \\
5  & 3.320 & $3.6\pm0.2 $ & $5.95\pm0.01$ & $2.27\pm0.01$ & $6542\pm33$ & $5.78\pm0.06$ & $0.311\pm0.003$ \\
6  & 3.029 & $7.0\pm0.4 $ & $13.13\pm0.09$ & $5.00\pm0.03$ & $8035\pm34$ & $13.26\pm0.12$ & $0.299\pm0.003$ \\
7  & 3.117 & $8.0\pm0.4 $ & $10.72\pm0.05$ & $4.09\pm0.02$ & $5381\pm17$ & $5.34\pm0.04$ & $0.607\pm0.005$ \\
8  & 3.132 & $7.1\pm0.4 $ & $4.44\pm0.05$ & $1.69\pm0.02$ & $3958\pm14$ & $1.81\pm0.02$ & $0.741\pm0.011$ \\
9  & 3.301 &$10.0\pm0.5 $ & $19.53\pm0.07$ & $7.44\pm0.02$ & $5185\pm19$ & $6.82\pm0.05$ & $0.866\pm0.007$ \\
10 & 3.227 & $8.8\pm0.4 $ & $7.95\pm0.20$ & $3.03\pm0.08$ & $4976\pm26$ & $3.90\pm0.07$ & $0.616\pm0.019$ \\
11 & 3.078 & $9.9\pm0.5 $ & $1.92\pm0.01$ & $0.73\pm0.01$ & $6446\pm52$ & $3.08\pm0.05$ & $0.188\pm0.003$ \\
12 & 3.376 & $4.9\pm0.2 $ & $9.14\pm0.04$ & $3.48\pm0.02$ & $7284\pm33$ & $9.00\pm0.09$ & $0.307\pm0.003$ \\
13 & 3.164 & $5.2\pm0.3 $ & $9.98\pm0.01$ & $3.80\pm0.01$ & $5660\pm31$ & $5.69\pm0.06$ & $0.530\pm0.006$ \\
14 & 3.126 & $1.9\pm0.1 $ & $4.42\pm0.01$ & $1.69\pm0.01$ & $5278\pm38$ & $3.22\pm0.05$ & $0.416\pm0.006$ \\
15 & 3.141 & $1.2\pm0.1 $ & $7.65\pm0.02$ & $2.91\pm0.01$ & $3319\pm4$ & $1.70\pm0.01$ & $1.361\pm0.005$ \\
16 & 3.142 & $2.2\pm0.1 $ & $4.39\pm0.01$ & $1.67\pm0.01$ & $7152\pm61$ & $5.88\pm0.10$ & $0.226\pm0.004$ \\
17 & 3.340 & $2.2\pm0.1 $ & $3.72\pm0.03$ & $1.42\pm0.01$ & $6814\pm33$ & $4.89\pm0.05$ & $0.230\pm0.003$ \\
18 & 3.265 & $7.6\pm0.4 $ & $5.98\pm0.02$ & $2.28\pm0.01$ & $4966\pm21$ & $3.34\pm0.03$ & $0.542\pm0.005$ \\
19 & 3.188 & $4.5\pm0.2 $ & $8.70\pm0.04$ & $3.32\pm0.01$ & $8736\pm58$ & $12.61\pm0.17$ & $0.209\pm0.003$ \\
20 & 3.395 & $5.6\pm0.3 $ & $10.47\pm0.01$ & $3.99\pm0.01$ & $7594\pm43$ & $10.51\pm0.12$ & $0.301\pm0.003$ \\
21 & 3.219 & $1.5\pm0.1 $ & $8.31\pm0.02$ & $3.16\pm0.01$ & $3322\pm6$ & $1.78\pm0.01$ & $1.412\pm0.007$ \\
22 & 3.176 & $4.5\pm0.2 $ & $9.36\pm0.03$ & $3.57\pm0.01$ & $9147\pm51$ & $14.37\pm0.16$ & $0.197\pm0.002$ \\
23 & 3.061 & $2.8\pm0.1 $ & $6.29\pm0.07$ & $2.39\pm0.03$ & $11518\pm40$ & $18.45\pm0.16$ & $0.103\pm0.001$ \\
24 & 3.062 & $9.7\pm0.5 $ & $3.56\pm0.01$ & $1.36\pm0.01$ & $3809\pm10$ & $1.49\pm0.01$ & $0.722\pm0.005$ \\
25 & 3.307 & $4.0\pm0.2 $ & $7.77\pm0.04$ & $2.96\pm0.02$ & $8552\pm41$ & $11.38\pm0.11$ & $0.207\pm0.002$ \\
26 & 3.181 & $7.1\pm0.4 $ & $118.32\pm0.44$ & $45.08\pm0.17$ & $5519\pm17$ & $20.07\pm0.13$ & $1.783\pm0.013$ \\
27 & 3.318 & $6.8\pm0.3 $ & $11.46\pm0.04$ & $4.37\pm0.02$ & $5244\pm22$ & $5.26\pm0.05$ & $0.659\pm0.006$ \\
28 & 3.344 & $2.6\pm0.1 $ & $3.77\pm0.01$ & $1.44\pm0.01$ & $5856\pm25$ & $3.64\pm0.03$ & $0.313\pm0.003$ \\
29 & 3.180 & $3.0\pm0.2 $ & $6.38\pm0.05$ & $2.43\pm0.02$ & $7896\pm39$ & $8.74\pm0.10$ & $0.221\pm0.003$ \\
30 & 3.357 & $1.2\pm0.1 $ & $8.43\pm0.03$ & $3.21\pm0.01$ & $3266\pm8$ & $1.73\pm0.01$ & $1.471\pm0.009$ \\
31 & 3.131 & $5.0\pm0.2 $ & $12.56\pm0.1$ & $4.78\pm0.04$ & $8700\pm55$ & $15.19\pm0.20$ & $0.250\pm0.004$ \\
32 & 3.069 & $1.8\pm0.1 $ & $3.16\pm0.01$ & $1.20\pm0.01$ & $7486\pm57$ & $5.41\pm0.08$ & $0.176\pm0.003$ \\
33 & 3.125 & $8.9\pm0.4 $ & $4.37\pm0.02$ & $1.67\pm0.01$ & $3579\pm6$ & $1.47\pm0.01$ & $0.900\pm0.006$ \\
34 & 3.223 & $1.1\pm0.1 $ & $1.8\pm0.02$ & $0.68\pm0.01$ & $7888\pm79$ & $4.46\pm0.10$ & $0.122\pm0.003$ \\
35 & 3.247 & $5.8\pm0.3 $ & $2.67\pm0.01$ & $1.02\pm0.01$ & $3296\pm5$ & $0.96\pm0.01$ & $0.842\pm0.004$ \\
36 & 3.197 & $3.3\pm0.2 $ & $3.21\pm0.02$ & $1.22\pm0.01$ & $3364\pm21$ & $1.10\pm0.01$ & $0.881\pm0.012$ \\
37 & 3.130 & $7.5\pm0.4 $ & $5.10\pm0.02$ & $1.94\pm0.01$ & $3245\pm9$ & $1.31\pm0.01$ & $1.177\pm0.008$ \\
38 & 3.348 & $8.3\pm0.4 $ & $9.10\pm0.02$ & $3.47\pm0.01$ & $4460\pm23$ & $3.37\pm0.03$ & $0.818\pm0.009$ \\
39 & 3.100 & $1.0\pm0.1 $ & $6.38\pm0.06$ & $2.43\pm0.02$ & $3281\pm5$ & $1.51\pm0.01$ & $1.278\pm0.014$ \\
40 & 3.342 & $7.7\pm0.4 $ & $7.96\pm0.04$ & $3.03\pm0.01$ & $4185\pm107$ & $2.76\pm0.14$ & $0.872\pm0.045$ \\
41 & 3.321 & $2.8\pm0.1 $ & $4.84\pm0.02$ & $1.84\pm0.01$ & $5496\pm28$ & $3.66\pm0.04$ & $0.400\pm0.004$ \\
42 & 3.042 & $8.5\pm0.4 $ & $6.06\pm0.02$ & $2.31\pm0.01$ & $3546\pm9$ & $1.71\pm0.01$ & $1.068\pm0.007$ \\
43 & 3.087 & $2.6\pm0.1 $ & $3.12\pm0.01$ & $1.19\pm0.01$ & $4950\pm91$ & $2.35\pm0.09$ & $0.401\pm0.015$ \\
44 & 3.197 & $3.0\pm0.2 $ & $6.68\pm0.05$ & $2.55\pm0.02$ & $6582\pm44$ & $6.22\pm0.09$ & $0.325\pm0.005$ \\
45 & 3.312 & $3.0\pm0.2 $ & $10.88\pm0.05$ & $4.15\pm0.02$ & $5411\pm135$ & $5.45\pm0.28$ & $0.604\pm0.030$ \\
46 & 3.385 & $7.9\pm0.4 $ & $11.30\pm0.01$ & $4.31\pm0.0$ & $5681\pm16$ & $6.12\pm0.04$ & $0.558\pm0.003$ \\
47 & 3.156 & $3.7\pm0.2 $ & $3.69\pm0.02$ & $1.40\pm0.01$ & $3529\pm11$ & $1.31\pm0.01$ & $0.854\pm0.008$ \\
48 & 3.197 & $3.7\pm0.2 $ & $3.90\pm0.01$ & $1.49\pm0.01$ & $5286\pm91$ & $3.02\pm0.11$ & $0.391\pm0.014$ \\
49 & 3.109 & $1.5\pm0.1 $ & $2.83\pm0.01$ & $1.08\pm0.01$ & $6275\pm57$ & $3.59\pm0.07$ & $0.239\pm0.004$ \\
50 & 3.197 & $3.4\pm0.2 $ & $5.17\pm0.02$ & $1.97\pm0.01$ & $5585\pm50$ & $3.91\pm0.07$ & $0.400\pm0.007$ \\
51 & 3.138 & $5.1\pm0.3 $ & $4.19\pm0.02$ & $1.60\pm0.01$ & $3279\pm10$ & $1.21\pm0.01$ & $1.051\pm0.008$ \\
52 & 3.144 & $4.6\pm0.2 $ & $5.40\pm0.03$ & $2.06\pm0.01$ & $3975\pm35$ & $2.03\pm0.04$ & $0.805\pm0.015$ \\
53 & 3.149 & $2.0\pm0.1 $ & $1.45\pm0.01$ & $0.55\pm0.01$ & $4101\pm24$ & $1.08\pm0.01$ & $0.408\pm0.006$ \\
54 & 3.109 & $3.6\pm0.2 $ & $2.75\pm0.03$ & $1.05\pm0.01$ & $3543\pm10$ & $1.13\pm0.01$ & $0.739\pm0.011$ \\
55 & 3.188 & $3.5\pm0.2 $ & $4.16\pm0.03$ & $1.58\pm0.01$ & $3815\pm21$ & $1.63\pm0.02$ & $0.774\pm0.01$ \\
56 & 3.089 & $5.8\pm0.3 $ & $6.87\pm0.04$ & $2.62\pm0.02$ & $3930\pm15$ & $2.25\pm0.02$ & $0.923\pm0.009$ \\
57 & 3.406 & $2.3\pm0.1 $ & $6.84\pm0.04$ & $2.61\pm0.02$ & $7283\pm172$ & $7.71\pm0.37$ & $0.268\pm0.013$ \\
58 & 3.037 & $5.9\pm0.3 $ & $8.61\pm0.05$ & $3.28\pm0.02$ & $4990\pm31$ & $4.09\pm0.05$ & $0.637\pm0.009$ \\
59 & 3.118 & $3.8\pm0.2 $ & $7.59\pm0.06$ & $2.89\pm0.02$ & $8271\pm79$ & $10.51\pm0.21$ & $0.218\pm0.005$ \\
60 & 3.183 & $4.4\pm0.2 $ & $2.74\pm0.02$ & $1.04\pm0.01$ & $4203\pm13$ & $1.58\pm0.01$ & $0.524\pm0.005$ \\
61 & 3.061 & $5.9\pm0.3 $ & $7.74\pm0.03$ & $2.95\pm0.01$ & $4366\pm24$ & $2.96\pm0.03$ & $0.791\pm0.009$ \B \\
\hline
\end{tabular}
\tablefoot{
\tablefoottext{a}{Quasar systemic redshift. The intrinsic uncertainty on this value is $\Delta z\sim0.007$.}
}

\end{table*}

\begin{table*}
\footnotesize \centering
\caption{Quasar properties for the faint sample (QSO MUSEUM III).}\label{tab:QSOfit-faint}
\begin{tabular}{|c|c|c|c|c|c|c|c|}
\hline
\T
ID & z\tablefootmark{a} & $L_{\rm Ly\alpha;peak}^{\rm QSO}$ & $L_\lambda(1350\,\AA)$ & $L_{\rm bol}$ & FWHM(\ion{C}{iv}) & $M_{\rm BH}$ & $\lambda_{\rm Edd}$ \\
   &  & [${ 10^{42}\, \rm erg\,s^{-1}\,\AA^{-1} }$] & [${ \rm 10^{45}\, erg\,s^{-1} }$] & [${ \rm 10^{46}\, erg\,s^{-1} }$] & [${\rm km\,s^{-1}}$] & [${\rm 10^{8}\, M_{sun}}$] &  \B \\
\hline
\T
62 & 3.051   & $1.5\pm0.1 $ & $0.49\pm0.06$ & $0.19\pm0.02$ & $3963\pm247$ & $1.67\pm0.25$ & $0.089\pm0.016$ \\
63 & 3.142   & $3.6\pm0.2 $ & $3.93\pm0.06$ & $1.50\pm0.02$ & $6806\pm296$ & $14.82\pm1.35$ & $0.080\pm0.007$ \\
64 & 3.064   & $2.4\pm0.1 $ & $1.54\pm0.08$ & $0.59\pm0.03$ & $3214\pm25$ & $2.01\pm0.06$ & $0.231\pm0.013$ \\
65 & 3.033   & $4.9\pm0.2 $ & $1.80\pm0.06$ & $0.69\pm0.02$ & $3497\pm97$ & $2.59\pm0.16$ & $0.211\pm0.014$ \\
66 & 3.139   & $7.0\pm0.4 $ & $12.25\pm0.09$ & $4.67\pm0.03$ & $6883\pm136$ & $27.68\pm1.12$ & $0.134\pm0.005$ \\
67 & 3.128   & $6.7\pm0.3 $ & $7.99\pm0.07$ & $3.04\pm0.03$ & $5450\pm59$ & $13.84\pm0.31$ & $0.175\pm0.004$ \\
68 & 3.062   &$12.0\pm0.6 $ & $1.52\pm0.05$ & $0.58\pm0.02$ & $3420\pm20$ & $2.26\pm0.05$ & $0.203\pm0.008$ \\
69 & 3.135   & $4.4\pm0.2 $ & $5.09\pm0.07$ & $1.94\pm0.03$ & $5517\pm130$ & $11.16\pm0.54$ & $0.138\pm0.007$ \\
70 & 3.150   & $4.9\pm0.2 $ & $1.11\pm0.04$ & $0.42\pm0.02$ & $2823\pm59$ & $1.30\pm0.06$ & $0.257\pm0.015$ \\
71 & 3.195   & $5.4\pm0.3 $ & $5.78\pm0.08$ & $2.2\pm0.03$ & $3958\pm58$ & $6.15\pm0.19$ & $0.284\pm0.009$ \\
72 & 3.106   & $1.2\pm0.1 $ & $0.76\pm0.06$ & $0.29\pm0.02$ & $5716\pm411$ & $4.38\pm0.71$ & $0.053\pm0.009$ \\
73 & 3.088   & $2.3\pm0.1 $ & $0.75\pm0.06$ & $0.28\pm0.02$ & $3944\pm210$ & $2.06\pm0.25$ & $0.109\pm0.015$ \\
74 & 3.183   & $3.4\pm0.2 $ & $2.39\pm0.07$ & $0.91\pm0.02$ & $4090\pm94$ & $4.11\pm0.20$ & $0.176\pm0.01$ \\
75 & 3.003   & $8.5\pm0.4 $ & $4.71\pm0.06$ & $1.79\pm0.02$ & $4517\pm89$ & $7.18\pm0.29$ & $0.198\pm0.008$ \\
76 & 3.199   &$18.0\pm0.9 $ & $9.64\pm0.11$ & $3.67\pm0.04$ & $3881\pm32$ & $7.75\pm0.14$ & $0.376\pm0.008$ \\
77 & 3.114   & $3.9\pm0.2 $ & $0.47\pm0.11$ & $0.18\pm0.04$ & $4218\pm538$ & $1.86\pm0.59$ & $0.077\pm0.027$ \\
78 & 3.098   & $2.3\pm0.1 $ & $2.05\pm0.07$ & $0.78\pm0.03$ & $4438\pm119$ & $4.46\pm0.26$ & $0.139\pm0.009$ \\
79 & 3.155   & $8.9\pm0.4 $ & $2.37\pm0.06$ & $0.90\pm0.02$ & $3513\pm25$ & $3.02\pm0.06$ & $0.237\pm0.007$ \\
80 & 3.032   &$24.0\pm1.2 $ & $9.9\pm0.06$ & $3.77\pm0.02$ & $3671\pm12$ & $7.04\pm0.05$ & $0.425\pm0.004$ \\
81 & 3.027   & $2.9\pm0.1 $ & $2.07\pm0.05$ & $0.79\pm0.02$ & $3692\pm141$ & $3.11\pm0.25$ & $0.202\pm0.017$ \\
82 & 3.042   & $0.7\pm0.0 $ & $0.34\pm0.08$ & $0.13\pm0.03$ & $3207\pm657$ & $0.90\pm0.48$ & $0.114\pm0.054$ \\
83 & 3.094   & $5.0\pm0.2 $ & $1.50\pm0.07$ & $0.57\pm0.02$ & $3193\pm106$ & $1.96\pm0.14$ & $0.232\pm0.019$ \\
84 & 3.046   & $3.8\pm0.2 $ & $4.12\pm0.06$ & $1.57\pm0.02$ & $5290\pm178$ & $9.18\pm0.64$ & $0.136\pm0.009$ \\
85 & 3.135   & $2.6\pm0.1 $ & $5.16\pm0.06$ & $1.97\pm0.02$ & $4843\pm149$ & $8.67\pm0.55$ & $0.180\pm0.011$ \\
86 & 3.190   & $7.5\pm0.4 $ & $7.29\pm0.06$ & $2.78\pm0.02$ & $3517\pm32$ & $5.49\pm0.10$ & $0.401\pm0.008$ \\
87 & 3.157   & $3.5\pm0.2 $ & $1.31\pm0.05$ & $0.50\pm0.02$ & $4544\pm205$ & $3.69\pm0.36$ & $0.107\pm0.011$ \\
88 & 3.035   & $2.5\pm0.1 $ & $2.56\pm0.05$ & $0.98\pm0.02$ & $4429\pm134$ & $5.00\pm0.32$ & $0.155\pm0.01$ \\
89 & 3.151   & $2.9\pm0.1 $ & $3.68\pm0.09$ & $1.40\pm0.03$ & $5327\pm230$ & $8.77\pm0.80$ & $0.127\pm0.012$ \\
90 & 3.049   & $8.6\pm0.4 $ & $3.73\pm0.06$ & $1.42\pm0.02$ & $3713\pm27$ & $4.29\pm0.07$ & $0.263\pm0.006$ \\
91 & 3.044   & $5.7\pm0.3 $ & $1.24\pm0.05$ & $0.47\pm0.02$ & $3086\pm22$ & $1.65\pm0.04$ & $0.226\pm0.011$ \\
92 & 3.119   & $5.3\pm0.3 $ & $3.26\pm0.07$ & $1.24\pm0.03$ & $3761\pm46$ & $4.10\pm0.11$ & $0.241\pm0.008$ \\
93 & 3.046   & $0.5\pm0.0 $ & $0.53\pm0.06$ & $0.20\pm0.02$ & $10286\pm944$ & $11.68\pm2.46$ & $0.014\pm0.003$ \\
94 & 3.183   & $5.7\pm0.3 $ & $2.36\pm0.05$ & $0.90\pm0.02$ & $4059\pm48$ & $4.02\pm0.11$ & $0.177\pm0.006$ \\
95 & 3.036   & $2.0\pm0.1 $ & $3.23\pm0.07$ & $1.23\pm0.03$ & $5170\pm243$ & $7.70\pm0.77$ & $0.127\pm0.012$ \\
96 & 3.117   & $8.6\pm0.4 $ & $2.92\pm0.06$ & $1.11\pm0.02$ & $3392\pm26$ & $3.15\pm0.06$ & $0.281\pm0.008$ \\
97 & 3.097   & $7.4\pm0.4 $ & $7.47\pm0.07$ & $2.85\pm0.03$ & $5376\pm356$ & $12.99\pm1.84$ & $0.174\pm0.023$ \\
98 & 3.075   & $4.3\pm0.2 $ & $1.41\pm0.05$ & $0.54\pm0.02$ & $3854\pm46$ & $2.76\pm0.08$ & $0.154\pm0.007$ \\
99 & 3.109   & $4.0\pm0.2 $ & $1.40\pm0.07$ & $0.53\pm0.03$ & $3262\pm38$ & $1.97\pm0.07$ & $0.215\pm0.013$ \\
100 & 3.055  & $2.6\pm0.1 $ & $3.53\pm0.06$ & $1.34\pm0.02$ & $9271\pm363$ & $25.96\pm2.13$ & $0.041\pm0.003$ \\
101 & 3.080  & $7.3\pm0.4 $ & $5.06\pm0.06$ & $1.93\pm0.02$ & $3288\pm28$ & $3.96\pm0.07$ & $0.387\pm0.008$ \\
102 & 3.196  & $8.2\pm0.4 $ & $4.53\pm0.07$ & $1.73\pm0.03$ & $3773\pm87$ & $4.91\pm0.23$ & $0.279\pm0.014$ \\
103 & 3.016  & $9.4\pm0.5 $ & $4.55\pm0.07$ & $1.73\pm0.02$ & $3510\pm20$ & $4.26\pm0.06$ & $0.323\pm0.006$ \\
104 & 3.140  &$13.0\pm0.7 $ & $3.04\pm0.05$ & $1.16\pm0.02$ & $3163\pm39$ & $2.79\pm0.07$ & $0.329\pm0.01$ \\
105 & 3.193  & $3.4\pm0.2 $ & $2.51\pm0.07$ & $0.96\pm0.03$ & $6504\pm198$ & $10.68\pm0.69$ & $0.071\pm0.005$ \\
106 & 3.191  & $3.0\pm0.2 $ & $3.24\pm0.06$ & $1.23\pm0.02$ & $4408\pm88$ & $5.61\pm0.23$ & $0.175\pm0.008$ \\
107 & 3.091  & $6.0\pm0.3 $ & $5.66\pm0.07$ & $2.16\pm0.03$ & $3946\pm47$ & $6.05\pm0.15$ & $0.283\pm0.008$ \\
108 & 3.086  & $5.8\pm0.3 $ & $5.70\pm0.06$ & $2.17\pm0.02$ & $4493\pm114$ & $7.87\pm0.41$ & $0.219\pm0.011$ \\
109 & 3.067  & $5.2\pm0.3 $ & $6.03\pm0.06$ & $2.30\pm0.02$ & $5238\pm92$ & $11.02\pm0.40$ & $0.166\pm0.006$ \\
110 & 3.120  & $1.7\pm0.1 $ & $1.54\pm0.07$ & $0.59\pm0.02$ & $5589\pm207$ & $6.09\pm0.49$ & $0.077\pm0.007$ \\
111 & 3.033  & $3.9\pm0.2 $ & $4.37\pm0.07$ & $1.67\pm0.03$ & $3167\pm33$ & $3.40\pm0.08$ & $0.389\pm0.01$ \\
112 & 3.111  & $4.4\pm0.2 $ & $2.32\pm0.06$ & $0.88\pm0.02$ & $2962\pm34$ & $2.12\pm0.06$ & $0.330\pm0.012$ \\
113 & 3.105  & $4.6\pm0.2 $ & $4.85\pm0.09$ & $1.85\pm0.03$ & $7382\pm324$ & $19.49\pm1.80$ & $0.075\pm0.007$ \\
114 & 3.142  & $2.5\pm0.1 $ & $1.65\pm0.05$ & $0.63\pm0.02$ & $3530\pm66$ & $2.51\pm0.11$ & $0.198\pm0.01$ \\
115 & 3.117  & $3.3\pm0.2 $ & $0.51\pm0.06$ & $0.20\pm0.02$ & $7592\pm378$ & $6.28\pm0.77$ & $0.025\pm0.004$ \\
116 & 3.070  & $8.0\pm0.4 $ & $3.73\pm0.07$ & $1.42\pm0.03$ & $4936\pm253$ & $7.58\pm0.82$ & $0.149\pm0.016$ \\
117 & 3.172  & $2.4\pm0.1 $ & $1.24\pm0.06$ & $0.47\pm0.02$ & $5155\pm174$ & $4.62\pm0.34$ & $0.081\pm0.007$ \\
118 & 3.085  & $2.2\pm0.1 $ & $5.72\pm0.07$ & $2.18\pm0.03$ & $9254\pm373$ & $33.43\pm2.81$ & $0.052\pm0.004$ \\
119 & 3.190  & $2.9\pm0.1 $ & $4.59\pm0.07$ & $1.75\pm0.03$ & $5232\pm185$ & $9.50\pm0.70$ & $0.146\pm0.011$ \\
120 & 3.407  & $2.8\pm0.1 $ & $0.47\pm0.07$ & $0.18\pm0.02$ & $6354\pm817$ & $4.17\pm1.27$ & $0.034\pm0.01$ \B \\
\hline
\end{tabular}
\tablefoot{
\tablefoottext{a}{Quasar systemic redshift. The intrinsic uncertainty on this value is $\Delta z\sim0.007$.}
}

\end{table*}

In this appendix we describe how we obtained 
the FWHM values of the broad \lya and \ion{C}{iv} line emissions and the monochromatic luminosity of the quasar at rest-frame 1350\,\AA, 
which are used to compute the black hole mass, Eddington ratio and bolometric luminosity in Section~\ref{sec:qso_phys_prop} using Equations~\ref{eq:logMBH},~\ref{eq:ldaEdd}~and~\ref{eq:Lbol}, respectively.

\begin{figure}
    \centering
    \includegraphics[width=\linewidth]{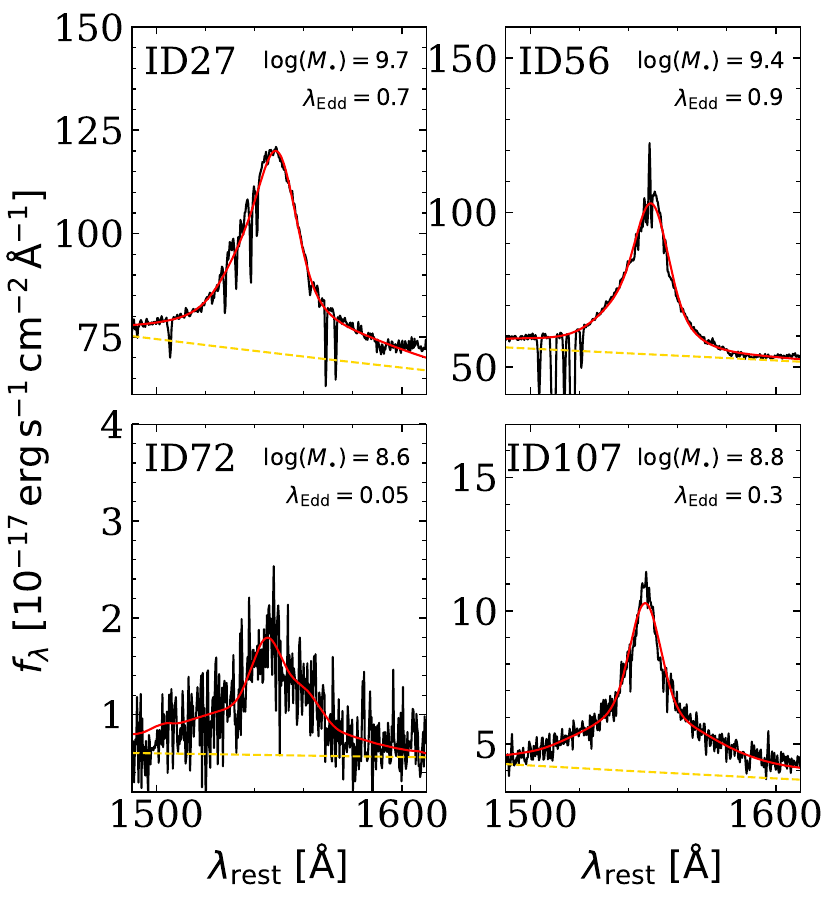}
    \caption{Four examples of \ion{C}{iv} and continuum fit using \textsc{PyQSOFit}. Each panel shows the 1-D spectrum (black) of one targeted quasar (ID in the top-left corner) together with the estimated continuum (yellow dashed line) and the best fit (red). The black hole mass and Eddington ratio obtained from this fit are listed in the top-right corner (see text for details). }
    \label{fig:CIVfit}
\end{figure}

To obtain those quantities we use the standard technique of modeling the 1-D quasar spectrum with a linear combination of a pseudo-continuum, broad and narrow emission lines (e.g., \citealt{Shen2019}). 
Specifically, we fit the spectra with 
\textsc{PyQSOFit} (\citealt{Guo2018}) which gives in output the aforementioned quantities. 
We fit for the following emission lines: Ly$\alpha$, \ion{N}{v}$\lambda1240\AA$, \ion{C}{iv}$\lambda1549\AA$, \ion{C}{iii}$\lambda1909\AA$.
For these lines we use the same fitting parameters and number of Gaussians as specified in previous works (see Table 1 in \citealt{Rakshit2020}).
Upon visual inspection of each obtained model spectrum, we (i) introduced the fit of \ion{Si}{iv}$\lambda1397\AA,1402\AA$, (ii) mask absorption features close or on top of the \ion{C}{iv} broad emission line, and (iii) refine the wavelength ranges for the continuum estimation, when needed and  re-run the fitting. 
The values obtained for ${\rm FWHM(\ion{C}{iv})}$ and $\lambda L_{\lambda}(1350\,\AA)$ 
for the full sample are reported in Tables~\ref{tab:QSOfit-bright}~and~\ref{tab:QSOfit-faint}, while we show four examples of our fit of the \ion{C}{iv} range in Figure~\ref{fig:CIVfit}. The same tables list also the black hole mass, Eddington ratio and bolometric luminosity computed in Section~\ref{sec:qso_phys_prop}. The values reported in the table are accompanied by their statistical uncertainties, which are much smaller than the potential systematic errors mentioned in that section.

\begin{figure}
    \centering
    \includegraphics[width=\linewidth]{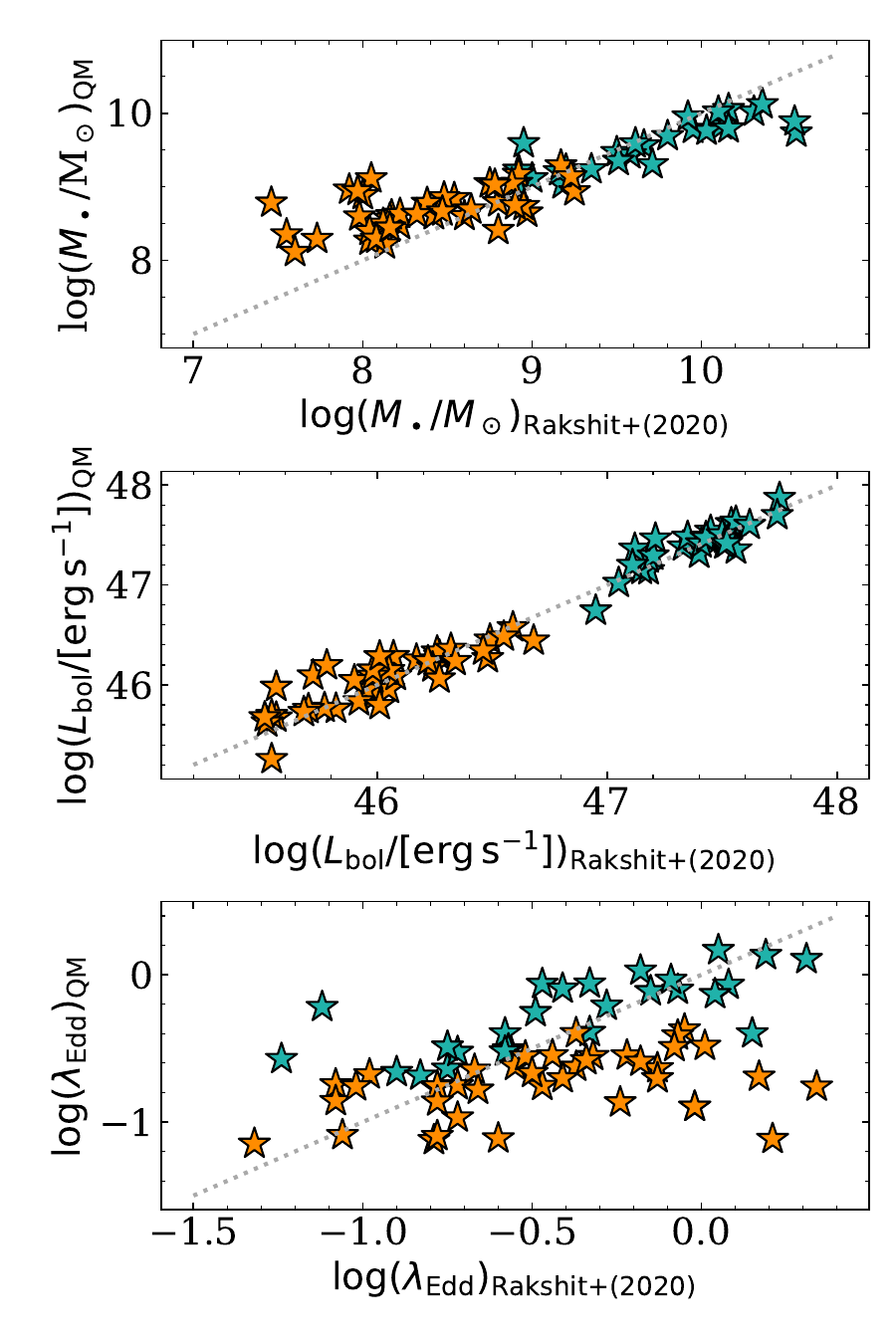}
    \caption{Comparison of SMBH properties with respect to a previous work. The obtained $M_{\rm BH}$, $L_{\rm bol}$ and $\lambda_{\rm Edd}$ are compared with those found in \citet{Rakshit2020} for a subsample of the QSOMUSEUM objects (see text for details). Datapoints for bright quasars studied in QSO MUSEUM I are shown in green, while those for fainter quasars added in this work are shown in orange. In each panel, the dotted grey line indicates the one to one relation.}
    \label{fig:comp}
\end{figure}

To verify the obtained $M_{\rm BH}$ and $\lambda_{\rm Edd}$, we compared them with the values from the automated fit of \citet{Rakshit2020} for the sources in common between the two studies and cataloged as good fits in that work. 
Figure~\ref{fig:comp} shows the result of this comparison.
The largest differences are found at the lowest and highest end of the black hole mass distribution, and the low end of the bolometric luminosities. These differences are due to the fact that \citet{Rakshit2020} fit >500000 sources and hence visual inspection of all the obtained models was not possible, while we improve the fit of several sources after visual inspection. 

\begin{figure}
    \centering
    \includegraphics[width=\linewidth]{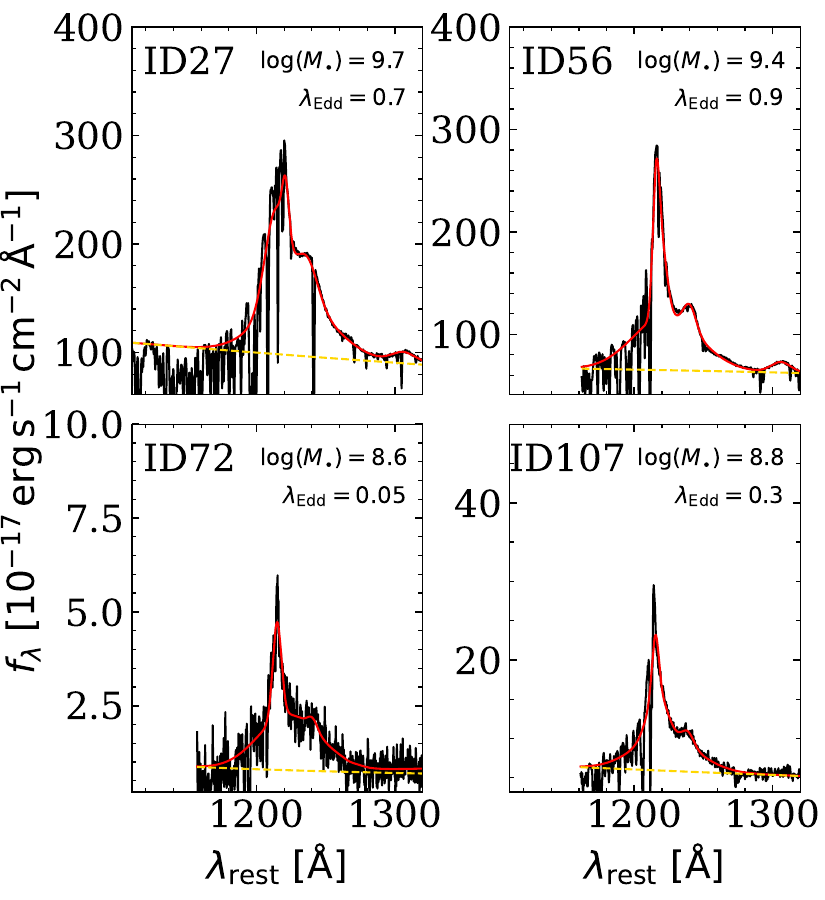}
    \caption{Four examples of \lya and continuum fit using \textsc{PyQSOFit}. Similar to figure~\ref{fig:CIVfit}, but for the \lya emission line of each quasar.}
    \label{fig:Lyafit}
\end{figure}

Regarding the calculation of the FWHM of the quasars' broad \lya emission, we used once again the values from the aforementioned fit using \textsc{PyQSOFit}. We show examples of the fit of the \lya line in Figure~\ref{fig:Lyafit}. As usually done in the literature (e.g., \citealt{Rakshit2020}), we use three Gaussians for the fit of the \lya line. We note that this approach might result in overestimation of the FWHM for the faint objects, for which the best fit does not pass through the peak of the observed emission (see e.g. ID 72 and 107 in Figure\ref{fig:Lyafit}). We stress that these values do not represent the FWHM of the intrinsic \lya emission, but of the observed line, which could be also affected by absorption from the intervening intergalactic medium (e.g., \citealt{Greig2024}).

\FloatBarrier

\section{QSO MUSEUM I}\label{sec:QSOMUSEUMI}

In this Appendix we report on all the observations and analysis of the bright quasar sample already published in QSO~MUSEUM~I \citep{FAB2019}.

\begin{table*}[!htbp]
\footnotesize \centering
\caption{Properties of the quasar and extended Ly$\alpha$ emission of the QSO MUSEUM I}
\label{tab:properties_QSOMUSEUM}
\begin{tabular}{|c|c|c|c|c|c|c|c|c|c|c|}
\hline
\T
ID & z$_{\rm Ly\alpha;peak}^{\rm QSO}$\tablefootmark{(a)} & z$_{\rm Ly\alpha;peak}^{\rm Neb}$\tablefootmark{(a)} & SB limit\tablefootmark{(b)} &
$L_{\rm Ly\alpha;peak}^{\rm QSO}$ & Luminosity & Area & ${\Delta v_{\rm gauss}}$ & ${\Delta v_{\rm m1}}$ & ${\rm FWHM_{gauss}}$ & $\Delta \lambda_{\rm m2}$ \\
 &  &  & [$10^{-18}\,{\rm cgs}$] & $10^{42}\,{ \rm erg\,s^{-1}\,\AA^{-1} }$ & $10^{42}\,{ \rm erg\,s^{-1} }$ & arcsec$^2$\ &
 ${\rm km\,s^{-1}}$ & ${\rm km\,s^{-1}}$ & ${\rm km\,s^{-1}}$ & ${\rm km\,s^{-1}}$ \B \\
\hline
\T
1 & 3.166 & 3.154 & 2.02 & 54.31$\pm$0.14 & 124.68$\pm$2.59 & 45$\pm$4 & 238$\pm$61 & 204$\pm$22 & 1950$\pm$146 & 884$\pm$13 \\
2 & 3.161 & 3.164 & 2.07 & 61.3$\pm$0.21 & 88.27$\pm$2.78 & 74$\pm$7 & 127$\pm$47 & 37$\pm$26 & 1478$\pm$113 & 482$\pm$25 \\
3 & 3.115 & 3.123 & 1.64 & 91.24$\pm$0.21 & 166.09$\pm$2.93 & 81$\pm$8 & 345$\pm$67 & 304$\pm$19 & 1974$\pm$163 & 925$\pm$11 \\
4 & 3.228 & 3.23 & 1.72 & 43.19$\pm$0.17 & 97.86$\pm$2.25 & 75$\pm$8 & 249$\pm$29 & 265$\pm$15 & 1110$\pm$68 & 458$\pm$10 \\
5 & 3.325 & 3.328 & 1.84 & 35.78$\pm$0.16 & 77.9$\pm$2.2 & 57$\pm$6 & 624$\pm$35 & 568$\pm$22 & 1351$\pm$85 & 519$\pm$18 \\
6 & 3.041 & 3.04 & 2.45 & 69.74$\pm$0.21 & 17.86$\pm$1.06 & 8$\pm$1 & 956$\pm$164 & 974$\pm$70 & 2006$\pm$391 & 1009$\pm$36 \\
7 & 3.118 & 3.121 & 2.11 & 79.65$\pm$0.2 & 105.13$\pm$2.58 & 39$\pm$4 & 369$\pm$108 & 359$\pm$27 & 1972$\pm$257 & 903$\pm$16 \\
8 & 3.125 & 3.131 & 1.23 & 71.06$\pm$0.09 & 162.74$\pm$1.79 & 137$\pm$14 & 511$\pm$17 & 487$\pm$8 & 1254$\pm$40 & 512$\pm$6 \\
9 & 3.306 & 3.313 & 2.19 & 101.44$\pm$0.27 & 152.41$\pm$3.33 & 66$\pm$7 & 444$\pm$37 & 441$\pm$21 & 1707$\pm$89 & 694$\pm$15 \\
10 & 3.245 & 3.245 & 1.92 & 88.05$\pm$0.19 & 322.09$\pm$3.39 & 161$\pm$16 & -75$\pm$20 & -119$\pm$8 & 1387$\pm$46 & 572$\pm$6 \\
11 & 3.068 & 3.1 & 2.03 & 9.87$\pm$0.1 & 40.12$\pm$1.04 & 40$\pm$4 & 2403$\pm$17 & 2415$\pm$11 & 752$\pm$39 & 277$\pm$9 \\
12 & 3.386 & 3.396 & 2.00 & 48.53$\pm$0.18 & 62.88$\pm$1.93 & 33$\pm$3 & 641$\pm$23 & 647$\pm$20 & 1135$\pm$54 & 449$\pm$15 \\
13 & 3.168 & 3.169 & 2.13 & 51.62$\pm$0.19 & 308.56$\pm$3.42 & 269$\pm$27 & 218$\pm$11 & 217$\pm$6 & 932$\pm$27 & 382$\pm$5 \\
14 & 3.139 & 3.127 & 2.90 & 18.7$\pm$0.15 & 42.61$\pm$1.88 & 20$\pm$2 & -123$\pm$93 & 24$\pm$50 & 1965$\pm$220 & 775$\pm$40 \\
15 & 3.145 & 3.142 & 1.87 & 118.43$\pm$0.18 & 235.68$\pm$3.01 & 104$\pm$10 & 47$\pm$17 & 68$\pm$9 & 1422$\pm$43 & 562$\pm$8 \\
16 & 3.186 & 3.19 & 1.94 & 21.83$\pm$0.14 & 73.87$\pm$2.4 & 56$\pm$6 & 170$\pm$66 & 127$\pm$33 & 1756$\pm$154 & 803$\pm$21 \\
17 & 3.349 & 3.351 & 1.79 & 22.27$\pm$0.13 & 142.92$\pm$2.23 & 115$\pm$12 & 76$\pm$13 & 72$\pm$8 & 876$\pm$30 & 354$\pm$5 \\
18 & 3.278 & 3.289 & 2.39 & 76.47$\pm$0.19 & 296.53$\pm$3.29 & 98$\pm$10 & 892$\pm$15 & 863$\pm$8 & 1290$\pm$34 & 512$\pm$6 \\
19 & 3.215 & 3.223 & 1.90 & 45.49$\pm$0.19 & 182.49$\pm$3.08 & 116$\pm$12 & 759$\pm$31 & 810$\pm$16 & 1687$\pm$74 & 607$\pm$14 \\
20 & 3.42 & 3.447 & 1.77 & 55.79$\pm$0.19 & 82.56$\pm$2.45 & 38$\pm$4 & 1090$\pm$97 & 1077$\pm$33 & 1833$\pm$234 & 953$\pm$17 \\
21 & 3.22 & 3.222 & 1.99 & 151.73$\pm$0.21 & 238.44$\pm$3.09 & 102$\pm$10 & 55$\pm$8 & 96$\pm$7 & 1004$\pm$20 & 400$\pm$5 \\
22 & 3.23 & 3.231 & 2.39 & 44.51$\pm$0.19 & 151.53$\pm$2.72 & 73$\pm$7 & 81$\pm$21 & 65$\pm$14 & 1285$\pm$49 & 531$\pm$10 \\
23 & 3.145 & 3.137 & 1.86 & 28.35$\pm$0.15 & 70.11$\pm$1.58 & 46$\pm$5 & -300$\pm$23 & -331$\pm$16 & 1278$\pm$55 & 479$\pm$13 \\
24 & 3.056 & 3.057 & 1.44 & 96.6$\pm$0.1 & 191.62$\pm$2.04 & 86$\pm$9 & 104$\pm$30 & 123$\pm$11 & 1882$\pm$71 & 762$\pm$8 \\
25 & 3.343 & 3.353 & 1.83 & 40.02$\pm$0.18 & 75.19$\pm$2.18 & 40$\pm$4 & 488$\pm$79 & 453$\pm$31 & 1869$\pm$187 & 866$\pm$20 \\
26 & 3.179 & 3.177 & 15.65 & 706.79$\pm$2.05 & 886.44$\pm$23.8 & 54$\pm$5 & -81$\pm$84 & -70$\pm$29 & 1946$\pm$201 & 933$\pm$17 \\
27 & 3.333 & 3.333 & 1.74 & 68.11$\pm$0.2 & 116.49$\pm$2.39 & 67$\pm$7 & -1$\pm$26 & -16$\pm$15 & 1252$\pm$64 & 498$\pm$11 \\
28 & 3.359 & 3.355 & 1.64 & 25.79$\pm$0.14 & 47.76$\pm$1.77 & 34$\pm$3 & 66$\pm$42 & 68$\pm$34 & 1662$\pm$99 & 667$\pm$25 \\
29 & 3.222 & 3.223 & 1.73 & 29.94$\pm$0.16 & 22.0$\pm$1.62 & 16$\pm$2 & 306$\pm$107 & 339$\pm$83 & 1924$\pm$243 & 904$\pm$51 \\
30 & 3.36 & 3.36 & 7.57 & 119.46$\pm$0.39 & 312.9$\pm$5.88 & 22$\pm$2 & -93$\pm$70 & -128$\pm$20 & 1866$\pm$165 & 918$\pm$11 \\
31 & 3.205 & 3.197 & 1.87 & 49.95$\pm$0.2 & 73.82$\pm$2.29 & 47$\pm$5 & -156$\pm$51 & -139$\pm$30 & 1702$\pm$123 & 689$\pm$23 \\
32 & 3.099 & 3.102 & 1.80 & 17.53$\pm$0.11 & 82.97$\pm$2.02 & 89$\pm$9 & 235$\pm$25 & 205$\pm$19 & 1337$\pm$59 & 540$\pm$13 \\
33 & 3.134 & 3.139 & 1.69 & 88.57$\pm$0.15 & 149.06$\pm$2.65 & 59$\pm$6 & -34$\pm$19 & -68$\pm$14 & 1561$\pm$47 & 641$\pm$10 \\
34 & 3.243 & 3.254 & 1.56 & 11.49$\pm$0.10 & 90.7$\pm$1.86 & 89$\pm$9 & 698$\pm$17 & 683$\pm$14 & 1181$\pm$40 & 474$\pm$11 \\
35 & 3.245 & 3.245 & 1.57 & 57.68$\pm$0.11 & 162.81$\pm$2.14 & 72$\pm$7 & 161$\pm$14 & 205$\pm$10 & 1373$\pm$34 & 538$\pm$7 \\
36 & 3.191 & 3.204 & 1.64 & 32.56$\pm$0.11 & 166.12$\pm$2.3 & 134$\pm$13 & 841$\pm$15 & 847$\pm$10 & 1212$\pm$37 & 479$\pm$7 \\
37 & 3.128 & 3.127 & 1.29 & 74.65$\pm$0.09 & 178.67$\pm$1.86 & 92$\pm$9 & 17$\pm$18 & -13$\pm$7 & 1287$\pm$43 & 539$\pm$5 \\
38 & 3.354 & 3.352 & 1.85 & 82.73$\pm$0.2 & 123.59$\pm$2.28 & 18$\pm$2 & -116$\pm$88 & -216$\pm$19 & 1868$\pm$216 & 1058$\pm$9 \\
39 & 3.097 & 3.097 & 2.30 & 103.2$\pm$0.17 & 166.4$\pm$2.41 & 66$\pm$7 & 160$\pm$15 & 249$\pm$9 & 1145$\pm$39 & 415$\pm$7 \\
40 & 3.339 & 3.341 & 1.82 & 77.42$\pm$0.18 & 140.41$\pm$2.56 & 73$\pm$7 & 59$\pm$20 & 54$\pm$15 & 1516$\pm$48 & 594$\pm$12 \\
41 & 3.335 & 3.338 & 1.74 & 27.63$\pm$0.16 & 65.31$\pm$1.79 & 50$\pm$5 & 233$\pm$34 & 236$\pm$20 & 1269$\pm$79 & 521$\pm$14 \\
42 & 3.035 & 3.033 & 2.20 & 84.84$\pm$0.14 & 154.21$\pm$2.29 & 61$\pm$6 & 51$\pm$24 & 94$\pm$10 & 1286$\pm$57 & 539$\pm$7 \\
43 & 3.125 & 3.124 & 1.82 & 26.27$\pm$0.12 & 92.2$\pm$1.76 & 69$\pm$7 & 178$\pm$21 & 197$\pm$14 & 1294$\pm$51 & 513$\pm$11 \\
44 & 3.229 & 3.234 & 1.93 & 29.52$\pm$0.15 & 45.23$\pm$1.9 & 29$\pm$3 & 681$\pm$91 & 829$\pm$46 & 1918$\pm$216 & 804$\pm$34 \\
45 & 3.362 & 3.37 & 1.90 & 29.61$\pm$0.16 & 23.91$\pm$1.57 & 19$\pm$2 & 427$\pm$89 & 505$\pm$57 & 1515$\pm$211 & 545$\pm$50 \\
46 & 3.391 & 3.4 & 1.93 & 79.13$\pm$0.21 & 135.04$\pm$3.14 & 93$\pm$9 & 460$\pm$42 & 506$\pm$19 & 1471$\pm$98 & 616$\pm$14 \\
47 & 3.155 & 3.16 & 2.01 & 37.23$\pm$0.13 & 372.29$\pm$2.57 & 156$\pm$16 & 551$\pm$6 & 535$\pm$3 & 835$\pm$14 & 351$\pm$2 \\
48 & 3.208 & 3.218 & 1.66 & 37.01$\pm$0.14 & 102.48$\pm$2.03 & 89$\pm$9 & 839$\pm$25 & 794$\pm$14 & 1233$\pm$59 & 511$\pm$9 \\
49 & 3.135 & 3.14 & 1.59 & 14.62$\pm$0.09 & 81.02$\pm$1.58 & 92$\pm$9 & 286$\pm$18 & 321$\pm$13 & 1143$\pm$42 & 414$\pm$12 \\
50 & 3.214 & 3.206 & 1.96 & 33.94$\pm$0.15 & 444.95$\pm$2.99 & 199$\pm$20 & -622$\pm$8 & -613$\pm$4 & 1066$\pm$18 & 424$\pm$3 \\
51 & 3.134 & 3.138 & 2.12 & 51.28$\pm$0.13 & 47.28$\pm$1.24 & 47$\pm$5 & 227$\pm$12 & 203$\pm$10 & 668$\pm$29 & 269$\pm$8 \\
52 & 3.139 & 3.137 & 1.70 & 46.3$\pm$0.14 & 55.56$\pm$1.64 & 33$\pm$3 & 0$\pm$27 & 68$\pm$22 & 1425$\pm$65 & 558$\pm$17 \\
53 & 3.153 & 3.167 & 1.71 & 19.63$\pm$0.09 & 52.39$\pm$1.46 & 71$\pm$7 & 816$\pm$22 & 827$\pm$17 & 1053$\pm$50 & 450$\pm$11 \\
54 & 3.109 & 3.114 & 1.96 & 35.8$\pm$0.11 & 100.94$\pm$2.03 & 82$\pm$8 & 254$\pm$20 & 294$\pm$15 & 1333$\pm$47 & 514$\pm$11 \\
55 & 3.193 & 3.19 & 1.97 & 34.7$\pm$0.13 & 75.38$\pm$1.89 & 32$\pm$3 & -41$\pm$36 & -2$\pm$25 & 1830$\pm$87 & 728$\pm$18 \\
56 & 3.091 & 3.098 & 2.21 & 57.74$\pm$0.16 & 200.0$\pm$2.49 & 110$\pm$11 & 428$\pm$12 & 414$\pm$8 & 1081$\pm$28 & 436$\pm$5 \\
57 & 3.454 & 3.464 & 3.41 & 22.98$\pm$0.22 & 32.82$\pm$2.05 & 17$\pm$2 & 197$\pm$54 & 174$\pm$47 & 1313$\pm$126 & 559$\pm$34 \\
58 & 3.05 & 3.047 & 2.07 & 59.12$\pm$0.18 & 82.02$\pm$1.49 & 47$\pm$5 & -31$\pm$12 & 21$\pm$7 & 733$\pm$29 & 270$\pm$6 \\
59 & 3.156 & 3.16 & 2.60 & 37.98$\pm$0.17 & 122.36$\pm$2.83 & 63$\pm$6 & 21$\pm$50 & 17$\pm$22 & 1612$\pm$113 & 701$\pm$14 \\
60 & 3.179 & 3.18 & 2.07 & 44.48$\pm$0.13 & 100.63$\pm$2.0 & 42$\pm$4 & 206$\pm$22 & 238$\pm$14 & 1315$\pm$53 & 549$\pm$11 \\
61 & 3.061 & 3.067 & 2.18 & 58.97$\pm$0.16 & 113.63$\pm$2.8 & 79$\pm$8 & 586$\pm$54 & 571$\pm$28 & 1998$\pm$129 & 759$\pm$23 \\
\hline
\end{tabular}
\tablefoot{
\tablefoottext{a}{Peak \lya redshift. The intrinsic uncertainty on this value is $\Delta z\sim0.001$.} 
\tablefoottext{b}{$1\sigma$ SB limit within 1\,arcsec$^2$ and a 30\,\AA\ narrow-band centered at the observed \lya wavelength in units of $10^{-18}\,{\rm erg\,s^{-1}\,cm^{-2}\,arcsec^{-2}}$.}
}
\end{table*}

\begin{figure*}
    \centering\includegraphics[width=0.9\textwidth]{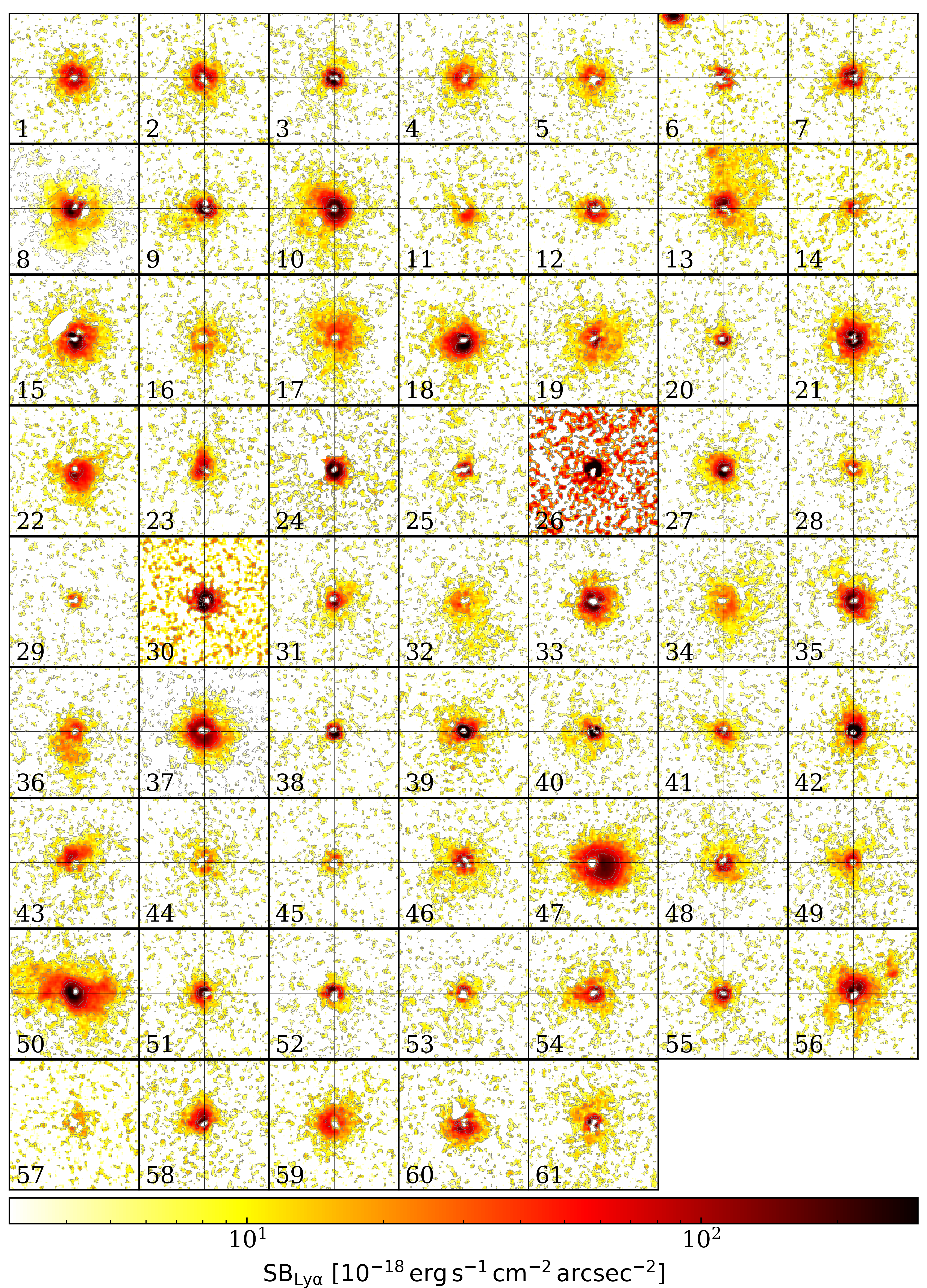}    
    \caption{\lya SB maps around 
    QSO MUSEUM I sample 
    after PSF- and continuum subtraction (see Section~\ref{sec:subtraction}) from 30\,\AA\ NBs centered at the peak \lya wavelength of the nebula. 
    All images 
    show maps with projected sizes of 
    $20"\times20"$ (about 150\,kpc\,x\,150\,kpc at the median redshift of the sample) centered on the quasar position. 
    In each map, the gray crosshair indicates the location of the quasar. The contours indicate levels of $[2, 4, 10, 20, 50]$ times the SB limit of each observation (see Table~\ref{tab:properties_QSOMUSEUM}).}
    \label{fig:SBmaps_QSOMUSEUM}
\end{figure*}
\begin{figure*}
    \centering
    \includegraphics[width=\linewidth]{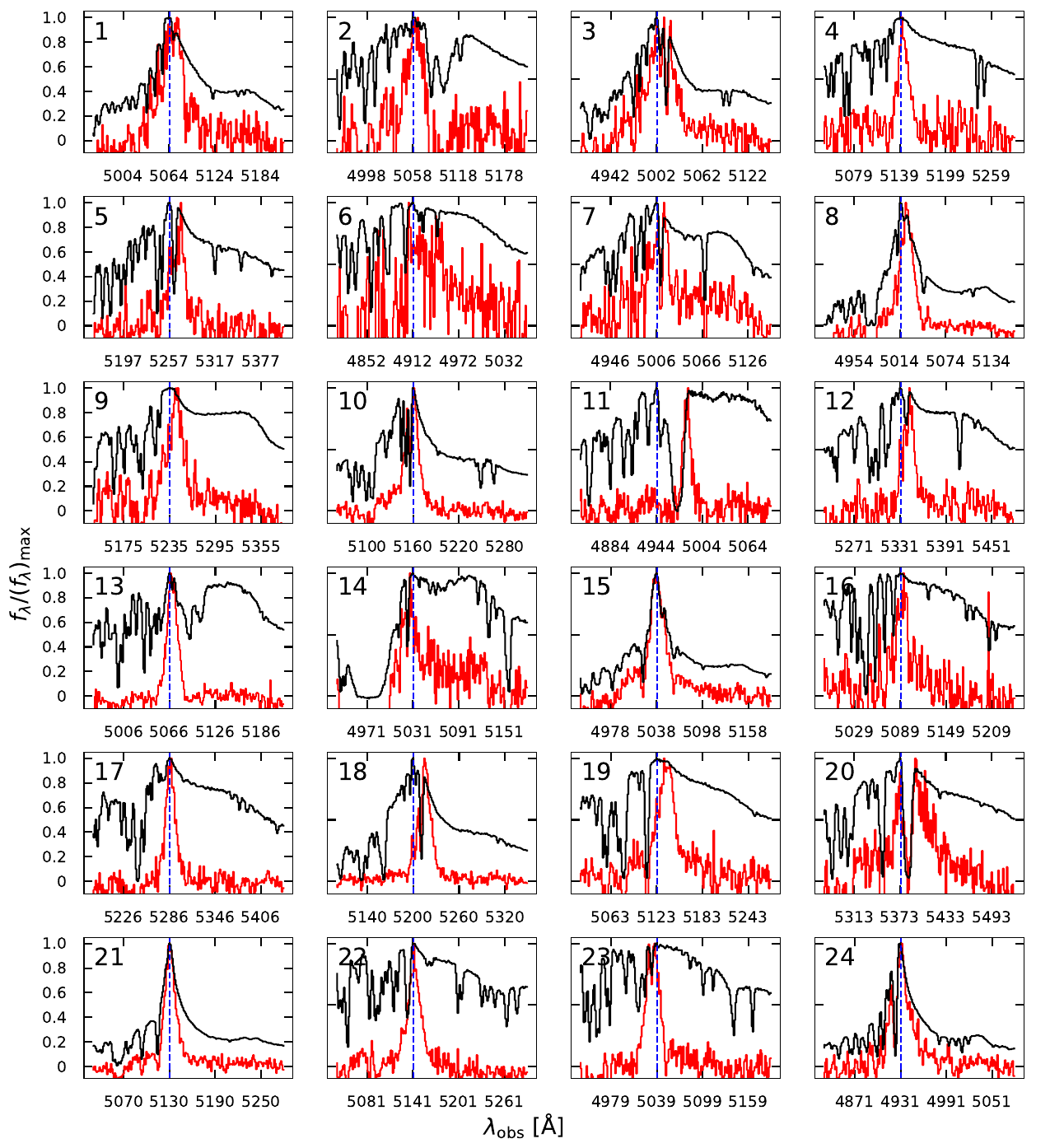}
    \caption{One-dimensional spectrum covering the 
    \lya line for ID 1-24 for the bright quasar sample presented in QSO MUSEUM I. The ID of each quasar is shown in the top left corner of each panel. Each panel shows a spectrum integrated from the MUSE datacube inside a 1.5'' radius aperture centered at the quasar location (black line) and the integrated spectrum of each detected nebula integrated from the PSF- and continuum-subtracted datacubes within the 2$\sigma$ isophotes from Figure~\ref{fig:SBmaps_QSOMUSEUM} (red line). The wavelength of the peak of the \lya emission of each quasar is indicated with a blue vertical line.     We mask the 1''$\times$1'' PSF normalisation region when we extract a spectrum using the PSF- and continuum-subtracted datacubes. All the spectra are shown as normalized to their peak emission to allow comparison to Figure~2 of \citet{FAB2019}.}
    \label{fig:spec_atlas01_bright}
\end{figure*}
\begin{figure*}
    \centering
    \includegraphics[width=\linewidth]{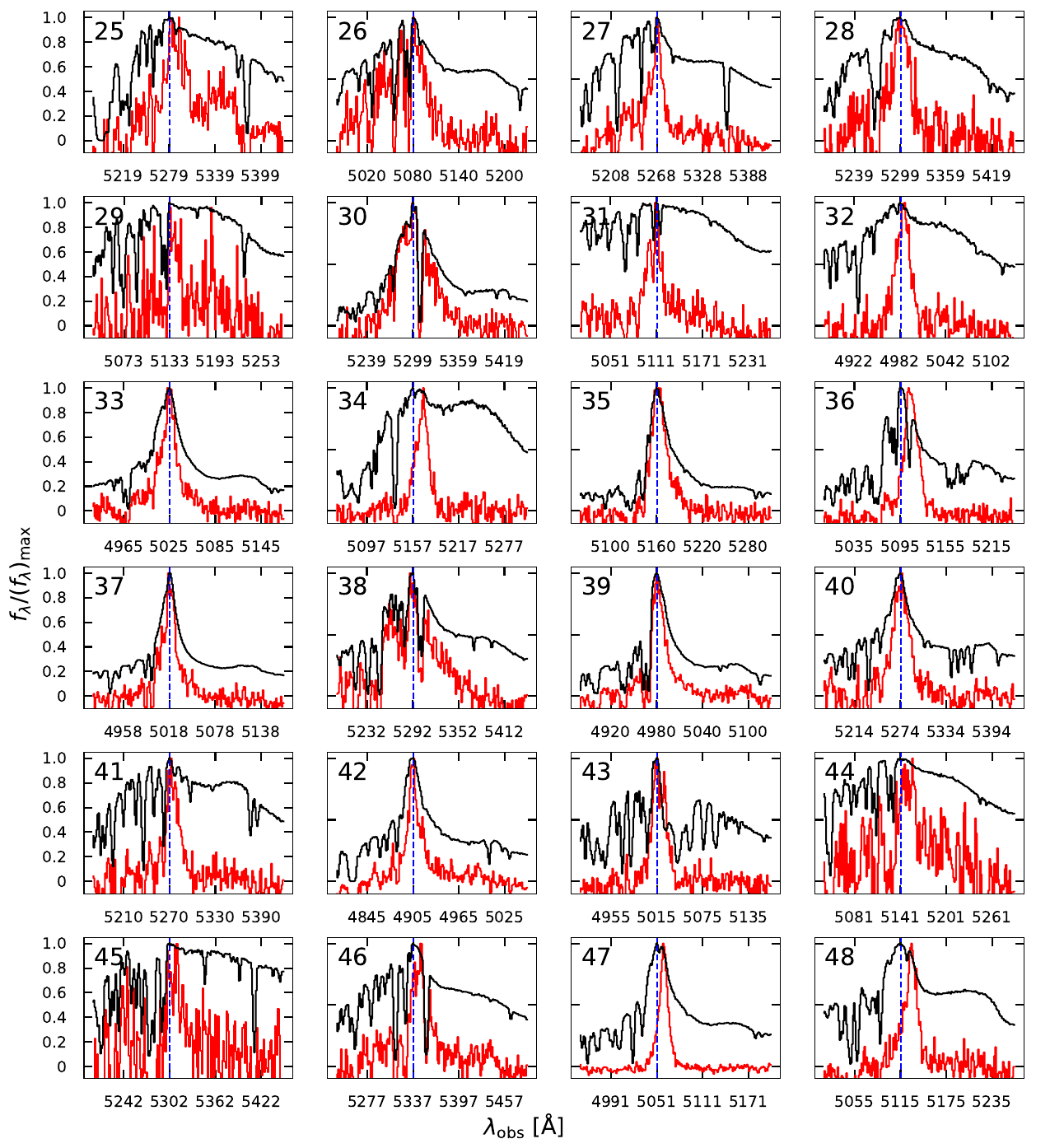}
    \caption{Same as Figure~\ref{fig:spec_atlas01_bright}, but for IDs 29-48.}
    \label{fig:spec_atlas02_bright}
\end{figure*}
\begin{figure*}
    \centering
    \includegraphics[width=\linewidth]{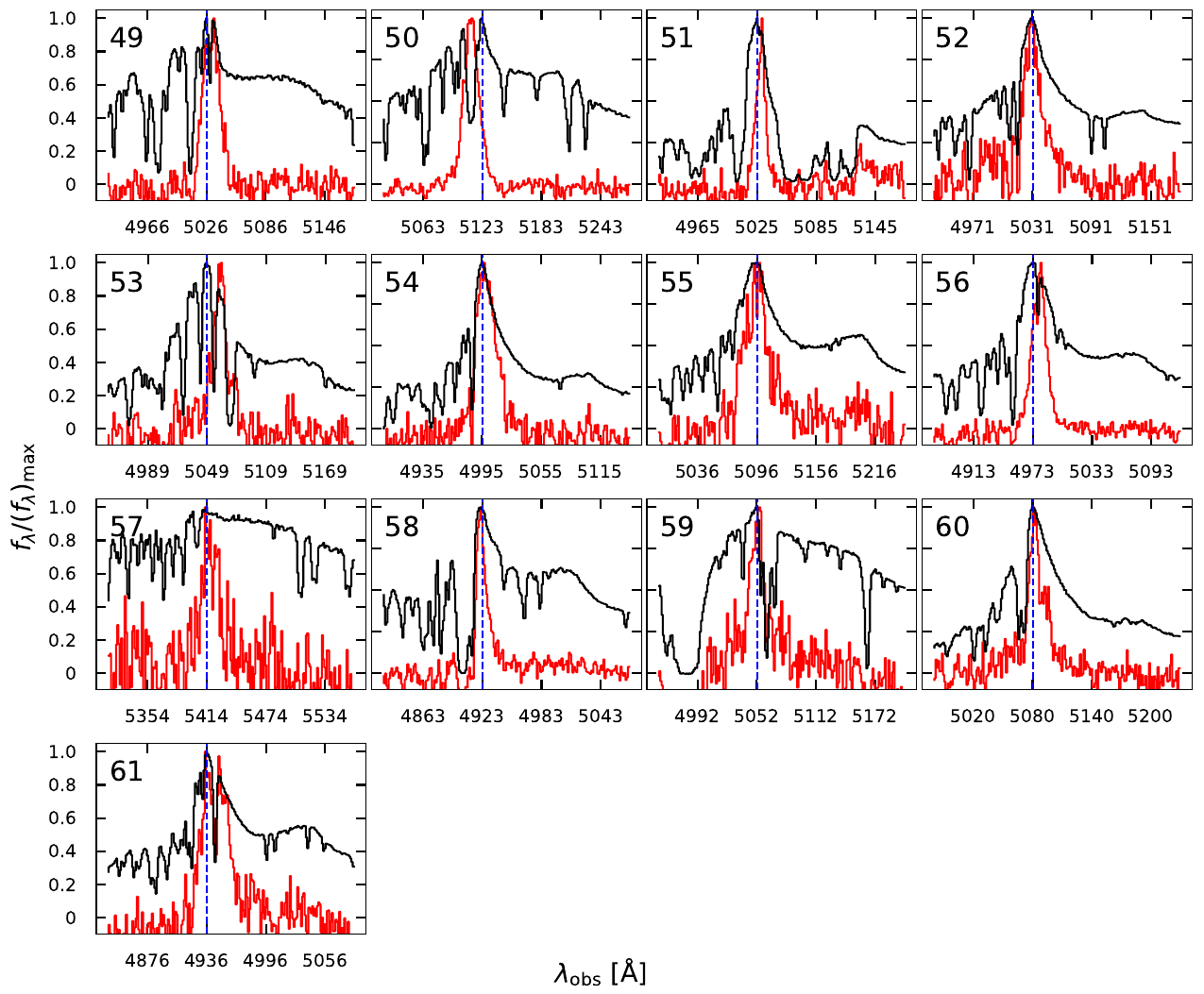}
    \caption{Same as Figure~\ref{fig:spec_atlas01_bright}, but for IDs 49-61.}
    \label{fig:spec_atlas03_bright}
\end{figure*}
\begin{figure*}
    \centering
    \includegraphics[width=0.9\textwidth]{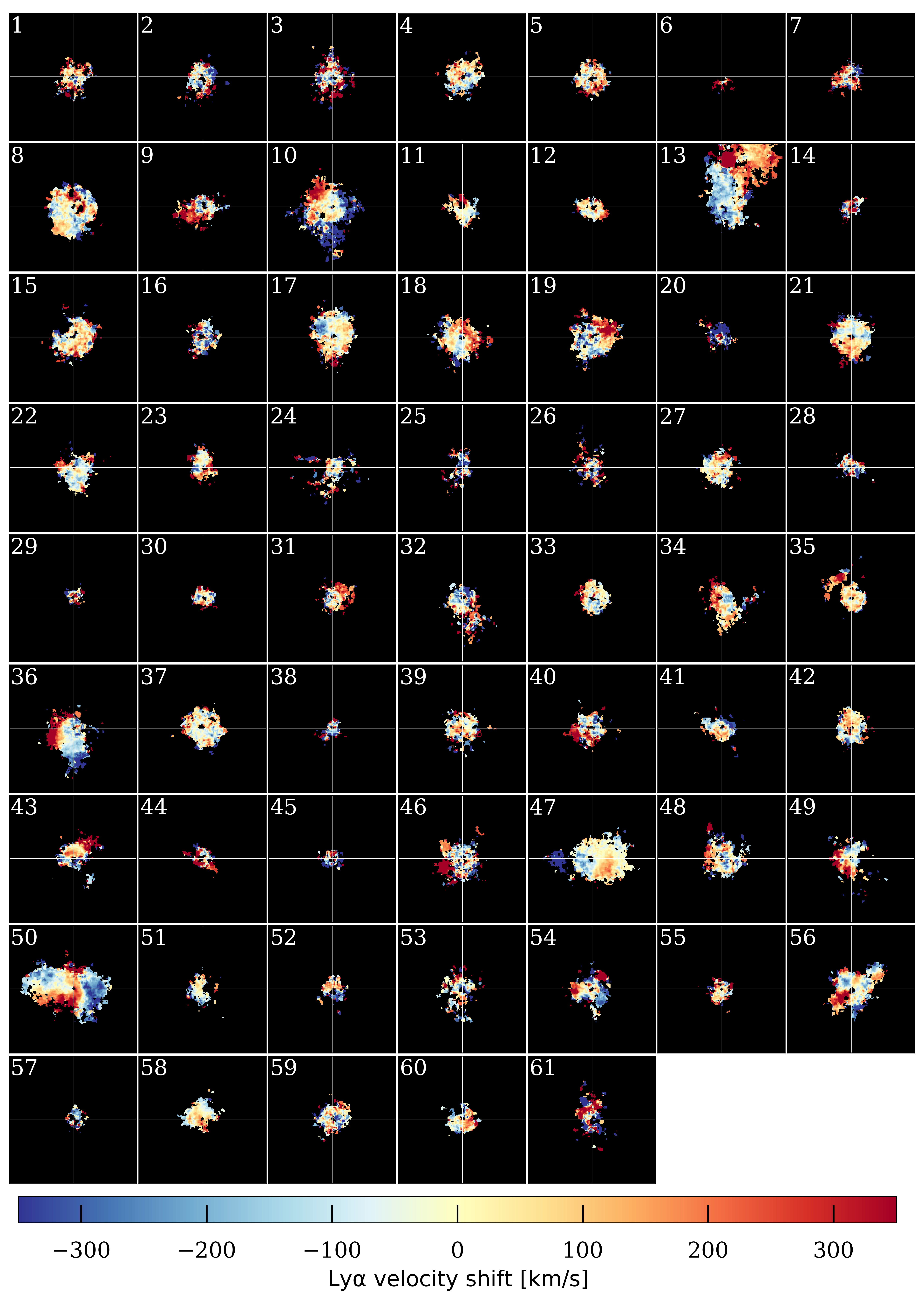}
    \caption{\lya Velocity shift maps for QSO MUSEUM I systems. Same as Figure~\ref{fig:velocity-shift} but for the bright sample.}
    \label{fig:velocity-shift_bright}
\end{figure*}

\begin{figure*}
    \centering
    \includegraphics[width=0.9\textwidth]{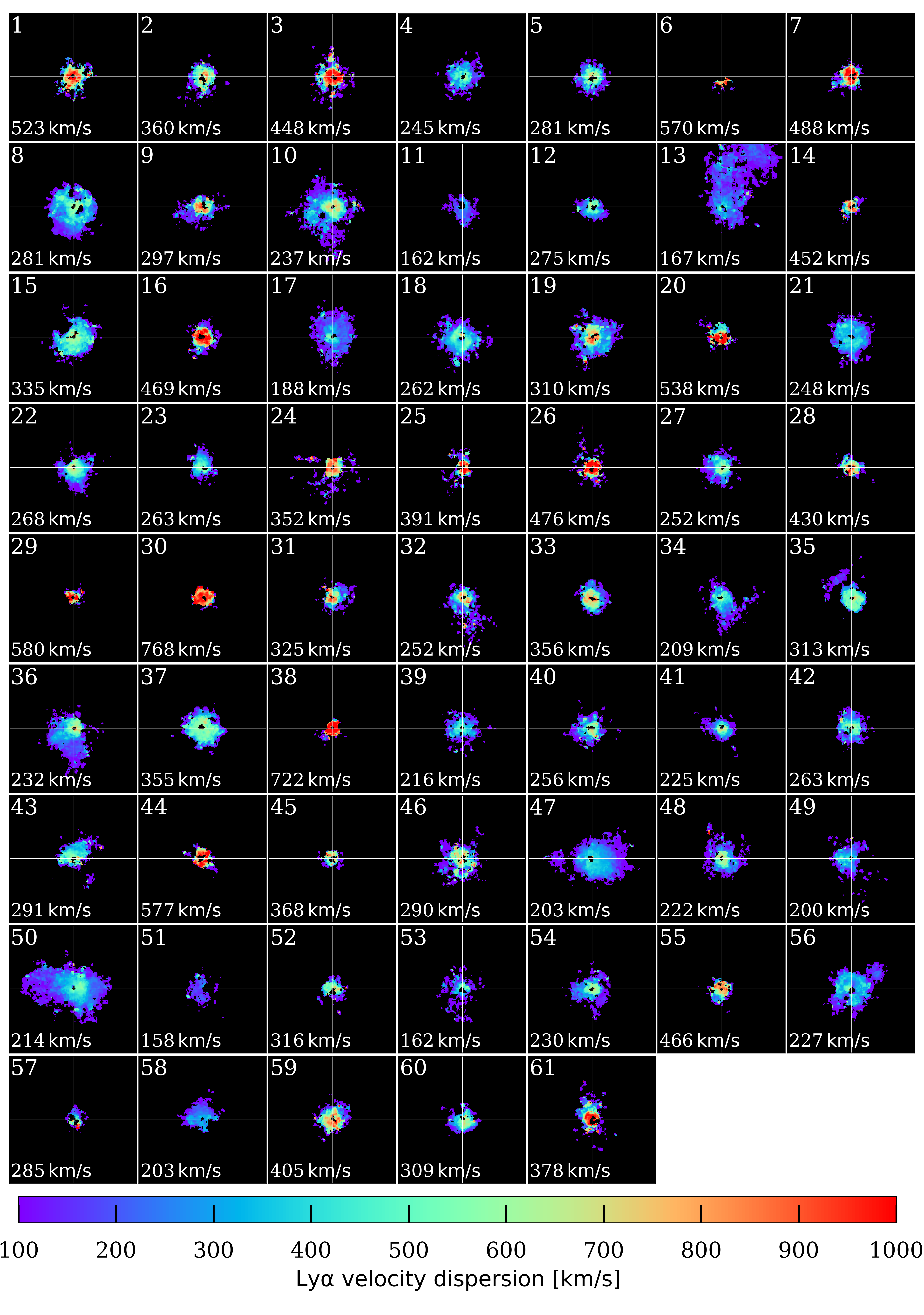}
    \caption{\lya Velocity dispersion maps for QSO MUSEUM I systems. Same as Figure~\ref{fig:velocity-dispersion} but for the bright sample.}
    \label{fig:velocity-dispersion_QSOMUSEUM}
\end{figure*}

\end{appendix}

\end{document}